\documentclass[twocolumn]{aastex63}
\usepackage{amsmath}
\usepackage{natbib}
\usepackage{float}
\usepackage{CJKutf8}
\usepackage{placeins}
\usepackage{subfigure}
\usepackage{graphicx}

\newcommand{\teff}{$T_\mathrm{eff}$}
\newcommand{\logg}{$\log{g}$}

\newcommand{\vsini}{$v\sin{i}$}
\newcommand{\kms}{km~s$^{-1}$}
\newcommand{\mjup}{M$_{\mathrm{Jup}}$}
\newcommand{\msun}{M$_{\odot}$}
\newcommand{\vovb}{$v_\mathrm{rot}$/$v_\mathrm{breakup}$}

\newcommand{\caltech}{Department of Astronomy, California Institute of Technology, Pasadena, CA 91125, USA}
\newcommand{\gps}{Division of Geological \& Planetary Sciences, California Institute of Technology, Pasadena, CA 91125, USA}
\newcommand{\ucsc}{Department of Astronomy \& Astrophysics, University of California, Santa Cruz, CA95064, USA}
\newcommand{\keck}{W. M. Keck Observatory, 65-1120 Mamalahoa Hwy, Kamuela, HI, USA}
\newcommand{\ucla}{Department of Physics \& Astronomy, 430 Portola Plaza, University of California, Los Angeles, CA 90095, USA}
\newcommand{\uclagps}{Department of Earth, Planetary, and Space Sciences, University of California, Los Angeles, CA 90095, USA}
\newcommand{\jpl}{Jet Propulsion Laboratory, California Institute of Technology, 4800 Oak Grove Dr.,Pasadena, CA 91109, USA}
\newcommand{\ucsd}{Center for Astrophysics and Space Sciences, University of California, San Diego, La Jolla, CA 92093}

\newcommand{\osu}{Department of Astronomy, The Ohio State University, 100 W 18th Ave, Columbus, OH 43210 USA}

\newcommand{\arizona}{James C. Wyant College of Optical Sciences, University of Arizona, Meinel Building 1630 E. University Blvd., Tucson, AZ. 85721}

\received{August 16, 2025}
\revised{December 6, 2025}
\accepted{February 4, 2026}

\submitjournal{AJ}

\shorttitle{KPIC Spin Survey}
\shortauthors{Hsu et al.}

\graphicspath{{./}{figures/}}

\begin{document}

\title{Distinct Rotational Evolution of Giant Planets and Brown Dwarf Companions}

\correspondingauthor{Chih-Chun Hsu}
\email{chsu@northwestern.edu}

\author[0000-0002-5370-7494]{Chih-Chun Hsu}
\affil{Center for Interdisciplinary Exploration and Research in Astrophysics (CIERA), Northwestern University,
1800 Sherman Ave, Evanston, IL, 60201, USA}

\author[0000-0003-0774-6502]{Jason J. Wang
\begin{CJK*}{UTF8}{gbsn}(王劲飞)\end{CJK*}}
\affil{Center for Interdisciplinary Exploration and Research in Astrophysics (CIERA), Northwestern University,
1800 Sherman Ave, Evanston, IL, 60201, USA}
\affil{Department of Physics and Astronomy, Northwestern University, 2145 Sheridan Rd, Evanston, IL 60208, USA}

\author[0000-0002-6618-1137]{Jerry W. Xuan}
\affiliation{\caltech}
\affiliation{\uclagps}
\affiliation{51 Pegasi b Fellow}

\author[0000-0003-0097-4414]{Yapeng Zhang}
\affiliation{\caltech}
\affiliation{51 Pegasi b Fellow}

\author[0000-0003-2233-4821]{Jean-Baptiste Ruffio}
\affiliation{\ucsd}
\affiliation{\caltech}

\author[0000-0002-8895-4735]{Dimitri Mawet}
\affiliation{\caltech}
\affiliation{\jpl}

\author[0000-0002-1392-0768]{Luke Finnerty}
\affiliation{\ucla}

\author[0000-0001-9708-8667]{Katelyn Horstman}
\affiliation{\caltech}

\author[0000-0003-1172-5755]{Julianne Cronin}
\affil{Center for Interdisciplinary Exploration and Research in Astrophysics (CIERA), Northwestern University,
1800 Sherman Ave, Evanston, IL, 60201, USA}
\affil{Department of Physics and Astronomy, Northwestern University, 2145 Sheridan Rd, Evanston, IL 60208, USA}

\author[0000-0002-6171-9081]{Yinzi Xin}
\affiliation{\caltech}

\author[0000-0003-1399-3593]{Ben Sappey}
\affiliation{\ucsd}

\author[0000-0002-1583-2040]{Daniel Echeverri}
\affiliation{\caltech}

\author[0000-0001-5213-6207]{Nemanja Jovanovic}
\affiliation{\caltech}

\author[0000-0002-6525-7013]{Ashley Baker}
\affiliation{\caltech}

\author{Randall Bartos}
\affiliation{\jpl}

\author[0000-0003-0787-1610]{Geoffrey A. Blake}
\affiliation{\gps}

\author[0000-0003-4737-5486]{Benjamin Calvin}
\affiliation{\caltech}
\affiliation{\ucla}

\author{Sylvain Cetre}
\affiliation{\keck}

\author[0000-0001-8953-1008]{Jacques-Robert Delorme}
\affiliation{\keck}

\author{Gregory W. Doppmann}
\affiliation{\keck}

\author[0000-0002-0176-8973]{Michael P. Fitzgerald}
\affiliation{\ucla}

\author[0000-0002-9936-6285]{Quinn M. Konopacky}
\affil{\ucsd}

\author[0000-0002-4934-3042]{Joshua Liberman}
\affiliation{Steward Observatory, University of Arizona, 933 N Cherry Ave, Tucson, AZ, USA 85719}
\affiliation{\arizona}

\author[0000-0002-2019-4995]{Ronald A. L\'opez}
\affiliation{\ucla}

\author[0000-0003-3165-0922]{Evan Morris} %
\affiliation{\ucsc}

\author{Jacklyn Pezzato}
\affiliation{\caltech}

\author{Tobias Schofield}
\affiliation{\caltech}

\author[0000-0001-6098-3924]{Andrew Skemer}
\affiliation{\ucsc}

\author[0000-0001-5299-6899]{J. Kent Wallace}
\affiliation{\jpl}

\author[0000-0002-4361-8885]{Ji Wang \begin{CJK*}{UTF8}{gbsn}(王吉)\end{CJK*}}
\affiliation{\osu}

\begin{abstract}
We present a rotational velocity ({\vsini}) survey of 32 stellar/substellar objects and giant planets using Keck/KPIC high-resolution spectroscopy, including 6 giant planets (2--7~{\mjup}) and 25 substellar/stellar companions (12--88~{\mjup}).
Adding companions with spin measurements from the literature, we construct a curated spin sample for 43 benchmark stellar/substellar companions and giant planets and 54 free-floating brown dwarfs and planetary mass objects.
We compare their spins, parameterized as fractional breakup velocities at 10~Myr, assuming constant angular momentum evolution.
We find the first clear evidence that giant planets exhibit distinct spins versus low-mass brown dwarf companions (10 to 40~{\mjup}) at 4--4.5~$\sigma$ significance assuming inclinations aligned with their orbits, while under randomly oriented inclinations the significance is at 1.6--2.1~$\sigma$.
Our findings hold when considering various assumptions about planets, and the mass ratio below 0.8\% gives a clean cut for rotation between giant planets and brown dwarf companions.
The higher fractional breakup velocities of planets can be interpreted as less angular momentum loss through circumplanetary disk braking during the planet formation phase.
Brown dwarf companions exhibit evidence of slower rotation compared to isolated brown dwarfs, while planets and planetary mass objects show similar spins.
Finally, our analysis of specific angular momentum versus age of 221 stellar/substellar objects below 0.1~{\msun} with spin measurements in the literature indicates that the
substellar objects of 5--40~{\mjup} retain much higher angular momenta compared to stellar and substellar objects of 40--100~{\mjup} after 10~Myr, when their initial angular momenta were set.

\end{abstract}

\keywords{Brown dwarfs (185), Exoplanet atmospheres (487), L dwarfs (894), Stellar rotation (1629), High resolution spectroscopy (2096), High angular resolution (2167)}

\section{Introduction} \label{sec:intro}

The known sample of directly imaged exoplanets is dominated by super-Jovians at separations of 2--250~au, reflecting the high-mass tail of planet formation \citep{Currie:2023ab}. 
At such large separations from their host stars, their formation can be through either core accretion \citep{Johansen:2017aa} (planet-like), disk instability \citep{Kuiper:1951aa, Cameron:1978aa} (planet-like or star-like), or cloud fragmentation \citep{Toomre:1964aa} (star-like).
These giant exoplanets also overlap with a population of brown dwarf companions at similar separations, likely formed through disk instability or cloud fragmentation.
These overlapping populations offer a unique opportunity to study planet formation and evolution since their atmospheres can be directly observed, with observational challenges due to the degeneracy between observational age, mass and luminosity of giant planets and brown dwarfs.

There have been several observational attempts to disentangle their formation, including measurements of orbital eccentricities \citep{Bowler:2020aa}, chemical abundances and metallicities \citep{Hoch:2023aa, Nasedkin:2024ab, Xuan:2024ab, Wang:2025aa}, and spins \citep{Bryan:2018aa, Bryan:2020ab}.
While the deuterium-burning mass limit ($\sim$13~{\mjup}) has been proposed to separate planets and brown dwarfs \citep{Burrows:2001aa, Baraffe:2003aa}, detections of deuterium in exoplanets and brown dwarfs are shown to be difficult and limited to only the nearest brown dwarfs \citep{Rowland:2024aa}.
\cite{Bowler:2020aa} showed evidence that giant planets are more likely to have low orbital eccentricities compared to brown dwarf companions, and the trends hold when taking into account different choices of Bayesian hyperpriors and updated smaller orbital eccentricities from observable-based priors \citep{Nagpal:2023aa, Do-O:2023aa} (See also an updated list of orbital eccentricities for our targets in Table~\ref{tab:sample}), and a recent stellar obliquity study shows tentative evidence that brown dwarf companions are more likely to show misalignments compared to giant planets \citep{Bowler:2023aa}.
Numerous spectroscopic studies have focused on measurements of atmospheric abundances and their isotopes \citep{Zhang:2021ae}.
\cite{Xuan:2024ab} conducted a high-resolution spectroscopic study of eight young substellar companions between 10--30~{\mjup}, and found that they all exhibit C/O ratios and metallicities consistent with their host stars. Three targets in their sample also show $^\mathrm{12}$CO/$^\mathrm{13}$CO isotope ratios consistent with the solar neighborhood.
The C/O ratios of directly imaged giant planets also appear to be solar/star-like, notably PDS 70 b \citep{Hsu:2024ac} (significantly different compared to the gas disk C/O $\sim$ 1; \citealp{Law:2024aa}), HR 8799 c \citep{Wang:2023aa}, and AF Lep b \citep{Denis:2025aa}.
These high-resolution spectroscopic studies are broadly consistent with C/O measurements with medium-resolution spectroscopy \citep{GRAVITY-Collaboration:2020aa, Molliere:2020aa, Ruffio:2021aa, Palma-Bifani:2023aa, Hoch:2023aa, Nasedkin:2024ab}.
While the metallicities for these planets might be enhanced, higher quality data and measurements of refractory species are required to validate these findings \citep{Wang:2023aa, Nasedkin:2024ab, Denis:2025aa, Wang:2025aa}.

Rotation could reflect the formation and evolution of giant planets and brown dwarf companions.
Theoretically, unmagnetized planets can spin up to 60--80\% of their breakup speed \citep{Dong:2021aa}. 
Planets are predicted to spin down due to disk braking \citep{Batygin:2018aa}, and later spin up after disks dissipate and their radii contract as they cool to around 1~R$_\mathrm{Jup}$ \citep{Baraffe:2003aa}, consistent with brown dwarf rotation surveys in very young clusters $<$10~Myr \citep{Scholz:2004aa, Scholz:2005aa, Herbst:2007aa}.
During the later stage of rotational evolution, previous works showed that rotation rates increase with age, consistent with constant angular momentum evolution, for both isolated substellar objects and companions reported in several spin surveys \citep{Zapatero-Osorio:2006aa, Schneider:2018aa, Bryan:2020ab, Vos:2022aa}.
Similar trends were found in the (mostly field) brown dwarfs over spectral types, and their spin-down timescales are longer for lower-mass and cooler spectral types \citep{Zapatero-Osorio:2006aa, Reiners:2008aa, Crossfield:2014aa, Tannock:2021aa, Hsu:2021aa}.

At the early stage of rotational evolution, spins of giant planets and substellar companions are predicted to vary as a function of mass.
The lower mass objects lose less angular momentum through disk braking as a result of the interplay between their weaker magnetic fields and weaker thermal ionization of the circumplanetary disk \citep{Batygin:2018aa, Ginzburg:2020aa}.
For the stellar/substellar regime, rotational periods of brown dwarfs in Taurus and Upper Scorpius indicate that high-mass brown dwarfs ($M$ $\gtrsim$50~{\mjup}) with disks rotate more slowly than those without a disk, and high-mass brown dwarfs rotate more quickly than stars \citep{Scholz:2015aa, Scholz:2018aa, Moore:2019aa}, consistent with the less efficient disk braking for lower-mass objects due to weaker magnetic field strengths \citep{Christensen:2009aa}.
However, prior to this study, high-resolution spectroscopic surveys were unable to confirm such mass-dependent rotation in the planetary-mass regime.
\cite{Bryan:2020ab} reported no correlations between the fractional breakup velocities and the mass ratios of 27 substellar and planetary-mass objects between 5--20~{\mjup} (while mostly 10--20~{\mjup}; see also \citealt{Xuan:2020aa}), and their sample exhibits rotation below $\sim$10\% breakup velocities.
\cite{Wang:2021aa} and \cite{Hsu:2024ab} compiled a representative set of rotation measurements from the literature {\vsini} and photometric rotational periods of 237 stellar/substellar objects below 0.1~{\msun} and found a tentative trend of the fractional breakup velocities in their samples, but the rotation versus mass trend remains inconclusive due to the small sample size of giant planets, in addition to relatively large measurement uncertainties.

In this study, we present a spin survey of 6 giant exoplanets (2--7~{\mjup}) and 26 stellar/substellar benchmark objects (12--88~{\mjup}) using Keck/KPIC high-resolution spectroscopy along with objects with literature spin measurements and show that giant planets exhibit distinct rotational evolution from brown dwarf companions.
In Section~\ref{sec:sample}, we outline our sample, observations and data reduction.
In Section~\ref{sec:ccf_detect}, we show our cross-correlation function detections.
In Section~\ref{sec:model}, we delineate our forward-modeling method to measure spins.
In Section~\ref{sec:vsini}, we evaluate our best-fit {\vsini} measurements.
In Section~\ref{sec:rotation}, we examine the spin versus mass trend in our sample and compare with the literature spin measurements.
We summarize our major findings in Section~\ref{sec:summarize}.

\section{Sample \& Observations \& Data Reduction} \label{sec:data}

\subsection{Sample} \label{sec:sample}

Our KPIC sample is summarized in Table~\ref{tab:sample}, including 32 exoplanets or stellar/substellar companions.
Our sample ranges from directly imaged exoplanets (AF Lep b, PDS 70 b, HR 8799 bcde) to lowest-mass stars slightly above the stellar/substellar boundary ($\sim$75--78.5~{\mjup}; \citealp{Baraffe:2003aa, Chabrier:2023aa}), with masses ranging from $\sim$2~M$_\mathrm{Jup}$ to $\sim$90~M$_\mathrm{Jup}$.
In this work, our definition of bona fide directly imaged exoplanets is determined by (1) mass $<$10~{\mjup} and (2) mass ratio $<$0.8\% (also see Section~\ref{sec:rotation}).
In addition to the Keck II telescope and site limiting targets to Dec $\gtrsim$ $-40^{\circ}$, our sample is limited by the KPIC design: the host stars are required to be brighter than $\sim$10~mag in H band, planets/companions must have projected separations $<4\arcsec$, the companion sensitivity of 1.3 $\times$ 10$^{-4}$ at 90~mas and 9.2 $\times$ 10$^{-6}$ at 420~mas separation from the star \citep{Wang:2024aa}.

\begin{figure}[ht]
    \centering
    \includegraphics[width=\columnwidth]{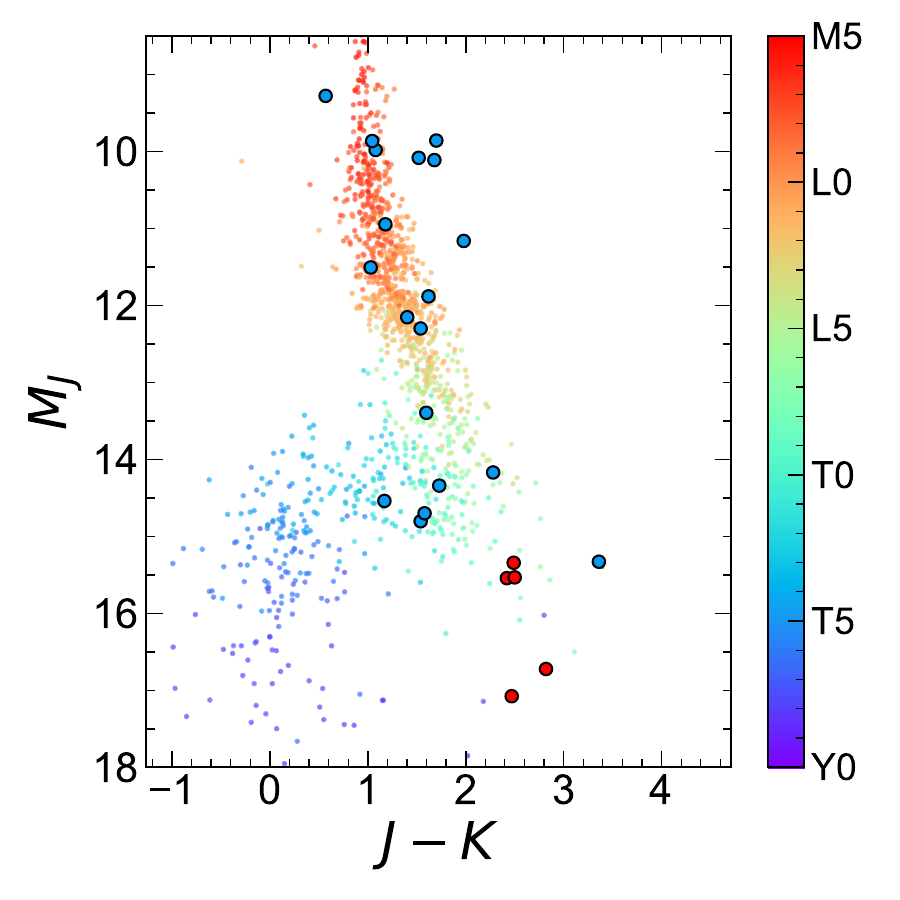}
    \caption{%
    Absolute $M_J$ versus $J-K$ color of our KPIC sample. In our KPIC sample, the directly imaged exoplanets are labeled in red circles, and the stellar/substellar companions are denoted in blue circles.
    The 100 pc M5--T9 dwarfs, compiled from the Ultracool Sheet \citep{Best:2024aa}, are depicted in small dots and color-coded by their spectral types.
    Note that PDS 70 b is not shown here since it lacks $K$-band photometry in the literature.
    Directly imaged planets and a few young substellar companions occupy the redder colors in the plot, and have been pointed out in the literature.
    }
    \label{fig:cmd}
\end{figure}

Our sample distributions are detailed below.
The ages in our sample span from 1--2~Myr (e.g., DH Tau B) to field ages (1--10~Gyr).
Twenty of our targets ($\sim$60\%) are young ($\leq$200~Myr), deviating from the field dwarf sequence as shown in Figure~\ref{fig:cmd}.
These targets range from mid-M ($\sim$3000~K) to L/T ($\sim$1200--1400~K) spectral types.
The directly imaged exoplanets are typically offset from the field dwarf sequence, occupying a similar color-magnitude space as the very red brown dwarf companion HD 206893 B ($J-K$ $\sim$ 3.4; \citealp{Ward-Duong:2021aa}).
Our targets are mostly below 30\% mass ratio of their host stars, except for the close stellar/substellar binary LP 349-25 AB (separation = 2.06 $\pm$ 0.04~au and mass ratio = 0.78; \citealp{Curiel:2024aa}).

\begin{figure}[ht]
    \centering
    \includegraphics[width=\columnwidth]{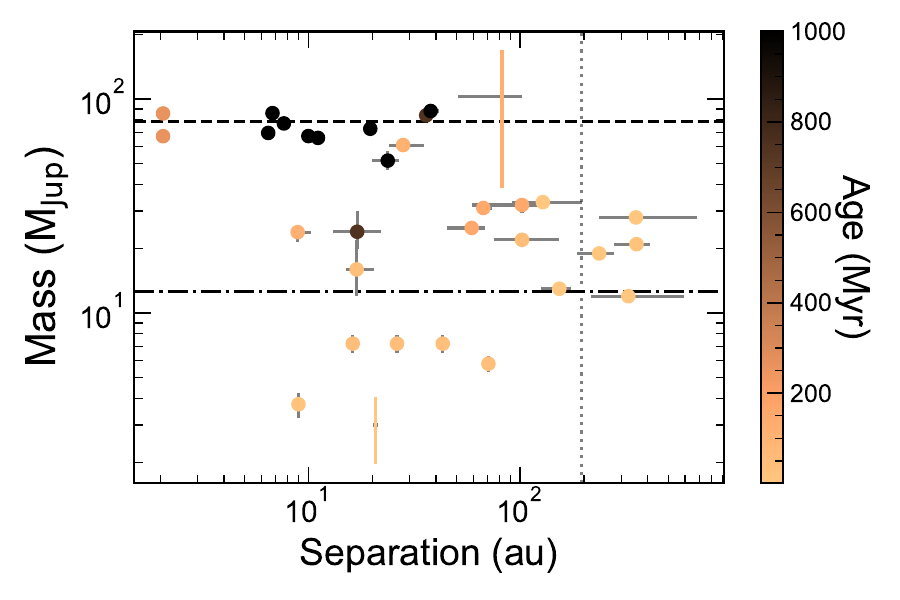}
    \includegraphics[width=\columnwidth]{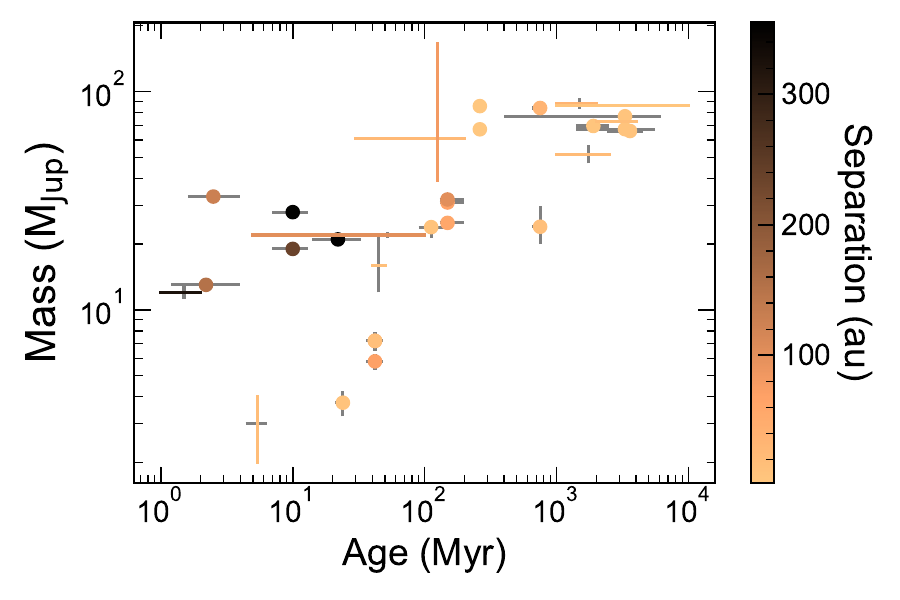}
    \caption{Our KPIC sample distributions. \textit{Top}: Distribution of our KPIC targets as a function of mass (M$_\mathrm{Jup}$) and separation (au), color-coded by age (Myr). 
    Hydrogen mass burning limit (78.6~M$_\mathrm{Jup}$; \citealp{Chabrier:2023aa}), deuterium mass burning limit (12.6~M$_\mathrm{Jup}$; \citealp{Saumon:1996aa, Burrows:1997aa}) are labeled in the horizontal dashed and dashed-dotted lines, respectively, and typical disk size (192~au; \citealp{Ansdell:2018aa}) is depicted in the vertical dotted line.
    \textit{Bottom}: Distribution of our KPIC targets as a function of mass (M$_\mathrm{Jup}$) and age (Myr), color-coded by separation (au). Note that the lack of brown dwarf companions at close separation is due to the so-called brown dwarf desert, reflecting the formation, not the observational bias.
    }
    \label{fig:sample}
\end{figure}

Figure~\ref{fig:sample} shows our sample distribution as a function of mass versus separation and mass versus age.
In the top panel of Figure~\ref{fig:sample}, our sample spans from the lowest-mass stellar and brown dwarf companions down to confirmed (young) imaged giant exoplanets at $\sim$10--100~au, but is limited in separation closer than $\sim$10~au and substellar companions ($<$50~M$_\mathrm{Jup}$).
Additionally, our companions with masses closer to the hydrogen-burning limit are mostly at field ages, reflecting the sparse sampling of field-age substellar objects below 50~M$_\mathrm{Jup}$ at close separation (10--100~au).
The lack of brown dwarf companions ($\sim$10--80~M$_\mathrm{Jup}$) at close separation is due to their formation or the so-called ``brown dwarf desert'' \citep{Grether:2006aa}.
The mass versus age distribution (bottom panel in Figure~\ref{fig:sample}) also reflects our detection limit at older ($>$100~Myr) giant planets ($<$10~M$_\mathrm{Jup}$) in the bottom right corner of the figure.
The lack of more massive young objects is partially due to our sample selection bias, shown in the upper left corner of the figure.

\begin{deluxetable*}{lcccccccc}
\tablewidth{700pt}
\tablecaption{KPIC Imaged Exoplanets and Companion Sample \label{tab:sample}} 
\tabletypesize{\scriptsize} 
\tablehead{ 
\colhead{Comp.} & 
\colhead{Comp. Mass} & 
\colhead{Comp. Radius} & 
\colhead{Separation} & 
\colhead{Inclination} & 
\colhead{Eccentricity} & 
\colhead{Age} & 
\colhead{Membership\tablenotemark{b}} & 
\colhead{Reference}
 \\
\colhead{} & %
\colhead{(M$_\mathrm{Jup}$)} & \colhead{(R$_\mathrm{Jup}$)} & \colhead{(au)} & \colhead{(deg)} & \colhead{} &\colhead{(Myr)} & \colhead{} & \colhead{} 
} 
\startdata
\hline
PDS 70 b & $2-4$ & $2-3$ & $20.8^{+0.6}_{-0.7}$ & $131.0 \pm 3$ & $0.17 \pm 0.06$ & $5.4 \pm 1$ & UCL & 55, 43, 54 \\ 
AF Lep b & $3.75 \pm 0.5$ & $1.3 \pm 0.15$ & $8.98 \pm 0.15$ & $57.5^{+0.6}_{-0.7}$ & $0.013^{+0.024}_{-0.010}$ & $24 \pm 3$ & BPMG & 3, 2 \\ 
HR 8799 b & $5.8 \pm 0.5$ & $1.2 \pm 0.1$ & $70.8 \pm 0.19$ & $26.8 \pm 2.3$ & $0.018^{+0.018}_{-0.013}$ & $42 \pm 6$ & Columba & 53, 3, 38 \\ 
HR 8799 c & $7.2 \pm 0.7$ & $1.2 \pm 0.1$ & $43.1 \pm 1.4$ & $26.8 \pm 2.3$ & $0.022^{+0.023}_{-0.017}$ & $42 \pm 6$ & Columba & 53, 3, 38 \\ 
HR 8799 d & $7.2 \pm 0.7$ & $1.2 \pm 0.1$ & $26.2 \pm 0.9$ & $26.8 \pm 2.3$ & $0.129^{+0.022}_{-0.025}$ & $42 \pm 6$ & Columba & 53, 3, 38 \\ 
HR 8799 e & $7.2 \pm 0.7$ & $1.2 \pm 0.1$ & $16.2 \pm 0.5$ & $26.8 \pm 2.3$ & $0.018^{+0.019}_{-0.028}$ & $42 \pm 6$ & Columba & 53, 3, 38 \\ 
DH Tau B & $12 \pm 4$ & $2.6 \pm 0.6$ & $324.0^{+270}_{-109}$ & $82.0^{+25}_{-27}$ & $0.61^{+0.30}_{-0.41}$ & $1-2$ & TAU & 56, 30, 62, 31 \\ 
ROXs 42 Bb & $13 \pm 5$ & $2.1 \pm 0.35$ & $153.0^{+21}_{-27}$ & $35.0 \pm 16$ & $0.23^{+0.22}_{-0.16}$ & $2.2^{+1.8}_{-1.0}$ & ROPH & 56, 11, 12, 62 \\ 
HIP 99770 B & $16 \pm 5$ & $1.0-1.1$ & $16.9^{+3.4}_{-1.9}$ & $148.0^{+13}_{-11}$ & $0.25^{+0.14}_{-0.16}$ & $40-50$ & ARG & 17, 63 \\ 
ROXs 12 B & $19 \pm 5$ & $2.2 \pm 0.35$ & $236.0^{+41}_{-50}$ & $128.0^{+27}_{-25}$ & $0.68^{+0.18}_{-0.24}$ & $10 \pm 3$ & USCO & 56, 12, 6, 62, 45, 11 \\ 
GSC 6214-210 B & $21 \pm 6$ & $1.55 \pm 0.25$ & $354.0^{+57}_{-76}$ & $112.0^{+18}_{-9}$ & $0.68^{+0.20}_{-0.34}$ & $22^{+11}_{-8}$ & $\pi$ Sco & 56, 44, 12, 62 \\ 
kap And B & $22 \pm 9$ & $1.35 \pm 0.25$ & $102.0^{+50}_{-27}$ & $148.0^{+11}_{-15}$ & $0.88^{+0.03}_{-0.04}$ & $5-100$ & Columba? & 42, 3, 62 \\ 
HD 206893 B & $22.7^{+2.5}_{-1.7}$ & $1.11 \pm 0.03$ & $8.9^{+1.4}_{-0.2}$ & $156.0^{+2}_{-18}$ & $0.27^{+0.04}_{-0.23}$ & $112^{+36}_{-22}$ & ARG? & 49 \\ 
HIP 21152 B & $24^{+6}_{-4}$ & $0.997 \pm 0.023$ & $17.0^{+5}_{-4}$ & $95.0^{+4}_{-2}$ & $0.36^{+0.37}_{-0.25}$ & $750 \pm 100$ & HYA & 23, 8 \\ 
2M0122-2439 B & $25 \pm 12$ & $1.2 \pm 0.2$ & $59.0^{+9}_{-14}$ & $103.0^{+16}_{-5}$ & $0.85^{+0.10}_{-0.42}$ & $149^{+51}_{-19}$ & ABDMG & 3, 56, 12, 62 \\ 
HIP 79098 AB b & $28 \pm 13$ & $2.6 \pm 0.6$ & $352.0^{+332}_{-117}$ & $110.0^{+24}_{-16}$ & $0.58^{+0.27}_{-0.37}$ & $10 \pm 3$ & USCO & 29, 56, 48, 62, 45 \\ 
CD-35 2722 B & $31 \pm 8$ & $1.25 \pm 0.15$ & $67.0 \pm 4$ & $151.0^{+15}_{-20}$ & $0.94 \pm 0.03$ & $149^{+51}_{-19}$ & ABDMG & 56, 7, 52, 57, 3 \\ 
RX J2351.5+3127 B & $32 \pm 6$ & $1.16 \pm 0.11\tablenotemark{a}$ & $102.0^{+27}_{-43}$ & $127.0^{+13}_{-14}$ & $0.46^{+0.25}_{-0.17}$ & $149^{+51}_{-19}$ & ABDMG & 7, 64, 3 \\ 
GQ Lup B & $33 \pm 10$ & $3.7 \pm 0.7$ & $128.0^{+68}_{-37}$ & $49.0^{+13}_{-20}$ & $0.41^{+0.25}_{-0.29}$ & $2.5^{+1.5}_{-0.9}$ & Lupus 1 & 56, 26, 39, 50, 37, 62 \\ 
HD 1160 B & $39-166$ & $1.4-1.8$ & $82.0^{+20}_{-31}$ & $85.0 \pm 5$ & $0.18^{+0.18}_{-0.12}$ & $115-135$ & Psc-Eri? & 40, 18, 51, 7, 36, 19 \\ 
HD 33632 Ab & $51.6^{+5.4}_{-4.8}$ & $0.86 \pm 0.1$ & $23.7^{+3.1}_{-3.8}$ & $42.0^{+6}_{-11}$ & $0.14^{+0.13}_{-0.09}$ & $1000-2500$ & field & 28, 25, 21, 16 \\ 
HD 984 B & $61 \pm 4$ & $1.8 \pm 0.6$ & $28.0^{+7}_{-4}$ & $120.8^{+1.8}_{-1.6}$ & $0.76 \pm 0.05$ & $30-200$ & field & 22, 64, 41 \\ 
HD 176535 B & $65.9^{+2}_{-1.7}$ & $0.9 \pm 0.03\tablenotemark{a}$ & $11.1 \pm 0.6$ & $49.0^{+3}_{-4}$ & $0.5 \pm 0.02$ & $3600^{+900}_{-1200}$ & field & 35, 64, 34 \\ 
LP 349-25 B & $67.1 \pm 0.5$ & $1.2^{+0.02}_{-0.03}$ & $2.06 \pm 0.04$ & $117.81 \pm 0.07$ & $0.054 \pm 0.001$ & $262 \pm 21$ & field\tablenotemark{c} & 15 \\ 
HD 4747 B & $67.2 \pm 1.8$ & $0.9 \pm 0.1$ & $10.0 \pm 0.2$ & $48.0 \pm 0.9$ & $0.7317 \pm 0.0014$ & $3300^{+2300}_{-1900}$ & field & 13, 64, 60 \\ 
HD 72946 B & $69.5 \pm 0.5$ & $0.79-0.89$ & $6.462^{+0.030}_{-0.029}$ & $63.14 \pm 0.017$ & $0.498 \pm 0.005$ & $1900^{+600}_{-500}$ & field & 1, 10 \\ 
HR 7672 B & $72.7 \pm 0.8$ & $0.9 \pm 0.1$ & $19.6 \pm 1$ & $97.4 \pm 0.4$ & $0.543 \pm 0.018$ & $2000-4000$ & field & 64, 9 \\ 
HD 112863 B & $77 \pm 3$ & $0.94 \pm 0.12\tablenotemark{a}$ & $7.66 \pm 0.13$ & $60.5 \pm 0.6$ & $0.346 \pm 0.004$ & $3300 \pm 2900$ & field & 46, 64 \\ 
RX J0342.5+1216 B & $84 \pm 11$ & $1.1 \pm 0.8$ & $36.0 \pm 10$ & $83.1^{+0.7}_{-2.2}$ & $0.41^{+0.27}_{-0.08}$ & $750 \pm 100$ & HYA & 64, 8, 20 \\ 
LP 349-25 A & $85.7 \pm 0.6$ & $1.3 \pm 0.03$ & $2.06 \pm 0.04$ & $117.81 \pm 0.07$ & $0.054 \pm 0.001$ & $262 \pm 21$ & field\tablenotemark{c} & 15 \\ 
HD 72780 B & $86 \pm 5$ & $1 \pm 0.05$\tablenotemark{a} & $6.77^{+0.13}_{-0.14}$ & $139.0 \pm 2$ & $0.6312^{+0.010}_{-0.009}$ & $1000-10000$ & field & 64 \\ 
HIP 55507 B & $88 \pm 3$ & $1.3 \pm 0.02$ & $37.8^{+3.5}_{-2.7}$ & $119.3 \pm 0.7$ & $0.4 \pm 0.04$ & $1000-2000$ & field & 61 \\ 
\hline
\multicolumn{9}{c}{Additional Literature Companions Defined in our Gold Standard Sample in Section~\ref{sec:rotation}} \\
\hline
YSES 1 c & $7.2 \pm 0.7$ & $1.7 \pm 0.2\tablenotemark{a}$ & $320.0 \pm 3$ & $90.6^{+1.1}_{-1.0}\tablenotemark{d}$ & \nodata & $3 \pm 3$ & LCC & 4, 46, 63, 58 \\ 
beta Pic b & $11.9 \pm 3$ & $1.4 \pm 0.1$ & $9.93 \pm 0.03$ & $89.0 \pm 0.01$ & $0.103 \pm 0.003$ & $24 \pm 3$ & BPMG & 32, 31, 3 \\ 
AB Pic b & $12.5 \pm 1.5$ & $1.45 \pm 0.08\tablenotemark{a}$ & $307.0^{+305}_{-115}$ & $98.0 \pm 12$ & \nodata & $24 \pm 3$ & BPMG & 24, 3, 63, 5 \\ 
YSES 1 b & $21.8 \pm 3$ & $2.9 \pm 0.5\tablenotemark{a}$ & $146.0^{+16}_{-10}$ & $90.6^{+1.1}_{-1.0}$ & $0.44^{+0.17}_{-0.18}$ & $3 \pm 3$ & LCC & 46, 63, 58 \\ 
\enddata
\tablerefs{(1) \cite{Balmer:2023aa}; (2) \cite{Balmer:2025aa}; (3) \cite{Bell:2015aa}; (4) \cite{Bohn:2020aa}; (5) \cite{Bonnefoy:2010aa}; (6) \cite{Bowler:2017aa}; (7) \cite{Bowler:2020aa}; (8) \cite{Brandt:2015aa}; (9) \cite{Brandt:2019aa}; (10) \cite{Brandt:2021ac}; (11) \cite{Bryan:2016aa}; (12) \cite{Bryan:2020ab}; (13) \cite{Crepp:2016aa}; (14) \cite{Crepp:2018aa}; (15) \cite{Curiel:2024aa}; (16) \cite{Currie:2020aa}; (17) \cite{Currie:2023aa}; (18) \cite{Curtis:2019aa}; (19) \cite{Do-O:2023aa}; (20) \cite{Do-O:2024aa}; (21) \cite{El-Morsy:2025aa}; (22) \cite{Franson:2022aa}; (23) \cite{Franson:2023aa}; (24) \cite{Gandhi:2025aa}; (25) \cite{Gibbs:2024aa}; (26) \cite{Ginski:2014aa}; (27) \cite{Hsu:2024ab}; (28) \cite{Janson:2019aa}; (29) Keck (unpublished); (30) \cite{Kenyon:1995aa}; (31) \cite{Lacour:2021aa}; (32) \cite{Landman:2024aa}; (33) \cite{Lewis:2024aa}; (34) \cite{Li:2023aa}; (35) \cite{Maire:2018aa}; (36) \cite{Marois:2008aa}; (37) \cite{Marois:2008aa}; (38) \cite{McElwain:2007aa}; (39) \cite{Mesa:2020aa}; (40) \cite{Meshkat:2015aa}; (41) \cite{Morris:2024aa}; (42) \cite{Muller:2018aa}; (43) \cite{Pearce:2019aa}; (44) \cite{Pecaut:2016aa}; (45) \cite{Rickman:2024aa}; (46) \cite{Roberts:2025aa}; (47) \cite{Sanghi:2023aa}; (48) \cite{Sappey:2025aa}; (49) \cite{Seifahrt:2007aa}; (50) \cite{Sutlieff:2024aa}; (51) \cite{Wahhaj:2011aa}; (52) \cite{Wang:2018aa}; (53) \cite{Wang:2020aa}; (54) \cite{Wang:2021aa}; (55) \cite{Wang:2021ac}; (56) \cite{Wang:2026aa}; (57) \cite{Wilcomb:2020aa}; (58) \cite{Wood:2023aa}; (59) \cite{Xuan:2022aa}; (60) \cite{Xuan:2024aa}; (61) \cite{Xuan:2024ab}; (62) \cite{Zuckerman:2019aa}; (63) This work. }
\tablenotetext{a}{Interpolated using \cite{Baraffe:2003aa} models from literature mass and ages, with uncertainties propagated by Monte Carlo sampling.}
\tablenotetext{b}{Membership abbreviations are based on definitions in \cite{Gagne:2018ab}: AB Doradus (ABDMG), Argus (ARG), $\beta$ Pictoris (BPMG), Hyades (HYA), Lower Centaurus Crux (LCC), $\rho$ Ophiuchi (ROPH), Taurus (TAU), Upper Centaurus Lupus (UCL), Upper Scorpius (USCO). 
The question mark ``?'' denotes that membership is uncertain, potentially in the field.}
\tablenotetext{c}{Isochrone ages from the BAHC models \citep{Baraffe:2015aa} reported in \cite{Curiel:2024aa}.}
\tablenotetext{d}{The orbital inclination is assumed to be the same as YSES 1 b.}
\end{deluxetable*}

\subsection{Sample Mass and Age}\label{sec:sample_mass_age}

The fundamental parameters of our sample, including masses of the companions and host stars, and the system ages, are mostly drawn from each individual study that characterized the systems in detail.
Here we discuss the main methodology to determine these parameters in these studies and their limitations or uncertainties.

\subsubsection{Sample Age} \label{sec:sample_age}

The ages of our KPIC sample are listed in Table~\ref{tab:sample}, and we discuss the adopted ages of our sample.
In short, most of the host stars in our KPIC sample are well-known and have been studied in detail. 
However, the age, the fundamental astrophysical parameter, is one of the most challenging parameters to measure. 
We assign the ages based on available individual studies, and cross-check with their stellar kinematics using the BANYAN $\Sigma$ tool \citep{Gagne:2018ab}, since the cluster ages are the most reliable age method.
Readers interested in comparing the reliability of different age methods are referred to the review from \cite{Soderblom:2010aa}.

The bona fide planets are PDS 70 b, AF Lep b, HR 8799 bcde, and YSES 1 c.
PDS 70 is a member of the Upper Centaurus Lupus with an age of 5.4$\pm$1~Myr \cite{Riaud:2006aa, Pecaut:2013aa}, consistent with the lifetime transition disk $<$10~Myr \citep{Pfalzner:2024aa}. 
AF Lep is a bona fide member of the beta Pic moving group \citep{Zuckerman:2004aa, Franson:2023ab}, with an age of 24$\pm$3~Myr \citep{Bell:2015aa}.
HR 8799 is a possible member of 42$\pm$6~Myr Columba \citep{Bell:2015aa} through astrometry and proper motions, which is the adopted age used in this work.
It is an F0 $\gamma$ Doradus variable (slowly variable) and $\lambda$ Bootis (metal-poor Fe with slightly oversolar abundances of C, N, O, and S) \citep{Gray:1999aa, Gray:2003aa} with a debris disk \citep{Sadakane:1986aa, Su:2009aa}.
Another independent age constraint is 33$^{+7}_{-13}$~Myr from its direct radius measurement using optical interferometry and photometry \citep{Baines:2012aa}, assuming it is still contracting toward the main sequence. Interested readers of the age debate of HR 8799 are referred to \cite{Baines:2012aa}.
YSES 1 (TYC 8998-760-1) is a young star in MELANGE-4 (27$\pm$3~Myr) near the Lower Centaurus Crux \citep{Wood:2023aa}. The membership of YSES 1 was determined using kinematics, color–magnitude, and rotation periods.

Next we discuss host stars with companions between $\sim$10--40~{\mjup}.
DH Tau is a bona fide member of Taurus with an age of 1--2~Myr \citep{Kenyon:1995aa}.
ROXs 42 B is a member of $\rho$ Ophiuchus \citep{Miret-Roig:2022aa} with the isochrone age of 2.2$^{+1.8}_{-1.0}$~Myr \citep{Xuan:2024ab}.
HIP 99770 is a bona fide member of the 40--50~Myr Argus association determined via kinematics \citep{Zuckerman:2019aa, Currie:2023aa}, which is the adopted age of this work.
However, based on the dynamical masses and luminosity of the companion HIP 99770 b, the age of HIP 99770 A is more consistent with an age of 115--200~Myr \citep{Currie:2023aa}.
ROXs 12 is a possible member of the 10$\pm$3~Myr Upper Scorpius association \citep{Pecaut:2016aa}, which is the adopted membership and age used in this study.
Its full kinematics provide a 99.2\% probability via BANYAN $\Sigma$ \citep{Gagne:2018ab} in the Upper Scorpius association.
While some studies assign ROXs 12 as a member of $<2$~Myr $\rho$ Ophiuchus \citep{Wilking:2008aa}, this study uses ages that are consistent with those used in previous works, including 7.6$^{+4.1}_{-2.5}$~Myr in \cite{Kraus:2014aa} and 6.5$^{+3.8}_{-2.6}$~Myr in \cite{Xuan:2024ab}.
GSC 6214-210 is a $\pi$ Sco subgroup in the Upper Scorpius \citep{Miret-Roig:2022aa}. We adopted the age of 22$^{+11}_{-8}$~Myr in \cite{Xuan:2024ab}, a conservative uncertainty compared to other ages in the literature \citep{Pearce:2019aa}.
The age of kap And is the most uncertain in our sample \citep{Wilcomb:2020aa}, and we adopted the age range of 5--100~Myr from \cite{Xuan:2024ab}. Its potential membership in Columba is consistent with an age of 47$^{+27}_{-40}$~Myr from \cite{Jones:2016aa} who used interferometric radius, bolometric luminosity, equatorial velocity and MESA models.
However, the BANYAN $\Sigma$ tool finds a low probability of 16\% in Columba using Gaia DR3 astrometry, proper motions, parallax \citep{Gaia-Collaboration:2023aa}, and radial velocity ($-12.7 \pm 0.8$~{\kms} from \citealp{Gontcharov:2006aa}).
HD 206893 does not have a clear membership with known young moving groups.
From BANYAN $\Sigma$ and Gaia DR2 and DR3, it has 60\% probability in the field and 40\% in the 40--50~Myr Argus association \citep{Zuckerman:2019aa}.
We adopted the age of 112$^{+36}_{-22}$~Myr from \cite{Sappey:2025aa}, which used the substellar evolution model to estimate ages based on HD 206893 B photometry and spectra, and is consistent with the age of 155$\pm$15~Myr in \cite{Hinkley:2023aa} using the mass and luminosity of HD 206893 B and \cite{Saumon:2008aa} substellar evolutionary models.
HIP 21152 is a confirmed member in Hyades \citep{Perryman:1998aa, Franson:2023aa}, with an age of 750$\pm$100~Myr \citep{Brandt:2015aa}.
2MASS J01225093–2439505 (2M0122-2439) is a confirmed member of AB Doradus \citep{Malo:2013aa}, with an age of $\sim$149$^{+51}_{-19}$~Myr \citep{Bell:2015aa}.
HIP 79098 AB is a member of 10$\pm$3~Myr Upper Scorpius \citep{Pecaut:2016aa, Janson:2019aa}.
CD-35 2722 and RX J2351.5+3127 are both confirmed members of AB Doradus \citep{Torres:2008aa, Wahhaj:2011aa, Malo:2013aa}, with an age of $\sim$149$^{+51}_{-19}$~Myr \citep{Bell:2015aa}.
GQ Lup is a member of Lupus 1 \citep{Hughes:1994aa, Neuhauser:2005aa}, and we adopted the age of 2.5$^{+1.5}_{-0.9}$~Myr from \cite{Xuan:2024ab}.

Finally, we discuss systems with companions of masses above $\sim$40~{\mjup}, covering high-mass brown dwarfs and lowest-mass stars.
Other than 1RXS J034231.8+121622 (RX J0342.5+1216) being a member of the Hyades \citep{Kuzuhara:2022aa} using BANYAN $\Sigma$ \citep{Gagne:2018ab} and HD 1160 as a candidate member of the $\sim$120~Myr Psc-Eri Stream \citep{Curtis:2019aa}, the remaining objects are not associated with any known young clusters or moving groups.
HD 33632 A is a field-age star, where we used the age of 1--2.5~Gyr from \cite{Currie:2020aa} constrained with activity, gyrochronology, and heavy element abundances.
HD 984 is not associated with any known young clusters, and has an age of 30--200~Myr from \cite{Meshkat:2015aa} using isochrones, coronal X-ray emission, and stellar rotation.
HD 176535 has an age of 3.6~Gyr \citep{Li:2023aa, Lewis:2024aa} constrained by isochrones and activity.
LP 349-25 AB has an age of 262$\pm$21~Myr from \cite{Curiel:2024aa} using the \cite{Baraffe:2015aa} models.
HD 4747 has an age of 3.3$^{+2.3}_{-1.9}$~Gyr using gyrochronology by \cite{Crepp:2016aa}.
HD 72946 has an inferred age of 1.9$^{+0.6}_{-0.5}$~Gyr by \cite{Brandt:2021ac} using X-ray and chromospheric activity indices.
HR 7672 has an age of 2--4~Gyr from \cite{Brandt:2019aa} using isochrones and activity–age relations.
HD 112863 has an isochrone age of 3.3$\pm$2.9~Gyr from \cite{Rickman:2024aa}.
HD 72780 is not associated with any known young clusters or moving group from BANYAN $\Sigma$, so we simply assume an age of 1--10~Gyr.
HIP 55507 has an age of 1--2~Gyr by \cite{Xuan:2024aa} constrained using a lack of Li and slow rotation.

\subsubsection{Sample Mass of Companions}\label{sec:sample_comp_mass}

Here we discuss the adopted masses of our companions, with the adopted masses summarized in Table~\ref{tab:sample}.
In general, dynamical masses are always preferred over the model-derived masses.
Dynamical masses are relatively model-independent but require long-term orbit monitoring \citep{Dupuy:2017aa} and/or astrometric accelerations from Gaia and Hipparcos \citep{Brandt:2019aa, Brandt:2021aa} (aka the Hipparcos–Gaia Catalog of Accelerations, or HGCA), along with possible radial velocities of the host stars and companions.

The companion dynamical masses include AF Lep b ($3.75 \pm 0.5$~{\mjup}; \citealp{Balmer:2025aa}), 
beta Pic b ($11.9 \pm 3$~{\mjup}; \citealp{Lacour:2021aa}),
HIP 99770 B ($16 \pm 5$~{\mjup}; \citealp{Currie:2023aa}), 
HIP 21152 B ($24^{+6}_{-4}$~{\mjup}; \citealp{Franson:2023aa}), 
HD 33632 Ab ($51.6^{+5.4}_{-4.8}$~{\mjup}; \citealp{Gibbs:2024aa}), 
HD 984 B ($61 \pm 4$~{\mjup}; \citealp{Franson:2022aa}), 
HD 176535 B ($65.9^{+2}_{-1.7}$~{\mjup}; \citealp{Lewis:2024aa}), 
LP 349-25 B ($67.1 \pm 0.5$~{\mjup}; \citealp{Curiel:2024aa}), 
HD 4747 B ($67.2 \pm 1.8$~{\mjup}; \citealp{Xuan:2022aa}), 
HD 72946 B ($69.5 \pm 0.5$~{\mjup}; \citealp{Balmer:2023aa}), 
HR 7672 B ($72.7 \pm 0.8$~{\mjup}; \citealp{Brandt:2019aa}), 
HD 112863 B ($77 \pm 3$~{\mjup}; \citealp{Rickman:2024aa}), 
LP 349-25 A ($85.7 \pm 0.6$~{\mjup}; \citealp{Curiel:2024aa}), 
HD 72780 B ($86 \pm 5$~{\mjup}; this work), and 
HIP 55507 B ($88 \pm 3$~{\mjup}; \citealp{Xuan:2024aa}).

The companion masses derived from substellar evolutionary models include PDS 70 b (2--4~{\mjup}; \citealp{Wang:2020aa}), HR 8799 b (5.8$\pm$0.5~{\mjup}) and cde (7.2$\pm$0.7~{\mjup}; \citealp{Wang:2018aa}), 
YSES 1 c ($7.2 \pm 0.7$~{\mjup}; \citealp{Wood:2023aa}),
DH Tau B ($12 \pm 4$~{\mjup}; \citealp{Xuan:2024ab}), 
AB Pic b ($12.5 \pm 1.5$~{\mjup}; \citealp{Bonnefoy:2010aa}), 
ROXs 42 Bb ($13 \pm 5$~{\mjup}; \citealp{Xuan:2024ab}), 
ROXs 12 B ($19 \pm 5$~{\mjup}; \citealp{Xuan:2024ab}), 
GSC 6214-210 B ($21 \pm 6$~{\mjup}; \citealp{Xuan:2024ab}), 
YSES 1 b ($21.8 \pm 3$~{\mjup}; \citealp{Wood:2023aa})
kap And B ($22 \pm 9$~{\mjup}; \citealp{Xuan:2024ab}), 
HD 206893 B ($22.7^{+2.5}_{-1.7}$~{\mjup}; \citealp{Sappey:2025aa}), 
2M0122-2439 B ($25 \pm 12$~{\mjup}; \citealp{Xuan:2024ab}), 
HIP 79098 AB b ($28 \pm 13$~{\mjup}; \citealp{Xuan:2024ab}), 
RX J2351.5+3127 B ($32 \pm 6$~{\mjup}; \citealp{Bowler:2020aa}), 
GQ Lup B ($33 \pm 10$~{\mjup}; \citealp{Xuan:2024ab}), and 
RX J0342.5+1216 B ($84 \pm 11$~{\mjup}; \citealp{Do-O:2024aa}).
Notably, HD 1160 B has a very uncertain mass of $39-166$~{\mjup} due to its uncertain ages \citep{Maire:2016aa, Mesa:2020aa}.

\subsubsection{Sample Mass of Host Stars}\label{sec:sample_host_mass}

As mass ratios are used to characterize our sample, we list our adopted host star masses below.
In general, we adopted the host star masses used in individual studies that discovered and/or constrained the companion masses \& orbital fits for consistency.
PDS 70 ($0.88 \pm 0.02$~{\msun}; \citealp{Wang:2021ab}), 
AF Lep ($1.2 \pm 0.06$~{\msun}; \citealp{Mesa:2023aa}), 
HR 8799 ($1.47^{+0.12}_{-0.17}$~{\msun}; \citealp{Sepulveda:2022aa}), 
DH Tau ($0.64 \pm 0.04$~{\msun}; \citealp{Kraus:2009aa}), 
ROXs 42 B ($0.89 \pm 0.08$~{\msun}; \citealp{Kraus:2014aa}), 
HIP 99770 ($1.85 \pm 0.19$~{\msun}; \citealp{Currie:2023aa}), 
ROXs 12 ($0.65^{+0.05}_{-0.09}$~{\msun}; \citealp{Bowler:2017aa}), 
GSC 6214-210 ($0.9 \pm 0.1$~{\msun}; \citealp{Bowler:2014aa}), 
kap And ($2.768 \pm 0.1$~{\msun}; \citealp{Jones:2016aa}), 
HD 206893 ($1.32 \pm 0.02$~{\msun}; \citealp{Delorme:2017aa}), 
HIP 21152 ($1.4 \pm 0.05$~{\msun}; \citealp{Franson:2023aa}), 
2M0122-2439 ($0.4 \pm 0.05$~{\msun}; \citealp{Bowler:2013aa}), 
HIP 79098 AB ($2.50^{+0.23}_{-0.17}$~{\msun}; \citealp{Anders:2019aa}), 
CD-35 2722 ($0.4 \pm 0.05$~{\msun}; \citealp{Wahhaj:2011aa}), 
RX J2351.5+3127 ($0.45 \pm 0.05$~{\msun}; \citealp{Bowler:2012aa}), 
GQ Lup ($1.05 \pm 0.07$~{\msun}; \citealp{MacGregor:2017aa}), 
HD 1160 ($2.2 \pm 0.1$~{\msun}; \citealp{Nielsen:2012aa}), 
HD 33632 ($1.1 \pm 0.1$~{\msun}; \citealp{Currie:2020aa}), 
HD 984 ($1.2 \pm 0.06$~{\msun}; \citealp{Meshkat:2015aa}), 
HD 176535 ($0.72 \pm 0.02$~{\msun}; \citealp{Delgado-Mena:2019aa, Reiners:2020aa, Lewis:2024aa}), 
LP 349-25 ($0.08188 \pm 0.00061$~{\msun}; \citealp{Curiel:2024aa}), 
HD 4747 ($0.82 \pm 0.04$~{\msun}; \citealp{Crepp:2016aa}), 
HD 72946 ($0.97 \pm 0.01$~{\msun}; \citealp{Balmer:2023aa}), 
HR 7672 ($0.96^{+0.04}_{-0.05}$~{\msun}; \citealp{Brandt:2019aa}), 
HD 112863 ($0.85 \pm 0.02$~{\msun}; \citealp{Rickman:2024aa}), 
RX J0342.5+1216 ($0.3 \pm 0.15$~{\msun}; \citealp{Do-O:2024aa}), 
HD 72780 ($1.22 \pm 0.176$~{\msun}; \citealp{Stassun:2019aa}), 
HIP 55507 ($0.67 \pm 0.02$~{\msun}; \citealp{Xuan:2024aa, Sebastian:2021aa, Stassun:2019aa, Anders:2022aa}), 
beta Pic ($1.75 \pm 0.03$~{\msun}; \citealp{Lacour:2021aa}), 
YSES 1 ($1.0 \pm 0.02$~{\msun}; \citealp{Bohn:2020aa}), 
AB Pic ($0.84 \pm 0.1$~{\msun}; \citealp{Stassun:2019aa}). 

The mass ratio of 0.8\% is adopted as the boundary between planets and brown dwarfs along with mass $<$ 10~{\mjup} (Section~\ref{sec:sample}).
Around this threshold, kap And B (0.76$\pm$0.31\%), HIP 99770 B (0.83$\pm$0.27\%), and HIP 79098 AB b (1.07$\pm$0.51\%) are around this boundary.

\subsection{Observations} \label{sec:observe}
The Keck Planet Imager and Characterizer (KPIC) \citep{Mawet:2017aa, Delorme:2021aa} is a series of instrument upgrades that couples the Keck II adaptive optics system \citep{van-Dam:2006aa, Wizinowich:2006aa} into NIRC2 and the NIRSPEC 2.0 high-resolution spectrometer ($\mathcal{R} \sim$ 35,000; \citealp{McLean:1998aa, Martin:2018aa}), using single-mode fibers that enables diffraction-limited, high-contrast, high-resolution spectroscopy for directly imaged exoplanets, stellar/substellar companions, and hot Jupiters (e.g, \citealp{Wang:2021aa, Xuan:2022aa, Finnerty:2023aa, Finnerty:2024aa}).
Since its commissioning in 2019, KPIC has undergone major upgrades in February 2022 and April 2024 \citep{Echeverri:2022aa, Echeverri:2024aa, Jovanovic:2025aa}, significantly mitigating the fringing contributions from KPIC dichroics \citep{Horstman:2025aa}.

While our first KPIC science observations were conducted on 2019 May 17 (UT), the KPIC observations used in this work span from 2020 July 1 (UT) to 2025 June 30 (UT), as summarized in Table~\ref{tab:observations}.
We used the ``Kband-new'' filter on NIRSPEC, covering a wavelength range of 1.91--2.55 $\micron$.
A typical KPIC sequence includes science observations of the companion, its host star, and a telluric calibration star (typically A0V stars) to provide an empirical spectral response function.
Two out of the four KPIC fibers were usually chosen to observe to serve as ``nodding``, typically in the ABAB or ABBA nodding patterns, to remove the sky and thermal backgrounds with pair subtraction, unless the companion was too far away such that nodding was not possible.
To offset the companion, we used \texttt{whereistheplanet}\footnote{\url{http://whereistheplanet.com/}} \citep{Wang:2021ac} astrometry predictions of the companions, with the astrometric references summarized in Table~\ref{tab:sample}.
For calibrations, typically one bright A0V star and an M giant star (e.g., HIP 81497 or HIP 95771) are observed for tracing and wavelength calibrations, respectively.
We also take thermal background calibration that matches the on-sky integration time of the targets, before or after the observing runs \citep{Wang:2021aa, Hsu:2024ab}.
In this study, we use the spectra reduced under thermal background subtraction for a uniform analysis of objects that lack multiple fibers in the data.
The only exception is HR 8799 b, where pair subtraction provides a much better detection.

We briefly discuss our KPIC on-sky performance based on the detections in our sample.
The KPIC tracking camera CRED2, operated with an $H$-band filter, can provide real-time tracking and offsetting from the host star to achieve stable injection of companion fluxes into one of the KPIC single-mode fibers, with host star $H$ ranging from $\sim$0 to 10~mag.
Our brightest and faintest detected host stars are G0V HR 7672 ($H$=4.43; \citealp{Gray:2006aa}) and M8Ve LP 349-25 A ($H$=9.97; \citealp{Reid:2003aa}), respectively.
In terms of $K$-band contrast, KPIC is capable of offsetting from near-brightness close binary (LP 349-25 AB; $\Delta K'$ = 0.26, \citealp{Forveille:2005aa}) to $\Delta K \sim 12$ (AF Lep b, HR 8799 b).
Because our companions are typically close to their host stars in angular separation, off-axis star light is usually also injected into the KPIC fiber positioned on the companions. Therefore, during observations, the raw flux counts or signal-to-noise ratios in the raw 2D companion spectra are not indicative of the true companion fluxes or detections.
Instead, we rely on the end-to-end throughput for a more reliable estimate of the on-sky performance, ranging from very poor ($\lesssim$ 1\%) to excellent $\sim$4.6\% (HIP 55507 A on 2023 May 2 reported in \citealp{Xuan:2024aa}), with over 3\% generally considered good conditions.

\begin{deluxetable*}{lcccccccc}
\tablewidth{700pt}
\tablecaption{KPIC Observations and Detections \label{tab:observations}} 
\tabletypesize{\scriptsize} 
\tablehead{ 
\colhead{Comp.} & 
\colhead{Host $K$} & 
\colhead{Comp. $K$} & 
\colhead{Date of Obs.} & 
\colhead{MJD\tablenotemark{a}} & 
\colhead{Airmass\tablenotemark{a}} & 
\colhead{$T_\mathrm{int}$\tablenotemark{b}} & 
\colhead{TP\tablenotemark{c}} & 
\colhead{CCF S/N\tablenotemark{d}}
 \\
\colhead{} & \colhead{(mag)} & \colhead{(mag)} & 
\colhead{(UT)} & \colhead{(day)} & 
\colhead{} & \colhead{(min)} & \colhead{(\%)} & \colhead{}
} 
\startdata
\hline
PDS 70 b & $8.54$ & $16.6$\tablenotemark{f} & 2024 May 23 & $60453.299$ & $2.36$ & $140$ & $1.8$ & $4.2$ \\ 
AF Lep b & $4.92$ & $16.7$ & 2024 Sep 27 & $60580.574$ & $1.36$ & $160$ & $2.9$ & $5.3$ \\ 
HR 8799 b & $5.24$ & $16.9$ & 2021 Oct 21 & $59508.305$ & $1.09$ & $290$ & $2.5$ & $7.7$ \\ 
HR 8799 c & $5.24$ & $16.1$ & 2023 Jul 31 & $60156.512$ & $1.07$ & $240$ & $2.8$ & $19.1$ \\ 
HR 8799 d & $5.24$ & $16.0$ & 2025 Jun 30 & $60856.574$ & $1.05$ & $100$ & $2.6$ & $13.0$ \\ 
HR 8799 e & $5.24$ & $15.9$ & 2022 Jul 21 & $59781.599$ & $1.04$ & $120$ & $3.7$ & $9.6$ \\ 
DH Tau B & $8.17$ & $14.1$ & 2022 Oct 12 & $59864.433$ & $1.4$ & $100$ & $2.0$ & $15.9$ \\ 
ROXs 42 Bb & $8.67$ & $15.0$ & 2020 Jul 1 & $59031.385$ & $1.47$ & $80$ & $1.5$ & $6.9$ \\ 
HIP 99770 B & $4.42$ & $15.6$ & 2023 Jun 16 & $60111.598$ & $1.11$ & $90$ & $3.4$ & $11.9$ \\ 
ROXs 12 B & $9.09$ & $14.1$ & 2020 Jul 3 & $59033.339$ & $1.43$ & $50$ & $2.6$ & $14.0$ \\ 
GSC 6214-210 B & $9.15$ & $14.9$ & 2023 Jun 23 & $60118.407$ & $1.41$ & $105$ & $3.5$ & $11.9$ \\ 
kap And B & $4.57$ & $14.3$ & 2022 Nov 12 & $59895.340$ & $1.21$ & $155$ & $2.8$ & $7.0$ \\ 
HD 206893 B & $5.59$ & $15.0$ & 2022 Jul 20 & $59780.443$ & $1.35$ & $120$ & $3.2$ & $7.8$ \\ 
HIP 21152 B & $5.33$ & $16.5$ & 2022 Nov 11 & $59894.479$ & $1.16$ & $320$ & $3.2$ & $6.2$ \\ 
2M0122-2439 B & $9.19$ & $14.5$ & 2024 Sep 27 & $60580.435$ & $1.47$ & $100$ & $3.5$ & $32.7$ \\ 
HIP 79098 AB b & $5.70$ & $14.1$ & 2022 Jul 18 & $59778.327$ & $1.45$ & $40$ & $3.8$ & $19.1$ \\ 
CD-35 2722 B & $7.04$ & $12.0$ & 2022 Nov 12 & $59895.522$ & $1.81$ & $80$ & $2.2$ & $28.1$ \\ 
RX J2351.5+3127 B & $8.96$ & $13.9$ & 2021 Jul 3 & $60579.420$ & $1.03$ & $60$ & $0.7$ & $8.1$ \\ 
GQ Lup B & $7.09$ & $13.3$ & 2021 Apr 24 & $59328.483$ & $1.78$ & $60$ & $1.9$ & $26.0$ \\ 
HD 1160 B & $7.04$ & $14.1$ & 2020 Sep 28 & $59120.510$ & $1.22$ & $60$ & $2.0$ & $7.3$ \\ 
HD 33632 Ab & $5.16$ & $15.3$ & 2021 Nov 20 & $59538.435$ & $1.28$ & $240$ & $2.4$ & $7.5$ \\ 
HD 984 B & $6.07$ & $12.2$ & 2022 Aug 8 & $59799.563$ & $1.12$ & $40$ & $3.9$ & $41.8$ \\ 
HD 176535 B & $7.17$ & $15.9$ & 2024 Sep 27 & $60580.255$ & $1.29$ & $120$ & $3.6$ & $13.0$ \\ 
LP 349-25 B & $9.56$ & $10.4$ & 2021 Sep 18 & $59475.354$ & $1.24$ & $40$ & $0.5$ & $22.1\tablenotemark{e}$ \\ 
HD 4747 B & $5.30$ & $14.3$ & 2020 Sep 28 & $59120.445$ & $1.67$ & $60$ & $1.7$ & $20.5$ \\ 
HD 72946 B & $5.49$ & $13.8$ & 2022 Nov 11 & $59894.650$ & $1.03$ & $40$ & $2.7$ & $10.1$ \\ 
HR 7672 B & $4.38$ & $13.0$ & 2020 Sep 28 & $59120.281$ & $1.02$ & $70$ & $2.6$ & $22.1$ \\ 
HD 112863 B & $6.78$ & $13.4$\tablenotemark{f} & 2024 May 26 & $60456.283$ & $1.12$ & $95$ & $3.1$ & $7.3$ \\ 
RX J0342.5+1216 B & $9.27$ & $13.0$ & 2024 Jul 16 & $60507.622$ & $1.58$ & $30$ & $2.4$ & $11.3$ \\ 
LP 349-25 A & $9.56$ & $9.56$ & 2021 Sep 18 & $59475.317$ & $1.24$ & $19$ & $0.5$ & $24.4\tablenotemark{e}$ \\ 
HD 72780 B & $6.23$ & \nodata & 2022 Nov 14 & $59897.627$ & $1.03$ & $105$ & $3.5$ & $16.6$ \\ 
HIP 55507 B & $6.61$ & $11.6$ & 2024 Dec 13 & $60657.628$ & $1.17$ & $100$ & $2.4$ & $29.9$ \\ 
\enddata
\tablerefs{All host star $K$ magnitudes are drawn from the 2MASS catalog \citep{Cutri:2003aa}. Companion $K$ magnitudes: (1) PDS 70 b: \cite{Stolker:2020aa}; (2) AF Lep b: \cite{De-Rosa:2023aa}; (3) HR 8799 b: \cite{Esposito:2013aa}; (4) HR 8799 c: \cite{Esposito:2013aa}; (5) HR 8799 d: \cite{Marois:2008aa}; (6) HR 8799 e: \cite{Marois:2010aa}; (7) DH Tau B: \cite{Itoh:2005aa}; (8) ROXs 42 Bb: \cite{Kraus:2014aa}; (9) HIP 99770 B: \cite{Currie:2023aa}; (10) ROXs 12 B: \cite{Bowler:2017aa}; (11) GSC 6214-210 B: \cite{Ireland:2011aa}; (12) kap And B: \cite{Bonnefoy:2014aa}; (13) HD 206893 B: \cite{Ward-Duong:2021aa}; (14) HIP 21152 B: \cite{Kuzuhara:2022aa, Franson:2023aa}; (15) 2M0122-2439 B: \cite{Bowler:2013aa}; (16) HIP 79098 AB b: \cite{Janson:2019aa}; (17) CD-35 2722 B: \cite{Wahhaj:2011aa}; (18) RX J2351.5+3127 B: \cite{Bowler:2015aa}; (19) GQ Lup B: \cite{Marois:2007aa}; (20) HD 1160 B: \cite{Nielsen:2012aa}; (21) HD 33632 Ab: \cite{Currie:2020aa}; (22) HD 984 B: \cite{Franson:2022aa}; (23) HD 176535 B: \cite{Lewis:2024aa}; (24) LP 349-25 B: \cite{Forveille:2005aa}; (25) HD 4747 B: \cite{Crepp:2016aa}; (26) HD 72946 B: \cite{Maire:2020aa}; (27) HR 7672 B: \cite{Best:2021aa}; (28) HD 112863 B: \cite{Rickman:2024aa}; (29) RX J0342.5+1216 B: \cite{Bowler:2015aa}; (30) LP 349-25 A: \cite{Cutri:2003aa}; (31) HD 72780 B: unpublished.}
\tablenotetext{a}{Reported in the middle of the total exposure.}
\tablenotetext{b}{Companion on-target integration time.}
\tablenotetext{c}{End-to-end peak (the 95$^\mathrm{th}$ percentile) throughput of the host star, reported with the highest measured throughput among all available host star frames.}
\tablenotetext{d}{Cross-correlation signal-to-noise ratio (S/N). See Section~\ref{sec:ccf_detect} for details.}
\tablenotetext{e}{Reported as either primary or companion-only signal-to-noise ratio (S/N) per spectral channel, inferred from the extracted flux and noise with removal of host star light measured in our forward-modeling method (Section~\ref{sec:model}).}
\tablenotetext{f}{Inferred from VLT/SPHERE $K$2 band (centered at 2.251~$\micron$).}
\end{deluxetable*}

\subsection{Data Reduction} \label{sec:reduce}
Our KPIC data were reduced using the KPIC Data Reduction Pipeline \citep{Wang:2021aa}\footnote{\url{https://github.com/kpicteam/kpic_pipeline}}.
The reduction procedures are detailed in \cite{Wang:2021aa} and \cite{Hsu:2024ab}, which we briefly describe below.

Our reduction procedure starts with extracting thermal background using integration times that match the observation and then tracing spectral orders from the A0V star spectral traces on all four KPIC fibers.
The 1D spectra can then be optimally extracted \citep{Horne:1986aa} with subtraction of thermal backgrounds.
We rely on early M giant spectra to calibrate the wavelength, using the PHOENIX ACES AGSS COND stellar atmosphere models \citep{Husser:2013aa}, assuming its associated effective temperature and surface gravity, and the Planetary Spectrum Generator telluric models \citep{Villanueva:2018aa}, varying with airmass and precipitable water vapor.

\section{Detection via Cross-correlation Function} \label{sec:ccf_detect}

Since our KPIC spectra contain both the star light and planet/companion fluxes, our detections of the planets or stellar/substellar companions rely on cross-correlation functions (CCFs) against molecular templates.
Different from the traditional cross-correlation method (e.g., \citealp{Konopacky:2013aa}), the cross-correlation referred to in this study is a least-squares-based optimization to forward-model the host star and planet/companion flux counts in the spectra at different radial velocity shifts.
The resulting planet/companion counts across different velocity shifts can resemble traditional CCF signals. 
We refer interested readers to \cite{Wang:2021aa}, which details the formulation and implementation for KPIC, following the original formalism of this approach using Keck/OSIRIS \citep{Ruffio:2019aa}.

In short, we cross-correlate the Sonora Bobcat atmosphere models of ro-vibrational CO and H$_2$O lines that correspond to the companion's effective temperature and surface gravity, against our KPIC spectra of the planets and companions \citep{Marley:2021aa, Wang:2021aa}.
We focus on NIRSPEC orders 31--33 (2.29–2.49~$\micron$) as these wavelength ranges cover the CO ($\nu$ = 2--0 and 3--1) overtone bands as well as H$_2$O \citep{Hsu:2024ab}.
These orders also provide robust wavelength calibrations, widely used in previous high-resolution $K$-band studies \citep{Blake:2010aa, Hsu:2021aa}.
The radial velocity shifts for CCFs range from $-1000$ to $+1000$~{\kms}, and the CCF noise is defined as the standard deviations of the last $\pm$200~{\kms} in the CCF wings \citep{Hsu:2024ab}.

\begin{figure*}
    \centering
    \subfigure{\includegraphics[width=0.19\textwidth]{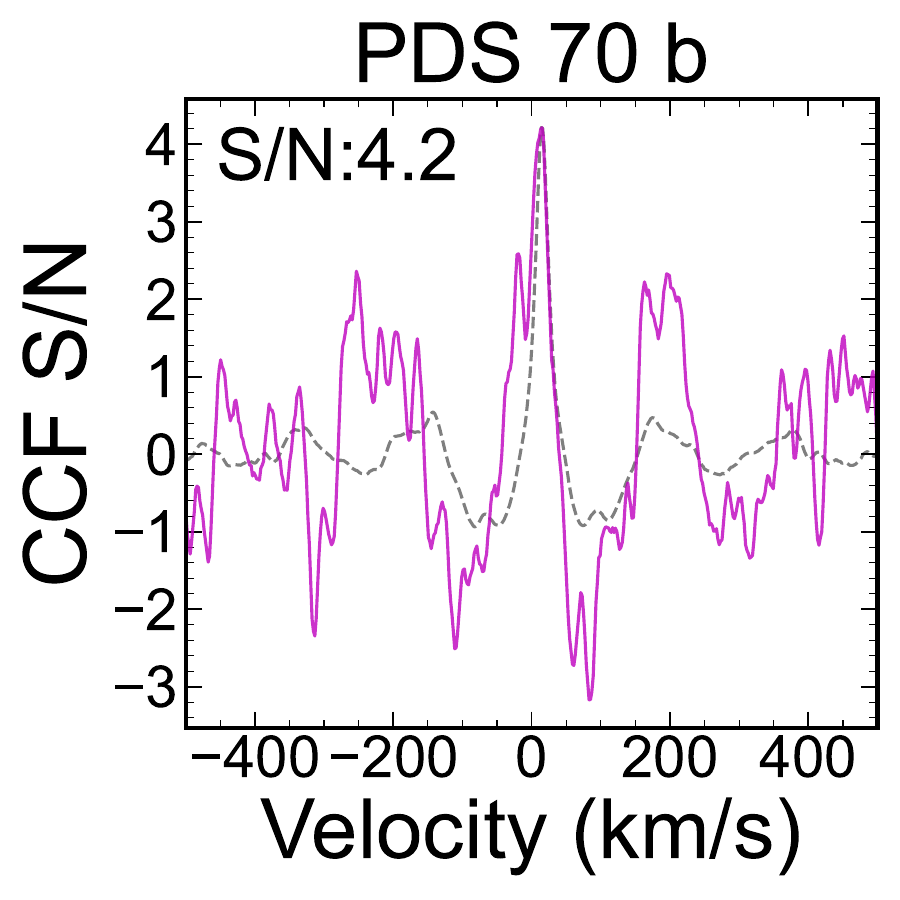}} 
    \subfigure{\includegraphics[width=0.19\textwidth]{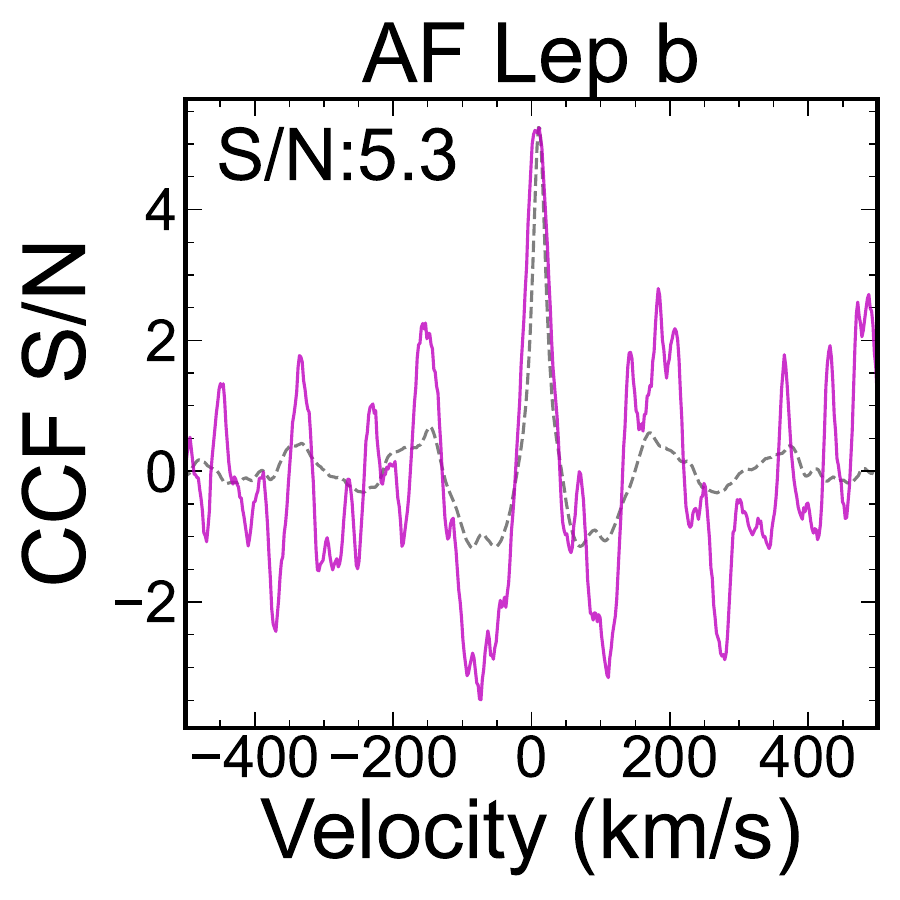}} 
    \subfigure{\includegraphics[width=0.19\textwidth]{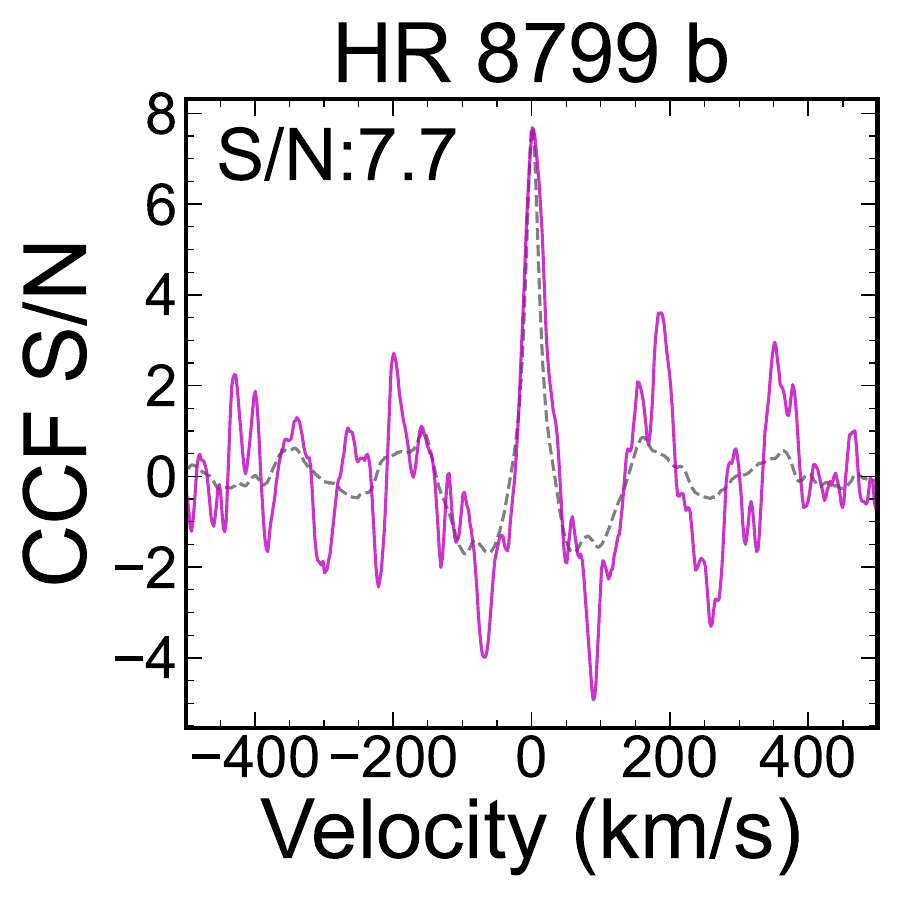}} 
    \subfigure{\includegraphics[width=0.19\textwidth]{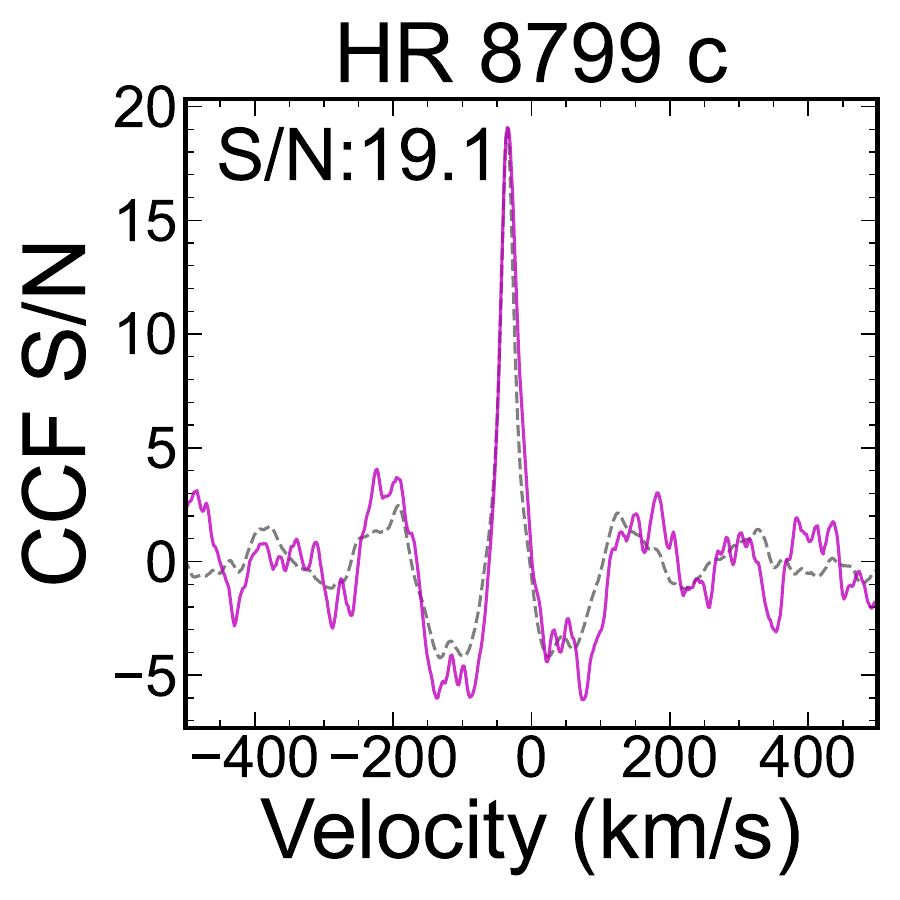}} 
    \subfigure{\includegraphics[width=0.19\textwidth]{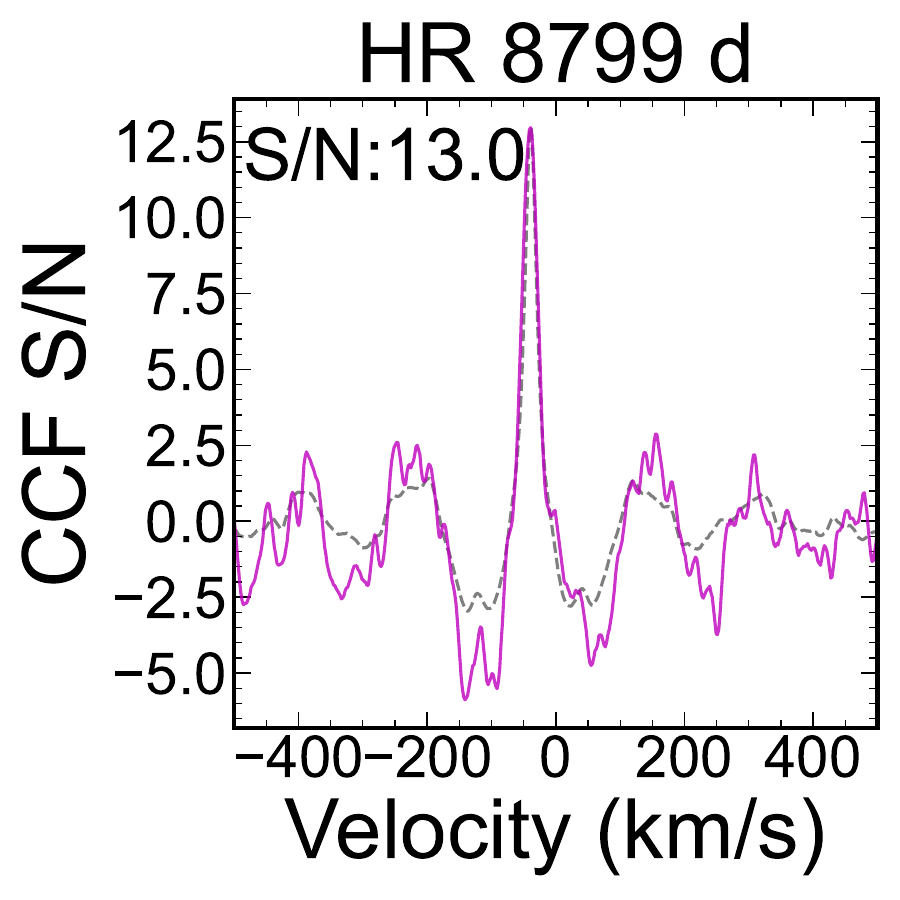}} 

    \subfigure{\includegraphics[width=0.19\textwidth]{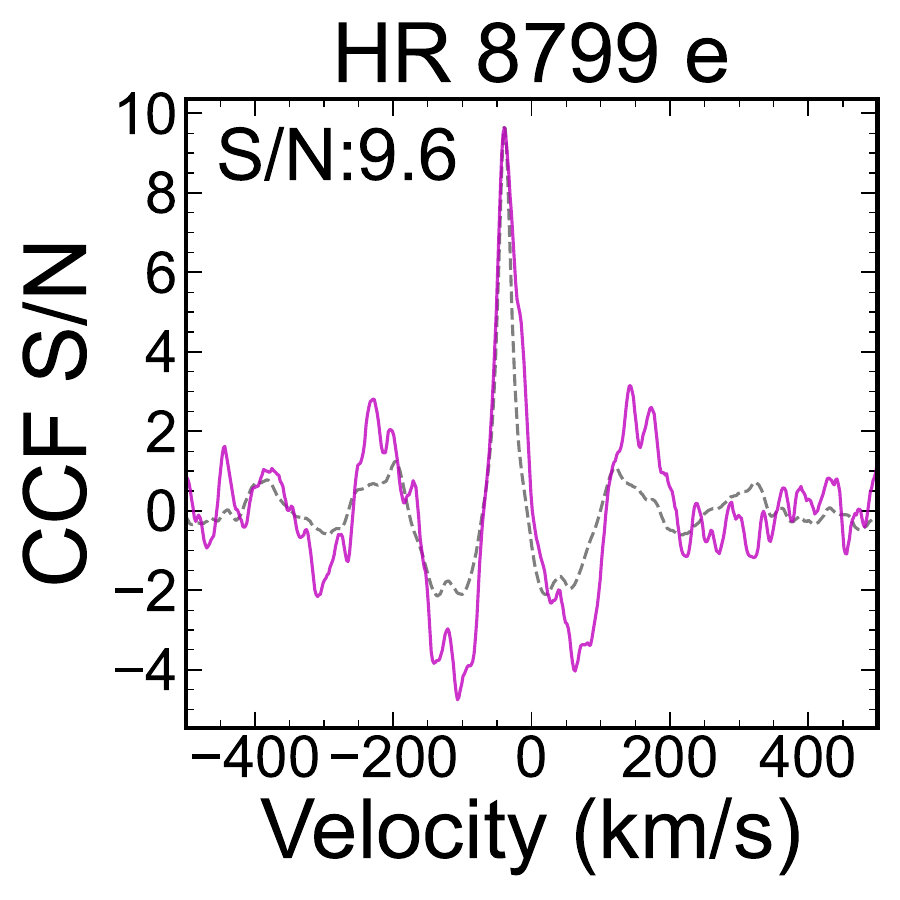}} 
    \subfigure{\includegraphics[width=0.19\textwidth]{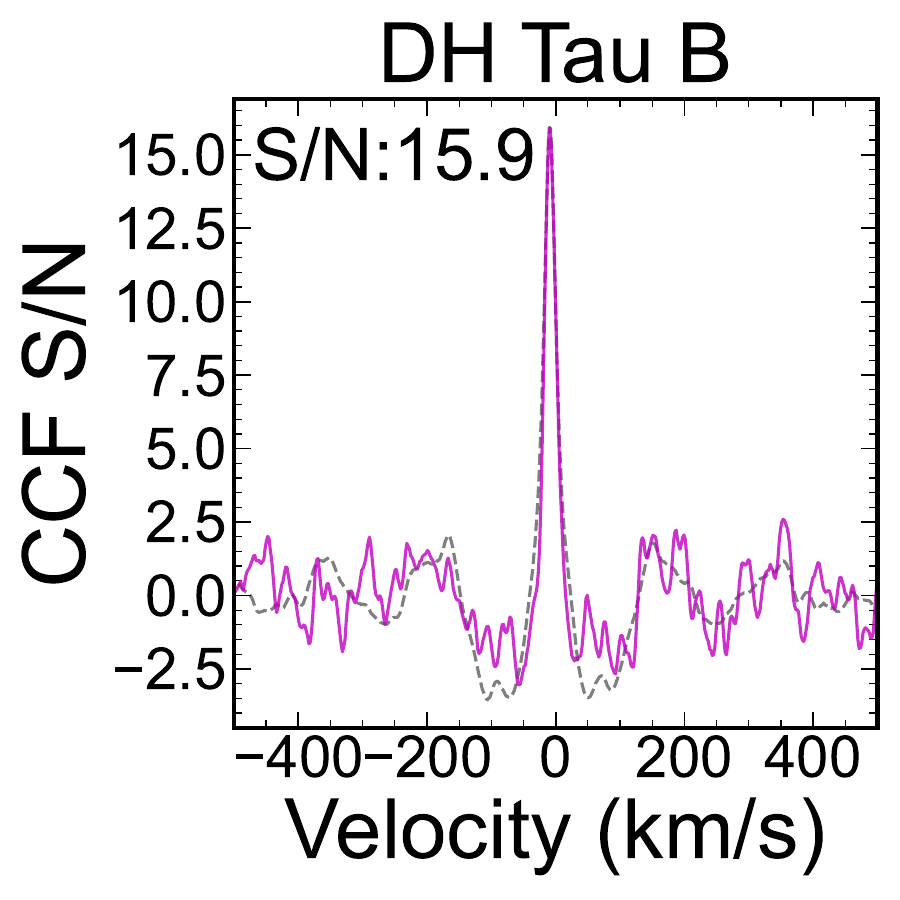}} 
    \subfigure{\includegraphics[width=0.19\textwidth]{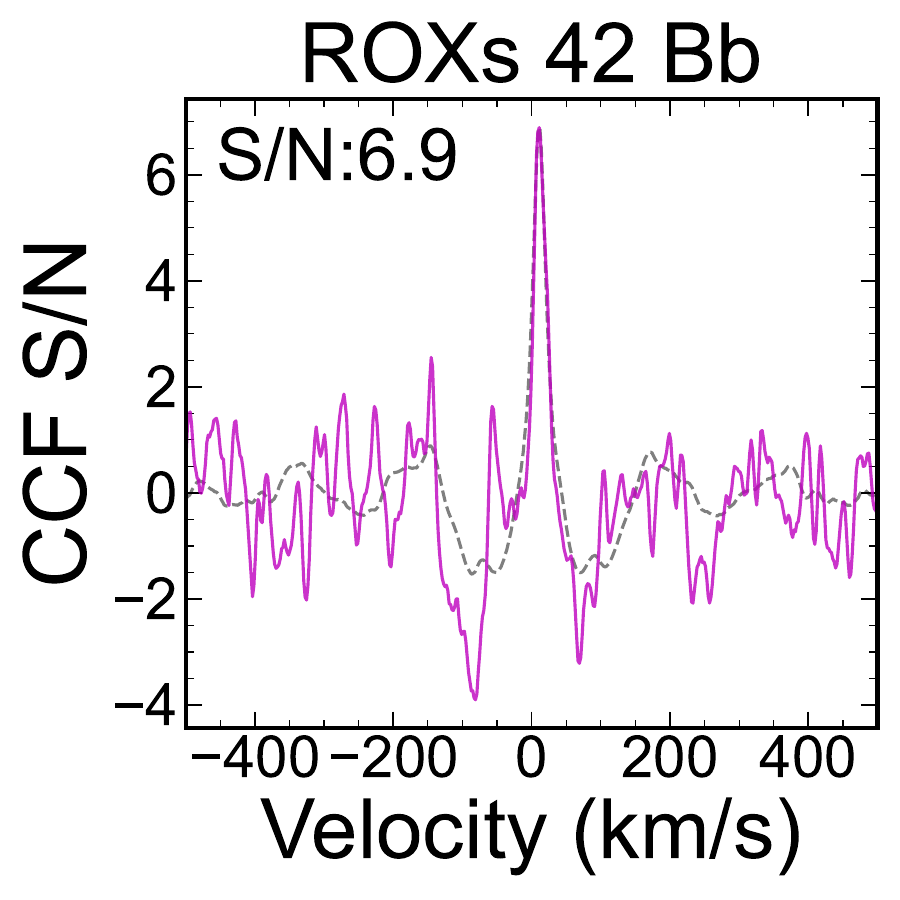}} 
    \subfigure{\includegraphics[width=0.19\textwidth]{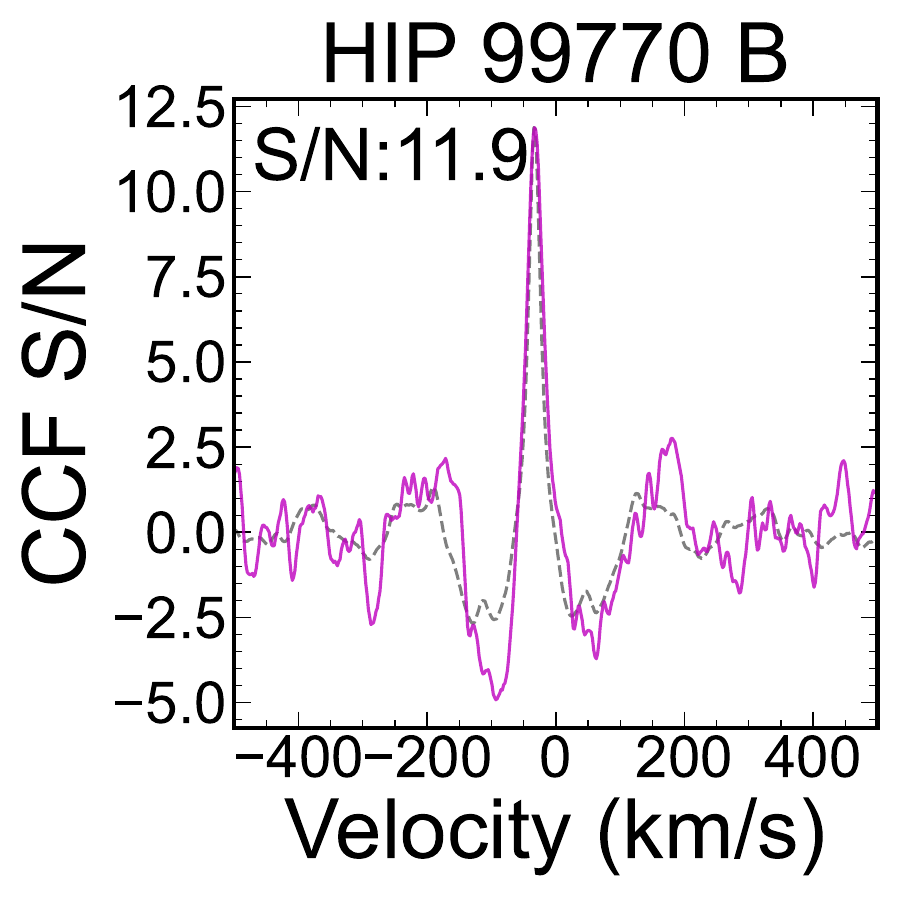}} 
    \subfigure{\includegraphics[width=0.19\textwidth]{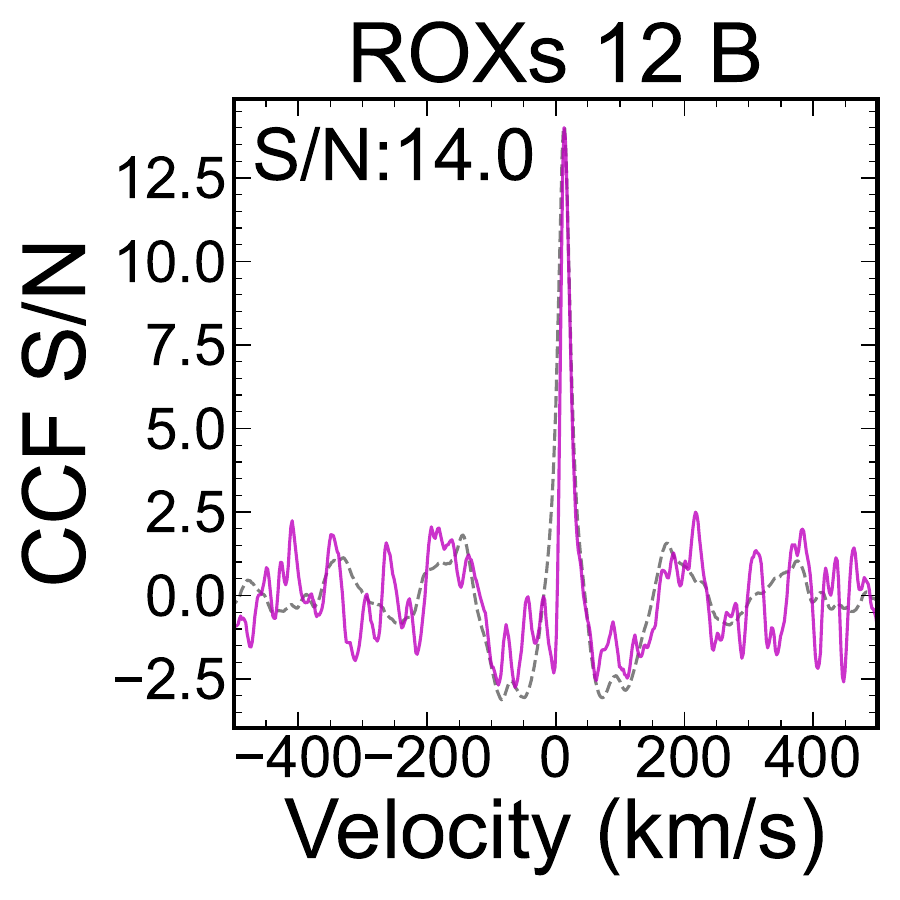}} 

    \subfigure{\includegraphics[width=0.19\textwidth]{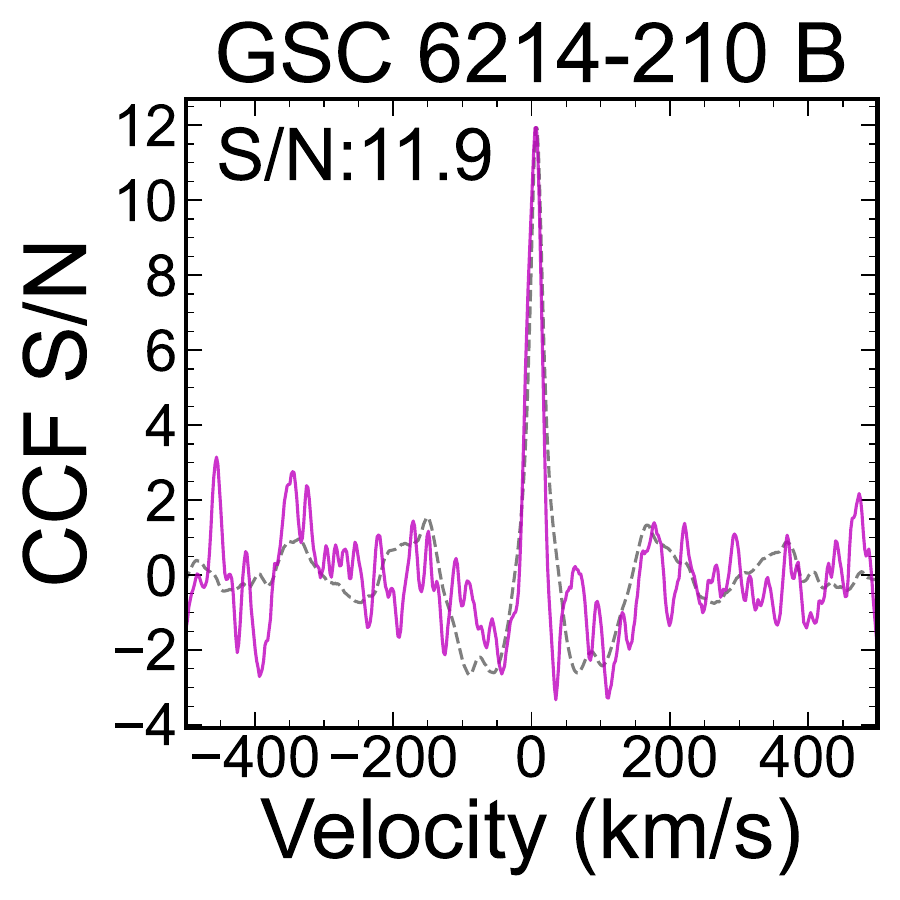}} 
    \subfigure{\includegraphics[width=0.19\textwidth]{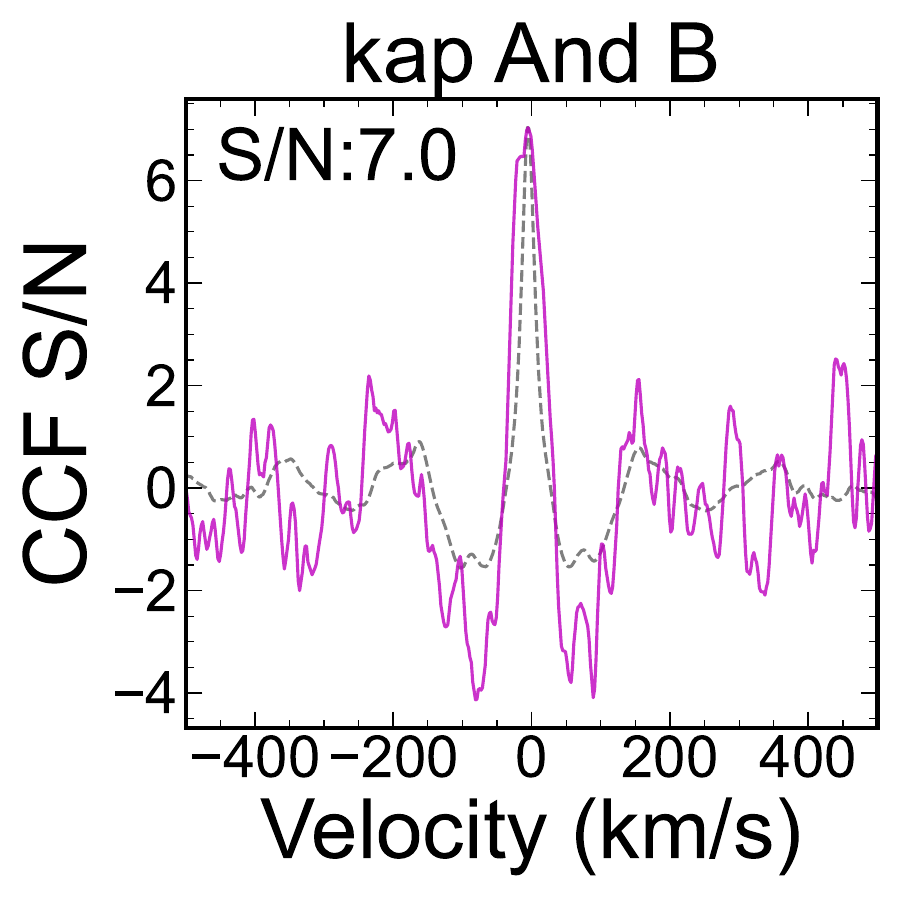}} 
    \subfigure{\includegraphics[width=0.19\textwidth]{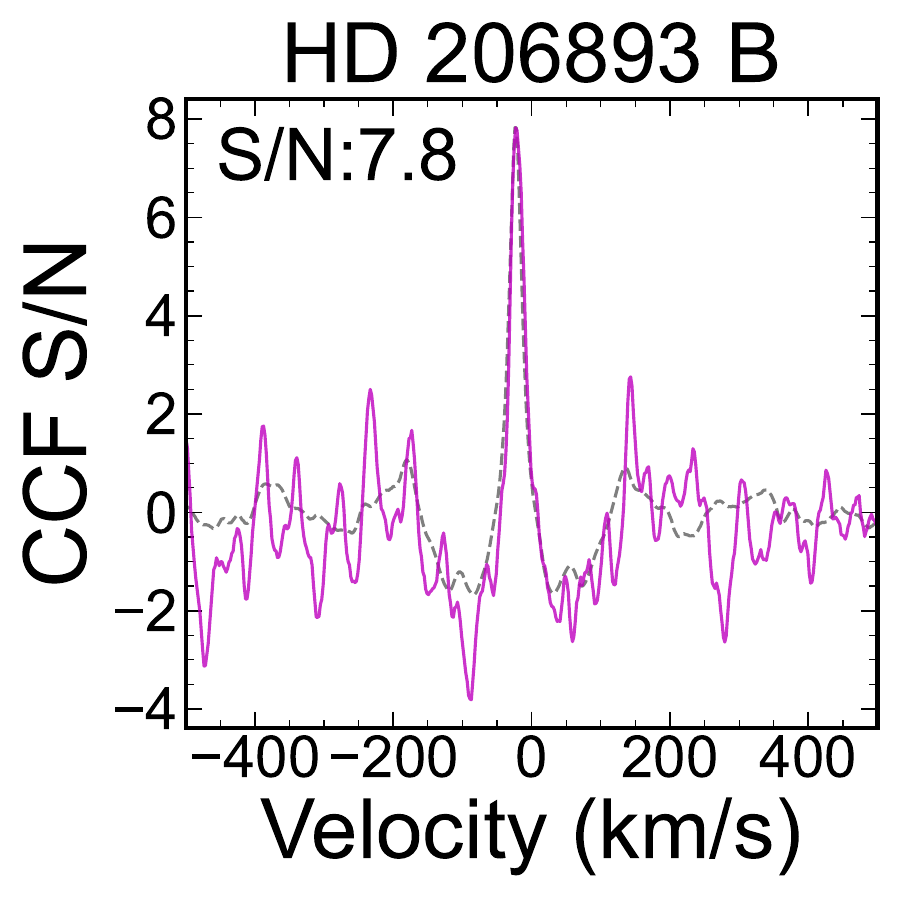}} 
    \subfigure{\includegraphics[width=0.19\textwidth]{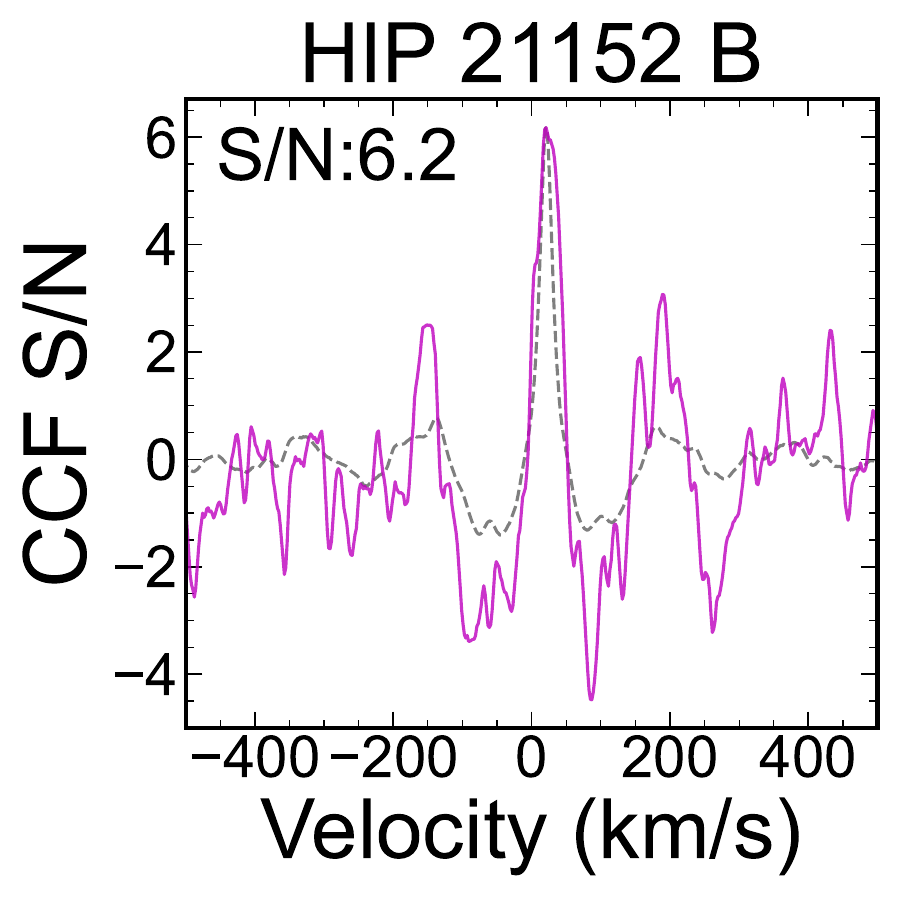}} 
    \subfigure{\includegraphics[width=0.19\textwidth]{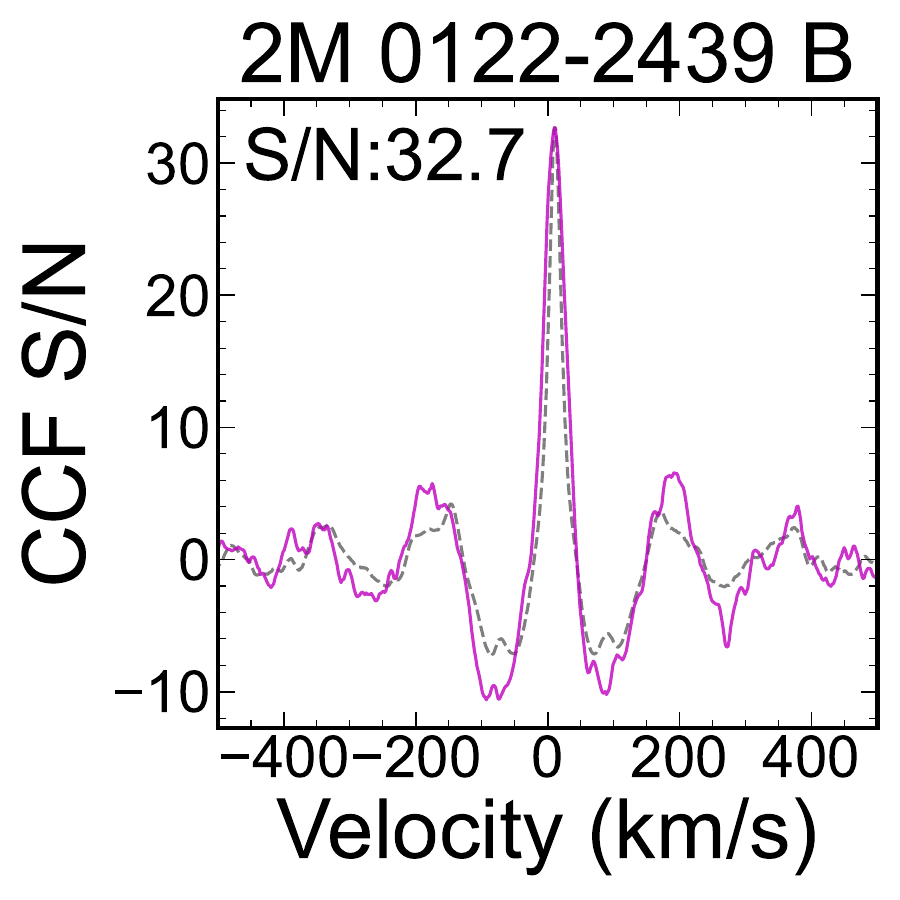}} 

    \subfigure{\includegraphics[width=0.19\textwidth]{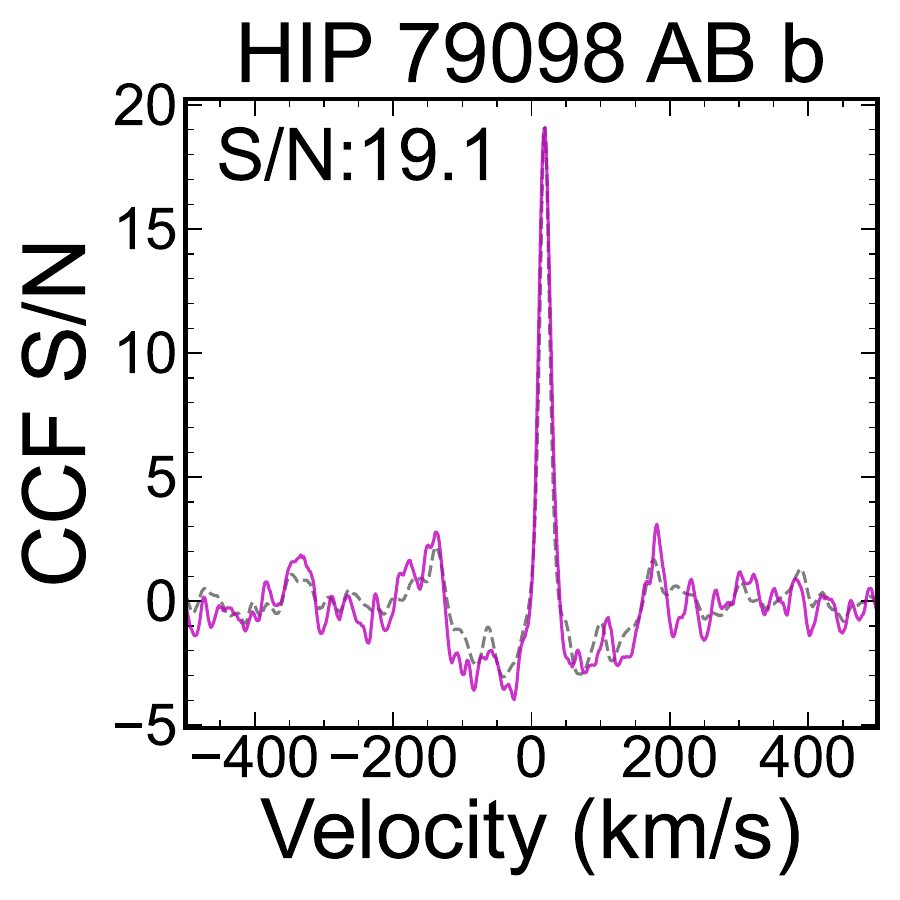}} 
    \subfigure{\includegraphics[width=0.19\textwidth]{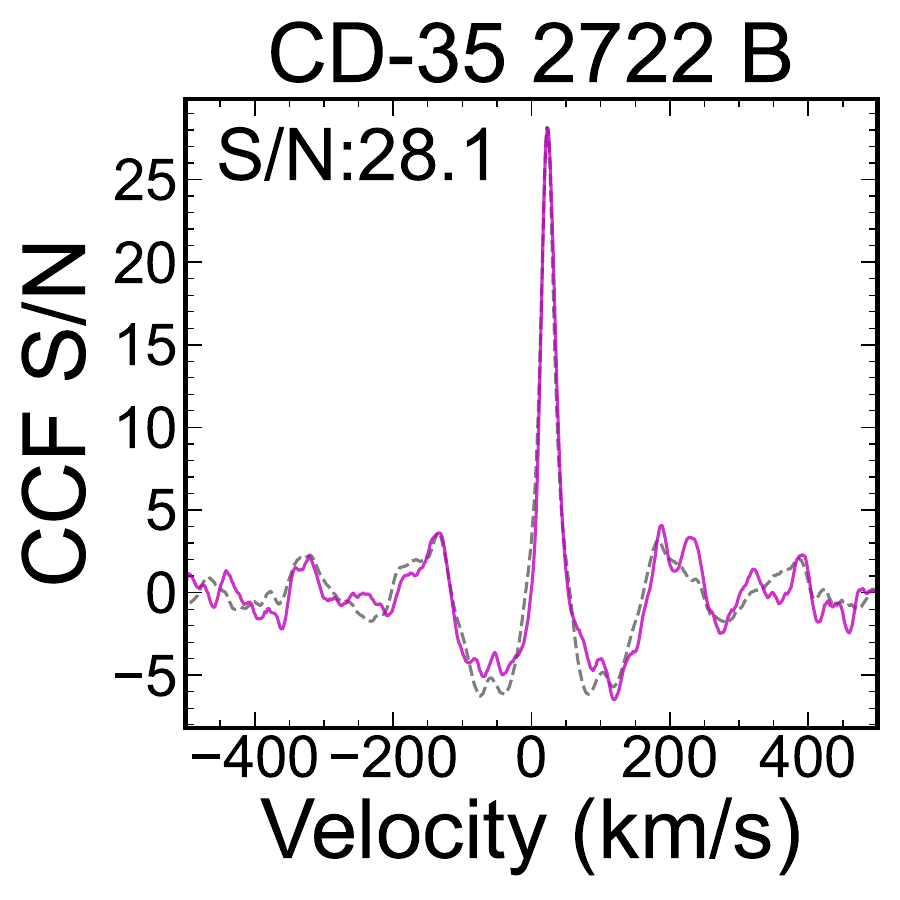}} 
    \subfigure{\includegraphics[width=0.19\textwidth]{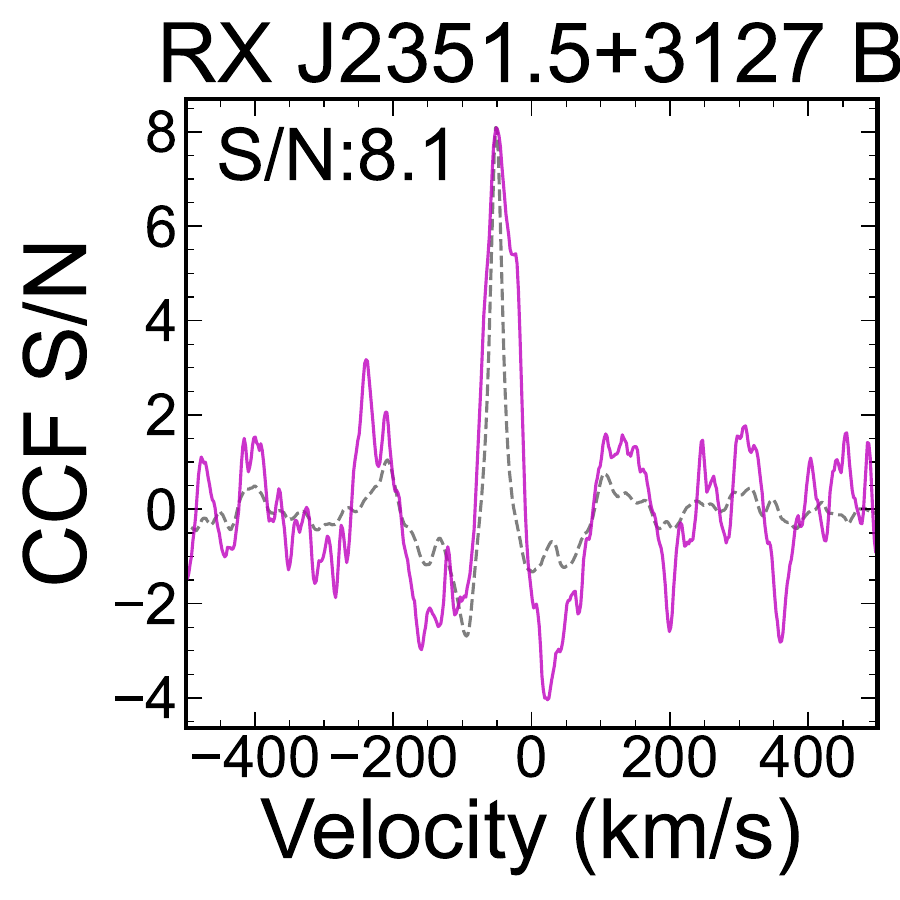}} 
    \subfigure{\includegraphics[width=0.19\textwidth]{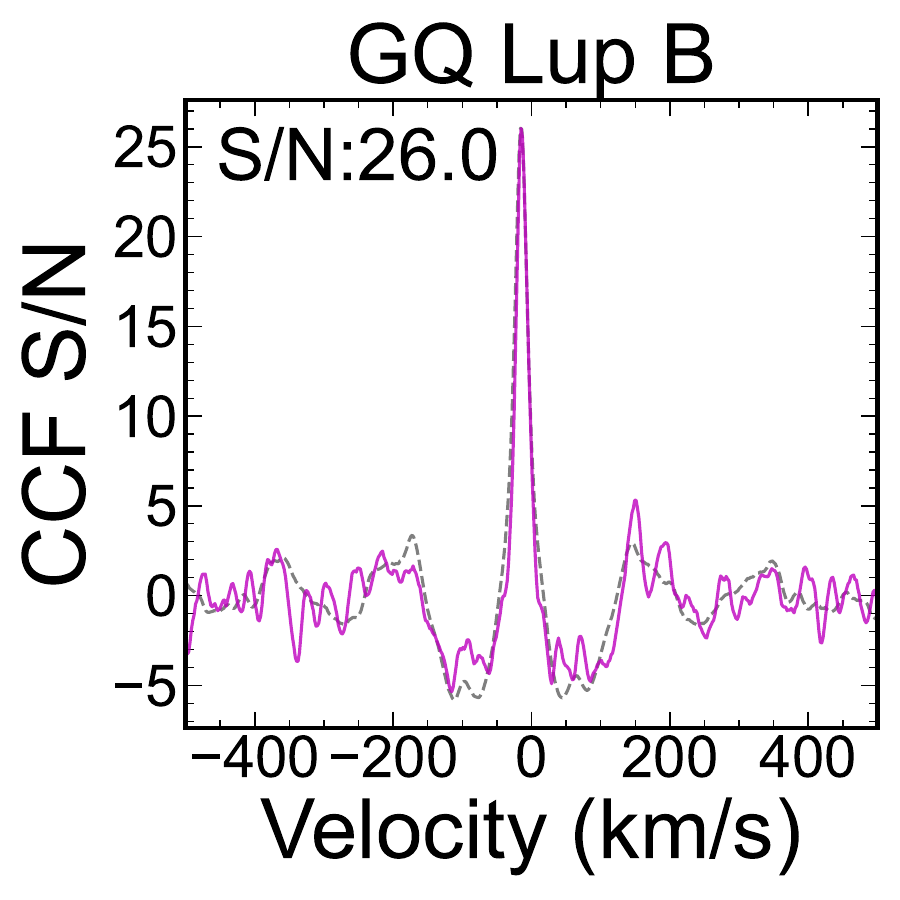}} 
    \subfigure{\includegraphics[width=0.19\textwidth]{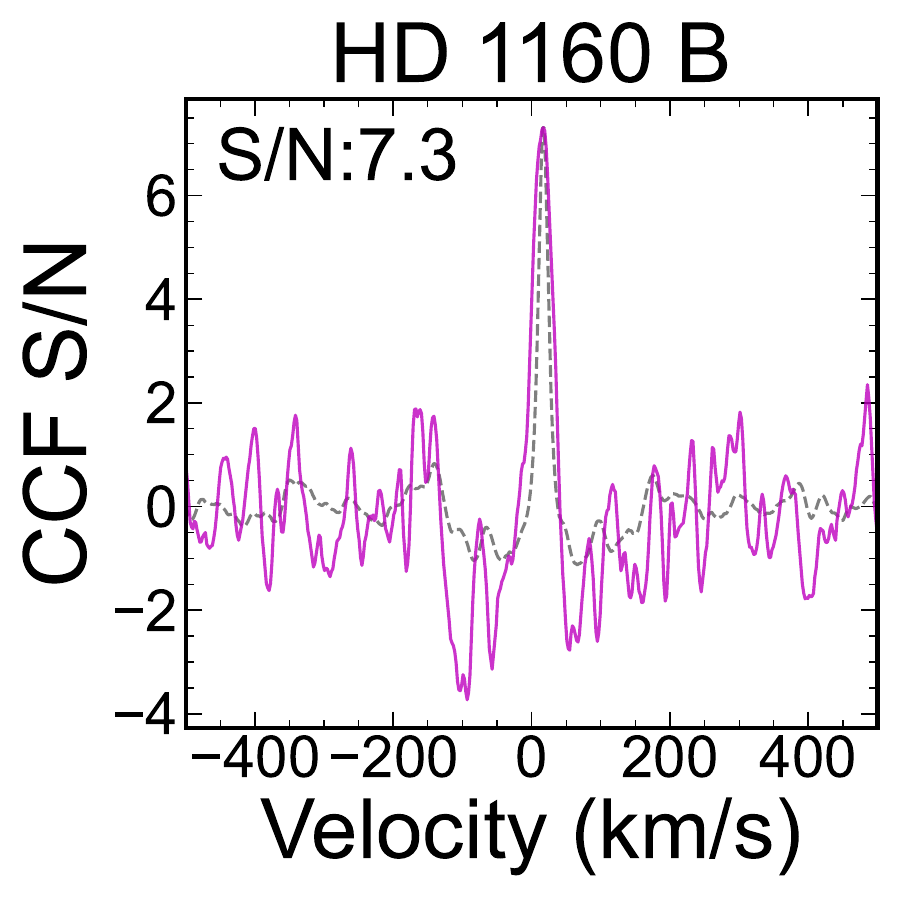}} 

    \subfigure{\includegraphics[width=0.19\textwidth]{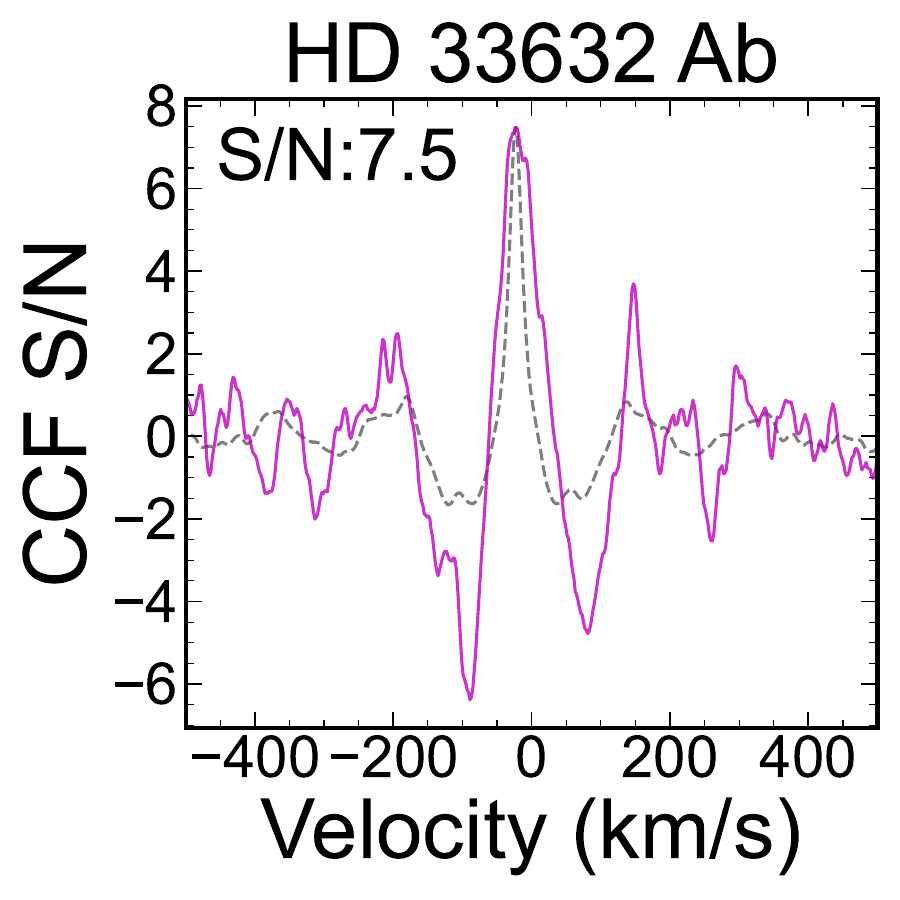}} 
    \subfigure{\includegraphics[width=0.19\textwidth]{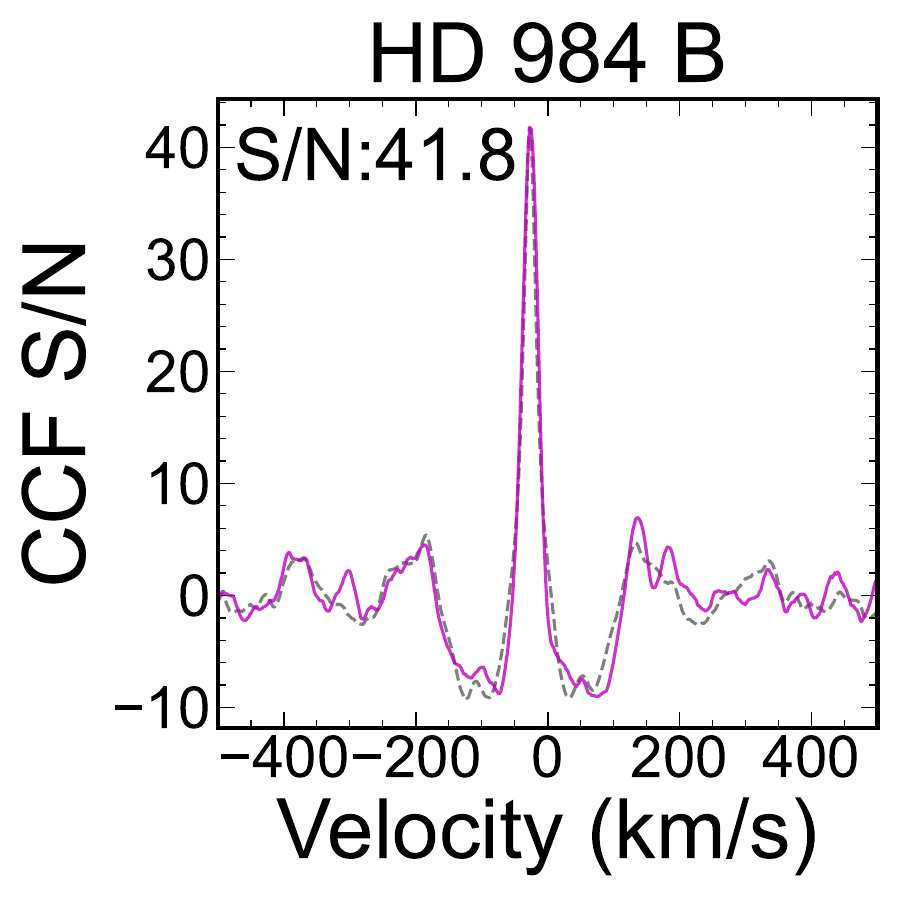}} 
    \subfigure{\includegraphics[width=0.19\textwidth]{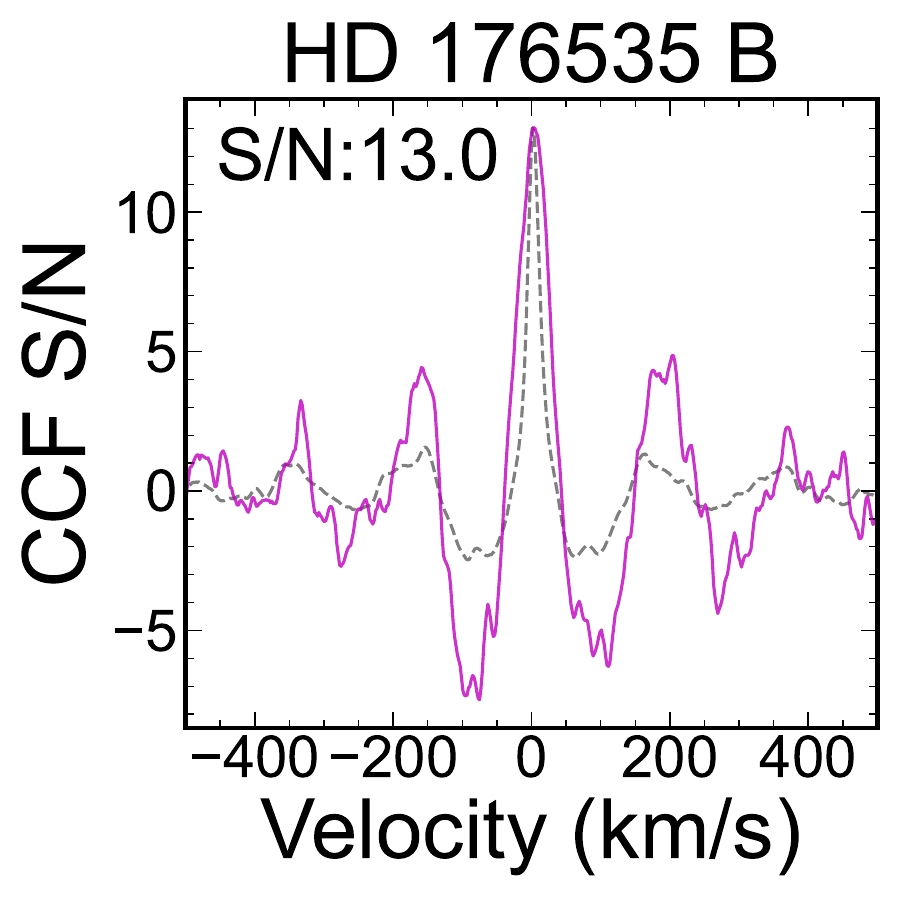}} 
    \subfigure{\includegraphics[width=0.19\textwidth]{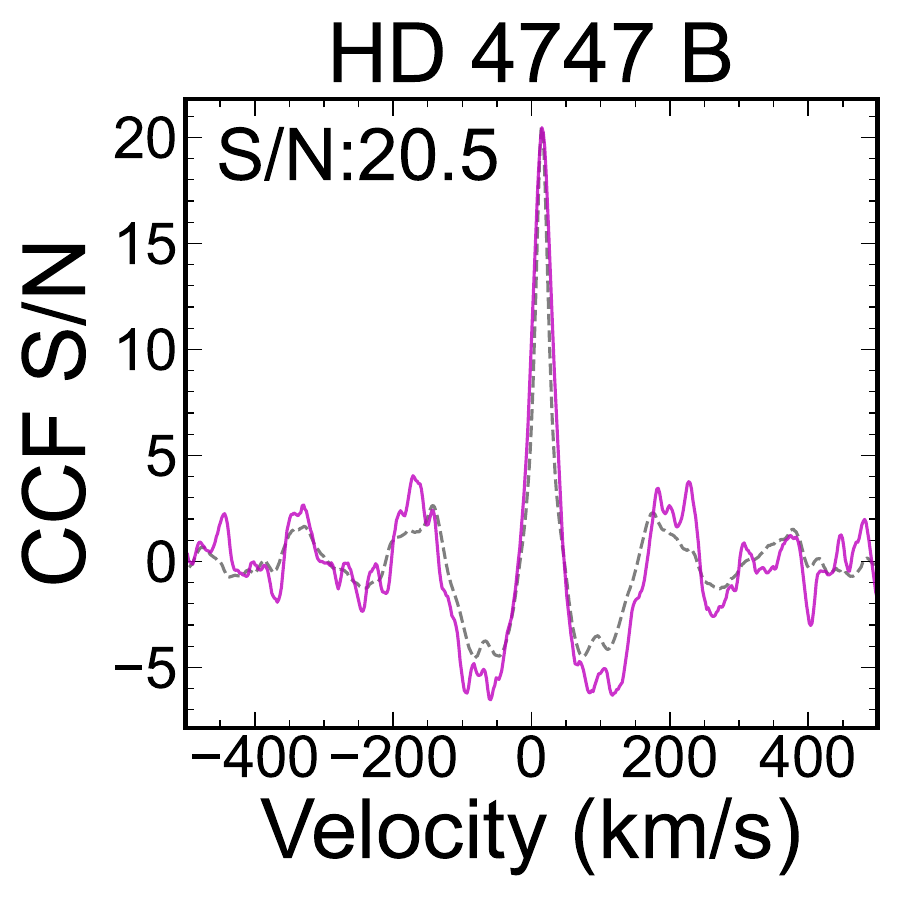}} 
    \subfigure{\includegraphics[width=0.19\textwidth]{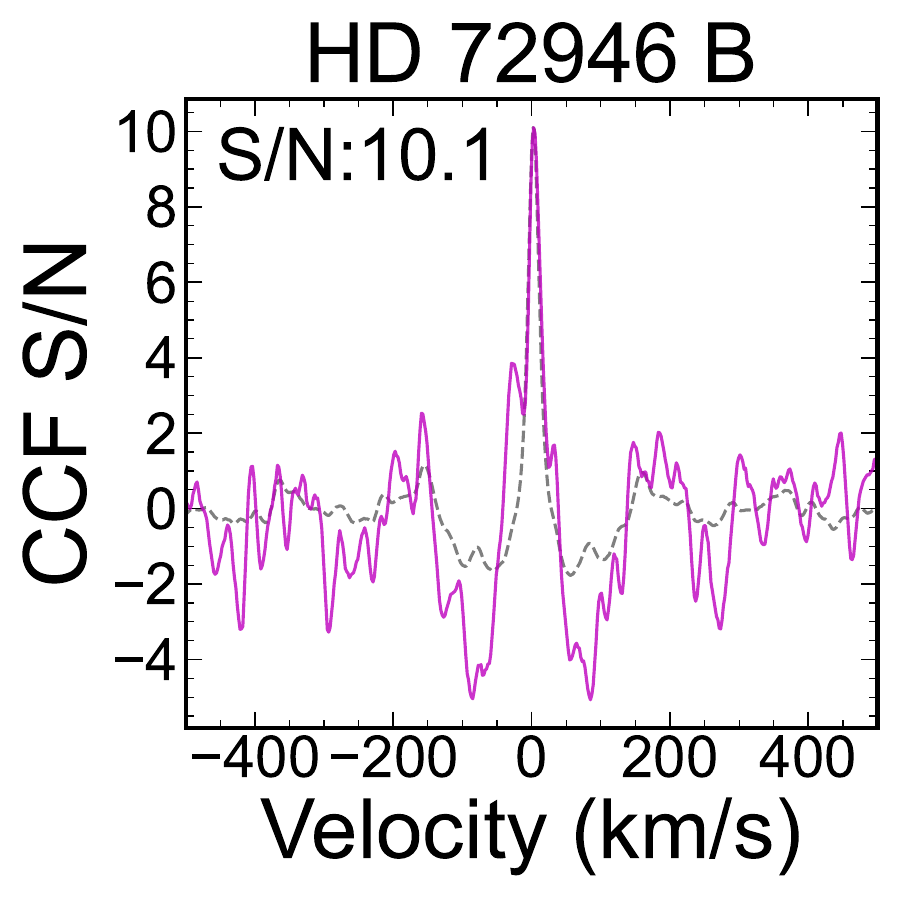}} 

    \subfigure{\includegraphics[width=0.19\textwidth]{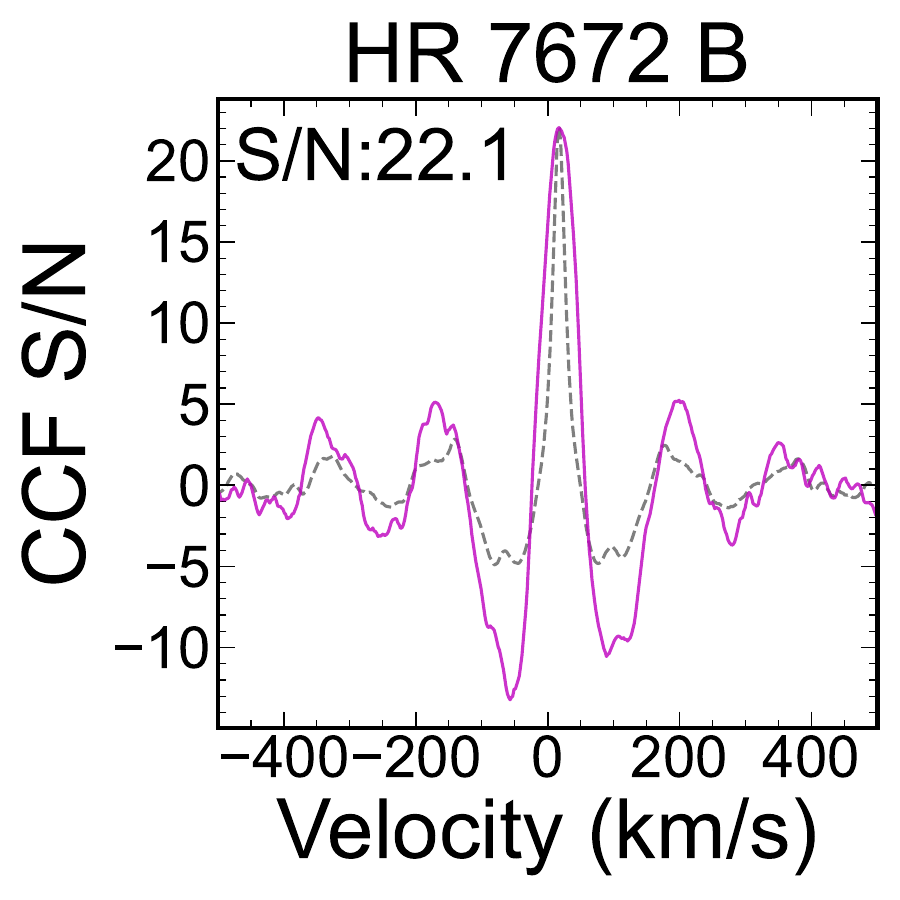}} 
    \subfigure{\includegraphics[width=0.19\textwidth]{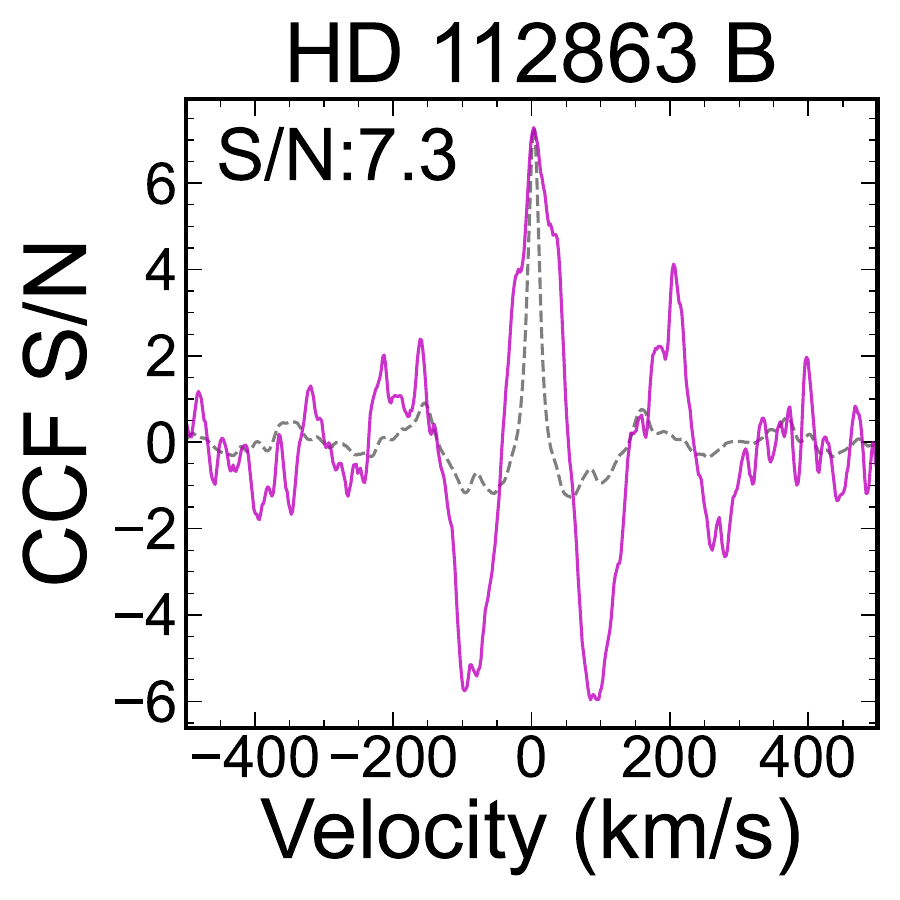}} 
    \subfigure{\includegraphics[width=0.19\textwidth]{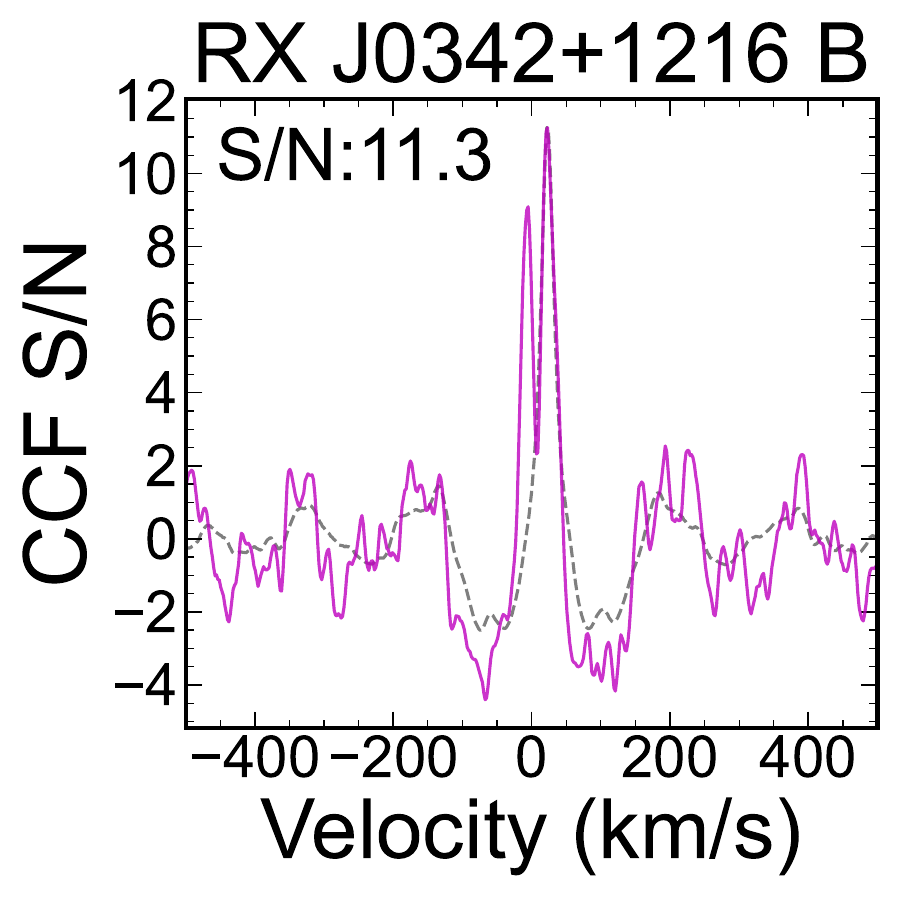}} 
    \subfigure{\includegraphics[width=0.19\textwidth]{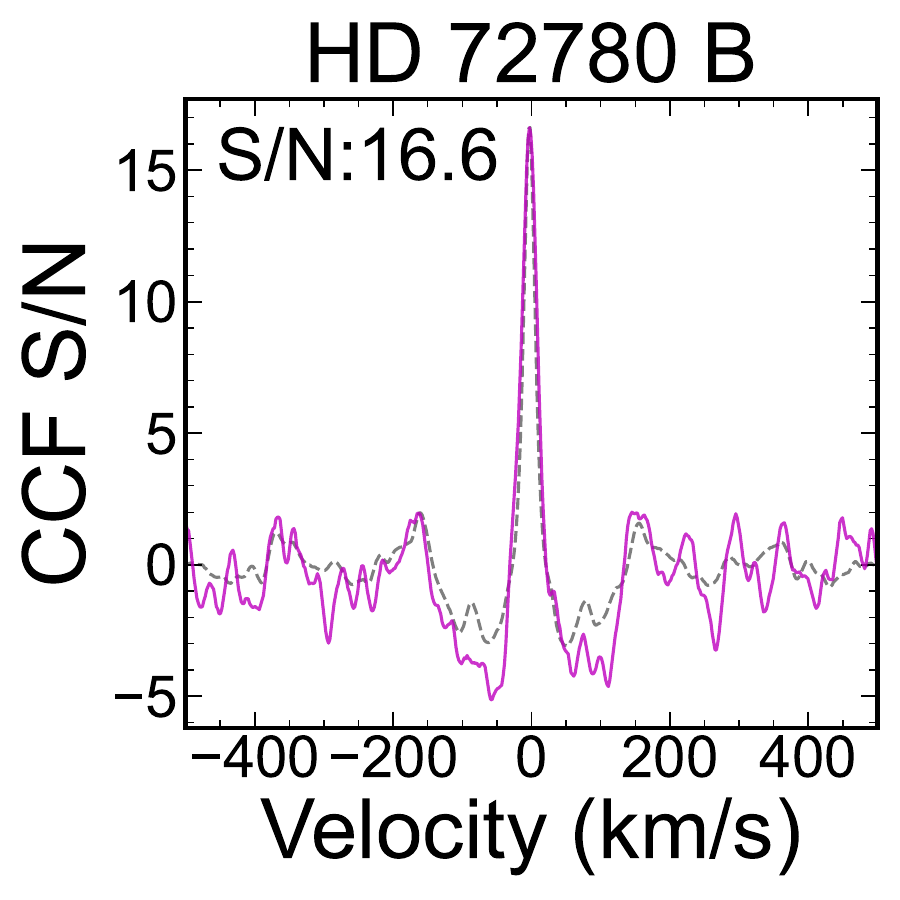}} 
    \subfigure{\includegraphics[width=0.19\textwidth]{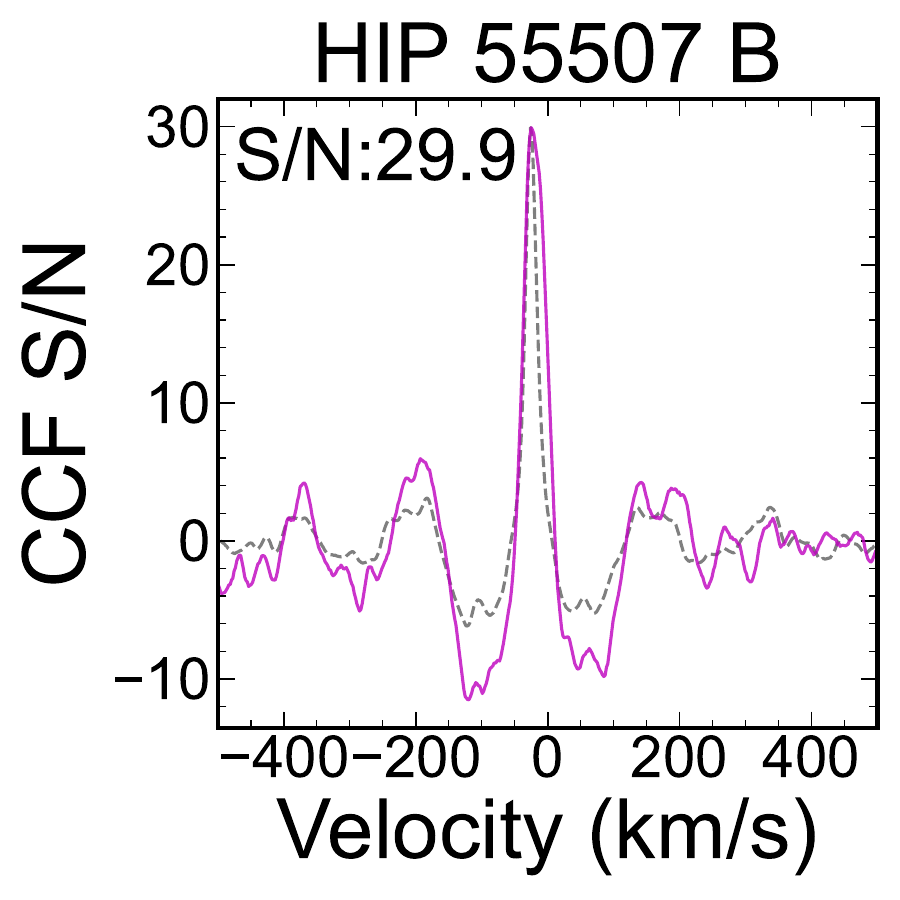}} 
    \caption{Cross-correlation function (CCF) detections for our KPIC sample against the Sonora Bobcat CO and H$_2$O template. The CCF and autocorrelation function of the template are labeled in magenta solid and grey dashed lines, respectively.
    }
    \label{fig:ccf}
\end{figure*}

The resulting CCFs of our full sample are illustrated in Figure~\ref{fig:ccf}, and Table~\ref{tab:observations} summarizes the CCF signal-to-noise ratios (SNRs).
Our CCF SNRs range from 4 to 42.
We note that the spectral inversion method, or retrievals, typically fits the data better and the corresponding templates would provide higher CCF signal-to-noise ratios \citep{Xuan:2022aa, Hsu:2024ab}. 
However, the CCFs used here are mainly used in detection, as the choice of spectral templates is not extremely sensitive for detection \citep{Agrawal:2023aa}.
The two targets in our sample not shown in our CCFs are M8 and M9 LP 349-25 AB.
LP 349-25 A is indeed the primary, and the companion LP 349-25 B has almost the same temperature and spectral type as the primary, hindering our CCF analysis. Nonetheless, their bulk atmospheric parameters can be independently measured in our forward modeling routine in Section~\ref{sec:model}.
Since our molecular templates are not rotationally broadened, the CCFs of fast rotators ({\vsini} $>$ 30~{\kms}) would appear broader compared to the molecular autocorrelation function, including kap And B, RX J2351.5+3127 B, HD 33632 Ab, HD 176535 B, HR 7672 B, and HD 112863 B.
It is noted that for these rotators the CCF S/N can be slightly improved if the templates are broadened to match the objects' rotation \citep{Hsu:2024ab}, but our main goal of CCF analysis is to ensure the detection.
We further quantify and measure their projected rotational velocities and atmospheric parameters in Section~\ref{sec:model}.

\section{Forward-modeling Method} \label{sec:model}

We used a forward-modeling method to derive the physical parameters of our targets \citep{Wang:2021aa, Hsu:2024ab}.
In short, we fit our KPIC companion on-axis spectra (31 targets), using star on-axis spectra and theoretical atmosphere models to derive companion atmospheric parameters.
The atmosphere models used in this study are BT-Settl CIFIST \citep{Baraffe:2015aa} or Sonora Bobcat \citep{Marley:2021aa} atmosphere model grids.
Our forward model composites 4 companion atmosphere parameters (effective temperature, surface gravity, radial velocity, and {\vsini}), 1 parameter for companion flux, and 12 other nuisance parameters, including star fluxes, line-spread function scale, and noise jitter term in each order, for a total of 17 parameters.
The best-fit parameters are derived under nested sampling with \texttt{DYNESTY} \citep{Speagle:2020aa} using 1000 live points and the default stopping criteria below $\epsilon = 10^{-3} ×(K-1)+0.01 = 1.009$, where $K$ = 1000 is the number of live points.
Interested readers are referred to Section 4 and equation~1 in \cite{Hsu:2024ab} for further details.

The only exception to this forward model in our fitting routine is M8 LP 349-25 A \citep{Forveille:2005aa}. Since it shares a similar mass and temperature to the companions in this work, we also derived its atmospheric parameters using \texttt{SMART} \citep{Hsu:2021aa, Hsu:2021ab} under Markov chain Monte Carlo (MCMC) sampling code \texttt{emcee} \citep{Foreman-Mackey:2013aa}.
The method is the same as PDS 70 A from \cite{Hsu:2024ac}, but here we utilized BT-Settl model grids \citep{Allard:2012ab}.
We used 50 walkers, 600 steps, with the last 400 steps as burn-ins using two steps of MCMC fittings, following \cite{Hsu:2021aa}, and the convergence occurred after the first 100 steps, verified by visual inspection.
The {\vsini} is largely insensitive to the choice of models \citep{Hsu:2024ab} thanks to the well-characterized CO bandhead at $\sim$2.3~{\micron}.
For the case of LP 349-25 A,
our {\vsini} value of 55.4 $\pm$ 1.3~{\kms} is fully consistent with the {\vsini} of 55.2 $\pm$ 1.2~{\kms} using the BT-Settl CIFIST models and the literature {\vsini} of 55 $\pm$ 2~{\kms} from \cite{Konopacky:2012aa}, obtained from Keck/NIRSPEC (NIRSPAO) spectra using the PHOENIX NextGen models \citep{Hauschildt:1999aa}.

\section{Projected Rotational Velocity Measurements} \label{sec:vsini}

\begin{deluxetable*}{lccccccccc}
\tablewidth{700pt}
\tablecaption{KPIC Sample Spin Measurements \label{tab:measurements}} 
\tabletypesize{\scriptsize} 
\tablehead{ 
\colhead{Comp.} & 
\colhead{{\vsini}} & 
\colhead{{\vsini}$_\mathrm{lit}$} & 
\colhead{$v_\mathrm{rot}$} & 
\colhead{$v_\mathrm{rot}$} & 
\colhead{$v_\mathrm{rot}/v_\mathrm{break}$} & 
\colhead{$v_\mathrm{rot}/v_\mathrm{break}$} & 
\colhead{$v_\mathrm{rot}/v_\mathrm{break}$} & 
\colhead{$v_\mathrm{rot}/v_\mathrm{break}$} & 
\colhead{Ref.\tablenotemark{b}}
 \\
\colhead{} & \colhead{({\kms})} & \colhead{({\kms})} & 
\colhead{({\tiny Uniform})} & \colhead{({\tiny Orbit})} &
\colhead{({\tiny Uniform})} & \colhead{({\tiny Orbit})} & \colhead{({\tiny Uniform, 10~Myr})} & \colhead{({\tiny Orbit, 10~Myr})} & \colhead{}
} 
\startdata
\hline
PDS 70 b & $<29$ & $<29$ & $40.6^{+64.2}_{-10.8}$ & $40.8^{+78.2}_{-11.2}$ & $0.91^{+1.38}_{-0.27}$ & $0.9^{+1.68}_{-0.26}$ & $0.41^{+0.9}_{-0.44}$ & $0.34^{+0.36}_{-0.35}$ & (7) \\ 
AF Lep b & $22^{+6.0}_{-4.0}$ & $12.2 \pm 2.5$ & $33.0^{+49.3}_{-11.5}$ & $26.0^{+8.0}_{-6.8}$ & $0.46^{+0.7}_{-0.16}$ & $0.37^{+0.11}_{-0.1}$ & $0.46^{+0.93}_{-0.19}$ & $0.35^{+0.11}_{-0.1}$ & (2) \\ 
HR 8799 b & $16^{+5.0}_{-4.0}$ & $31.2 \pm 11.9$ & $24.1^{+35.4}_{-9.1}$ & $33.2^{+79.4}_{-16.0}$ & $0.26^{+0.39}_{-0.1}$ & $0.36^{+0.85}_{-0.17}$ & $0.23^{+0.37}_{-0.08}$ & $0.35^{+0.12}_{-0.11}$ & (12) \\ 
HR 8799 c & $8.5^{+1.4}_{-1.5}$ & $8.1 \pm 4.0$ & $12.6^{+26.4}_{-3.9}$ & $18.5^{+32.3}_{-8.0}$ & $0.12^{+0.24}_{-0.04}$ & $0.18^{+0.33}_{-0.08}$ & $0.11^{+0.19}_{-0.03}$ & $0.17^{+0.04}_{-0.03}$ & (12) \\ 
HR 8799 d & $13.5^{+1.8}_{-1.7}$ & $10.1 \pm 2.8$ & $20.0^{+32.7}_{-5.8}$ & $29.9^{+4.9}_{-4.0}$ & $0.2^{+0.31}_{-0.06}$ & $0.29^{+0.05}_{-0.04}$ & $0.17^{+0.34}_{-0.05}$ & $0.26 \pm 0.05$ & (12) \\ 
HR 8799 e & $14.9^{+1.6}_{-1.9}$ & $15 \pm 2.6$ & $21.7^{+44.4}_{-6.5}$ & $32.3^{+76.8}_{-14.0}$ & $0.21^{+0.43}_{-0.06}$ & $0.31^{+0.74}_{-0.13}$ & $0.19^{+0.34}_{-0.06}$ & $0.29^{+0.06}_{-0.05}$ & (12) \\ 
DH Tau B & $5.5^{+1.6}_{-2.4}$ & $5.7 \pm 0.9$ & $8.5^{+14.0}_{-4.1}$ & $6.3^{+3.2}_{-3.0}$ & $0.1^{+0.17}_{-0.05}$ & $0.07^{+0.04}_{-0.03}$ & $0.12^{+0.2}_{-0.07}$ & $0.09^{+0.06}_{-0.04}$ & (16) \\ 
ROXs 42 Bb & $5.4^{+4.3}_{-3.6}$ & $4.4 \pm 1.8$ & $8.1^{+15.1}_{-6.8}$ & $9.8^{+20.7}_{-7.3}$ & $0.08^{+0.16}_{-0.06}$ & $0.1^{+0.21}_{-0.07}$ & $0.09^{+0.2}_{-0.07}$ & $0.11^{+0.15}_{-0.09}$ & (16) \\ 
HIP 99770 B & $6.5^{+2.3}_{-2.9}$ & $3.5 \pm 1.9$ & $9.9^{+17.3}_{-4.7}$ & $12.9^{+10.2}_{-6.3}$ & $0.06^{+0.11}_{-0.03}$ & $0.08^{+0.07}_{-0.04}$ & $0.05^{+0.08}_{-0.02}$ & $0.06^{+0.06}_{-0.03}$ & (17) \\ 
ROXs 12 B & $3.4^{+2.1}_{-2.2}$ & $3.6 \pm 1.4$ & $5.3^{+8.7}_{-3.3}$ & $4.8^{+5.3}_{-3.1}$ & $0.04^{+0.07}_{-0.03}$ & $0.04^{+0.05}_{-0.02}$ & $0.04^{+0.08}_{-0.03}$ & $0.04^{+0.05}_{-0.03}$ & (16) \\ 
GSC 6214-210 B & $9.6^{+3.5}_{-4.5}$ & $11.6 \pm 1.9$ & $14.8^{+23.7}_{-7.6}$ & $10.8^{+5.9}_{-5.4}$ & $0.1^{+0.17}_{-0.05}$ & $0.07^{+0.04}_{-0.03}$ & $0.08^{+0.15}_{-0.04}$ & $0.06^{+0.04}_{-0.03}$ & (16) \\ 
kap And B & $40.2^{+2.9}_{-4.5}$ & $39.4 \pm 1.3$ & $54.7^{+93.6}_{-14.4}$ & $74.3^{+57.8}_{-20.8}$ & $0.35^{+0.61}_{-0.13}$ & $0.47^{+0.4}_{-0.17}$ & $0.31^{+0.67}_{-0.14}$ & $0.37^{+0.47}_{-0.16}$ & (16) \\ 
HD 206893 B & $9.7^{+2.9}_{-3.5}$ & $9.3 \pm 2.3$ & $14.0^{+27.1}_{-5.8}$ & $23.0^{+41.1}_{-10.6}$ & $0.07^{+0.14}_{-0.03}$ & $0.12^{+0.21}_{-0.06}$ & $0.05^{+0.1}_{-0.02}$ & $0.08^{+0.14}_{-0.04}$ & (11) \\ 
HIP 21152 B & $12.6^{+3.7}_{-4.0}$ & \nodata & $19.1^{+27.8}_{-7.9}$ & $12.8^{+3.9}_{-4.4}$ & $0.09^{+0.14}_{-0.04}$ & $0.06 \pm 0.02$ & $0.06^{+0.11}_{-0.03}$ & $0.04^{+0.02}_{-0.01}$ &  \\ 
2M0122-2439 B & $24.2 \pm 1.0$ & $19.6 \pm 2.9$ & $34.1^{+56.8}_{-8.7}$ & $25.4^{+2.7}_{-1.5}$ & $0.19^{+0.37}_{-0.07}$ & $0.14^{+0.06}_{-0.03}$ & $0.14^{+0.28}_{-0.07}$ & $0.09^{+0.05}_{-0.02}$ & (16) \\ 
HIP 79098 AB b & $5.2^{+0.9}_{-1.1}$ & $4 \pm 1.0$ & $7.7^{+14.7}_{-2.5}$ & $6.0^{+2.0}_{-1.2}$ & $0.06^{+0.11}_{-0.02}$ & $0.04^{+0.03}_{-0.01}$ & $0.06^{+0.11}_{-0.03}$ & $0.04^{+0.03}_{-0.02}$ & (16) \\ 
CD-35 2722 B & $11.2 \pm 0.4$ & \nodata & $15.1^{+22.6}_{-3.5}$ & $23.3^{+30.8}_{-8.3}$ & $0.08^{+0.11}_{-0.02}$ & $0.11^{+0.17}_{-0.04}$ & $0.05^{+0.09}_{-0.02}$ & $0.08^{+0.12}_{-0.03}$ &  \\ 
RX J2351.5+3127 B & $30.4 \pm 3.1$ & \nodata & $43.6^{+84.4}_{-12.4}$ & $37.9^{+11.7}_{-5.9}$ & $0.21^{+0.39}_{-0.07}$ & $0.17^{+0.06}_{-0.03}$ & $0.13^{+0.26}_{-0.04}$ & $0.11^{+0.04}_{-0.02}$ &  \\ 
GQ Lup B & $6.5^{+1.2}_{-1.6}$ & $6.4 \pm 0.4$ & $9.4^{+17.9}_{-3.2}$ & $9.0^{+5.4}_{-2.6}$ & $0.08^{+0.15}_{-0.03}$ & $0.07^{+0.05}_{-0.02}$ & $0.09^{+0.15}_{-0.04}$ & $0.08^{+0.07}_{-0.03}$ & (16) \\ 
HD 1160 B & $21.5^{+2.3}_{-2.2}$ & \nodata & $31.2^{+51.9}_{-8.9}$ & $21.8^{+2.2}_{-2.4}$ & $0.1^{+0.16}_{-0.04}$ & $0.06^{+0.02}_{-0.01}$ & $0.07^{+0.12}_{-0.03}$ & $0.04 \pm 0.01$ &  \\ 
HD 33632 Ab & $53 \pm 3.0$ & $53 \pm 3.0$ & $74.1^{+141.8}_{-18.9}$ & $77.9^{+22.9}_{-12.5}$ & $0.23^{+0.44}_{-0.06}$ & $0.24^{+0.07}_{-0.04}$ & $0.12^{+0.21}_{-0.04}$ & $0.13^{+0.04}_{-0.03}$ & (6) \\ 
HD 984 B & $14.3^{+0.17}_{-0.15}$ & $12.72 \pm 0.03$ & $19.8^{+31.2}_{-5.0}$ & $16.7 \pm 0.4$ & $0.08^{+0.13}_{-0.02}$ & $0.07 \pm 0.01$ & $0.06^{+0.1}_{-0.02}$ & $0.05 \pm 0.02$ & (1) \\ 
HD 176535 B & $37.6 \pm 3.3$ & \nodata & $54.5^{+94.8}_{-15.3}$ & $50.0^{+5.3}_{-4.8}$ & $0.15^{+0.26}_{-0.04}$ & $0.14 \pm 0.02$ & $0.08^{+0.17}_{-0.02}$ & $0.07 \pm 0.01$ &  \\ 
LP 349-25 B & $79.6^{+2.4}_{-2.6}$ & $83 \pm 3.0$ & $114.5^{+197.6}_{-31.9}$ & $110.2^{+223.8}_{-27.3}$ & $0.37^{+0.61}_{-0.1}$ & $0.35^{+0.71}_{-0.09}$ & $0.02^{+0.04}_{-0.01}$ & $0.02 \pm 0.0$ & (9) \\ 
HD 4747 B & $15.9^{+1.8}_{-1.6}$ & $13.2 \pm 1.5$ & $23.1^{+42.4}_{-6.7}$ & $22.3^{+31.2}_{-5.9}$ & $0.06^{+0.11}_{-0.02}$ & $0.06^{+0.09}_{-0.02}$ & $0.03^{+0.06}_{-0.01}$ & $0.03^{+0.01}_{-0.0}$ & (14) \\ 
HD 72946 B & $43.8^{+3.4}_{-3.1}$ & \nodata & $62.2^{+120.8}_{-16.7}$ & $58.4^{+105.9}_{-12.7}$ & $0.16^{+0.31}_{-0.04}$ & $0.15^{+0.27}_{-0.03}$ & $0.08^{+0.16}_{-0.02}$ & $0.06 \pm 0.0$ &  \\ 
HR 7672 B & $41.5^{+0.7}_{-0.6}$ & $45 \pm 0.5$ & $53.0^{+94.8}_{-12.4}$ & $53.5^{+91.7}_{-12.7}$ & $0.14^{+0.25}_{-0.03}$ & $0.14^{+0.24}_{-0.03}$ & $0.08^{+0.14}_{-0.02}$ & $0.06 \pm 0.01$ & (13) \\ 
HD 112863 B & $61.1^{+3.0}_{-3.7}$ & \nodata & $85.6^{+136.9}_{-21.8}$ & $81.2^{+115.2}_{-18.5}$ & $0.23^{+0.35}_{-0.06}$ & $0.21^{+0.31}_{-0.05}$ & $0.11^{+0.21}_{-0.03}$ & $0.09^{+0.02}_{-0.01}$ &  \\ 
RX J0342.5+1216 B & $32.6^{+1.1}_{-1.2}$ & $28.4 \pm 1.3$ & $45.6^{+83.6}_{-11.9}$ & $32.9^{+1.1}_{-1.2}$ & $0.14^{+0.22}_{-0.05}$ & $0.09 \pm 0.03$ & $0.08^{+0.15}_{-0.06}$ & $0.05 \pm 0.04$ & (3) \\ 
LP 349-25 A & $55.4 \pm 1.3$ & $55 \pm 2.0$ & $80.1^{+165.2}_{-22.9}$ & $79.5^{+163.4}_{-22.1}$ & $0.23^{+0.48}_{-0.07}$ & $0.23^{+0.48}_{-0.06}$ & $0.01^{+0.02}_{-0.0}$ & $0.01 \pm 0.0$ & (9) \\ 
HD 72780 B & $7.7^{+1.0}_{-1.3}$ & \nodata & $11.1^{+18.6}_{-3.2}$ & $11.3^{+18.1}_{-3.5}$ & $0.03^{+0.05}_{-0.01}$ & $0.03^{+0.05}_{-0.01}$ & $0.01^{+0.03}_{-0.0}$ & $0.02 \pm 0.0$ &  \\ 
HIP 55507 B & $4.6 \pm 0.3$ & $5.5 \pm 0.3$ & $6.7^{+14.1}_{-1.9}$ & $6.7^{+11.6}_{-1.9}$ & $0.02^{+0.04}_{-0.01}$ & $0.02^{+0.03}_{-0.01}$ & $0.01^{+0.02}_{-0.0}$ & $0.01 \pm 0.0$ & (15) \\ 
YSES 1 c\tablenotemark{a} & \nodata & $11.3 \pm 2.1$ & $15.8^{+28.9}_{-4.6}$ & $11.4^{+2.0}_{-2.2}$ & $0.19^{+0.33}_{-0.06}$ & $0.13 \pm 0.03$ & $0.2^{+0.36}_{-0.07}$ & $0.14 \pm 0.03$ & (18) \\ 
beta Pic b\tablenotemark{a} & \nodata & $19.9 \pm 1.0$ & $27.8^{+53.0}_{-7.1}$ & $19.9 \pm 1.0$ & $0.24^{+0.43}_{-0.07}$ & $0.16^{+0.03}_{-0.02}$ & $0.22^{+0.42}_{-0.08}$ & $0.15^{+0.04}_{-0.03}$ & (10) \\ 
AB Pic b\tablenotemark{a} & \nodata & $3.7 \pm 1.0$ & $5.7^{+9.4}_{-2.2}$ & $5.4^{+9.2}_{-1.9}$ & $0.05^{+0.08}_{-0.02}$ & $0.04^{+0.07}_{-0.02}$ & $0.04^{+0.06}_{-0.01}$ & $0.03 \pm 0.01$ & (4) \\ 
YSES 1 b\tablenotemark{a} & \nodata & $5.34 \pm 0.14$ & $7.4^{+14.9}_{-1.9}$ & $5.3 \pm 0.1$ & $0.06^{+0.13}_{-0.02}$ & $0.05^{+0.01}_{-0.0}$ & $0.07^{+0.15}_{-0.02}$ & $0.05 \pm 0.01$ & (18) \\ 
\enddata
\tablerefs{(1) \cite{Costes:2024aa}; (2) \cite{Denis:2025aa}; (3) \cite{Do-O:2024aa}; (4) \cite{Gandhi:2025aa}; (5) \cite{Gontcharov:2006aa}; (6) \cite{Hsu:2024ab}; (7) \cite{Hsu:2024ac}; (8) \cite{Kharchenko:2007aa}; (9) \cite{Konopacky:2012aa}; (10) \cite{Landman:2024aa}; (11) \cite{Sappey:2025aa}; (12) \cite{Wang:2021aa}; (13) \cite{Wang:2022aa}; (14) \cite{Xuan:2022aa}; (15) \cite{Xuan:2024aa}; (16) \cite{Xuan:2024ab}; (17) \cite{Zhang:2024aa}; (18) \cite{Zhang:2024ac}.}
\tablenotetext{a}{Additional high-resolution spectroscopic measurements from the literature. See Section~\ref{sec:rotation} for details.}
\tablenotetext{b}{References for the literature {\vsini} measurements. See Section~\ref{sec:rotation} for details.}
\end{deluxetable*}

Our best-fit {\vsini} measurements are summarized in Table~\ref{tab:measurements}.
The radial velocities derived in this study will be reported in our KPIC multi-epoch radial velocity survey (K. Horstman et al. in prep.).
To ensure the robustness and consistency of our fitting results, we compare our derived {\vsini} values and discuss them as follows.

\begin{figure}
    \centering
    \includegraphics[width=\columnwidth]{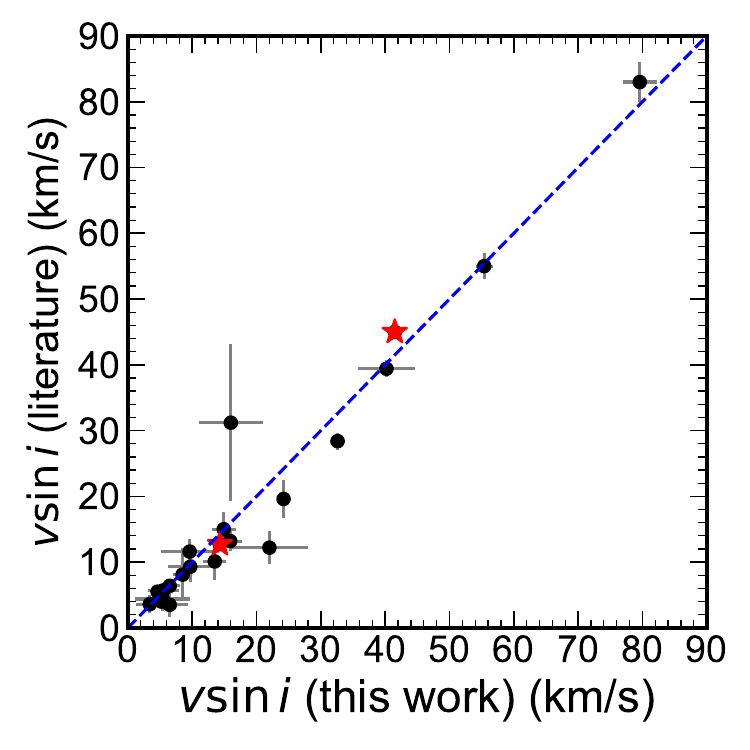}
    \caption{Comparison of our {\vsini} measurements with literature. Our {\vsini} measurements versus literature measurements are depicted in black dots. The blue dashed line represents the perfect agreement. The outliers ($>$2.5~$\sigma$) are labeled in red stars. See Section~\ref{sec:vsini} for details.
    }
    \label{fig:literature_vsini}
\end{figure}

Our {\vsini} measurements are largely consistent with the literature measurements ($\sim$90\% of our sample). 
Figure~\ref{fig:literature_vsini} compares our {\vsini} measurements with the literature measurements.
Excluding PDS 70b (which has only an upper limit of $<$29~{\kms}) and HD 33632 Ab (which uses the same fitting routine), 21 targets have independent literature {\vsini} measurements (out of 32 total targets).
19 out of 21 targets are consistent within 2.5~$\sigma$.
Two outlier targets, HR 7672 B and HD 984 B, are discussed below.

HR 7672 B is a relatively fast rotator ({\vsini} = $41.5^{+0.7}_{-0.6}$~{\kms}), which is at 3.5~{\kms} and 4.1~$\sigma$ discrepancy compared to the {\vsini} measurement of $45 \pm 0.5$~{\kms} reported in \cite{Wang:2022aa}.
\cite{Wang:2022aa} adopted a retrieval framework using more flexible pressure-temperature (PT) profiles from \cite{Piette:2020aa}. 
The PT profiles from \cite{Piette:2020aa} model the profiles 8 at pressure levels (100.0, 33.3, 10.0, 3.3, 1.0, 0.1, 0.001, 0.00001 bar), which show slight deviations from self-consistent PHOENIX PT profiles (see their Figures 7 and 12).
However, their retrievals using PHOENIX profiles give {\vsini} = $39.6 \pm 1.4$~{\kms}, which is consistent with our measurement at the 1.2$\sigma$ level.
\cite{Delorme:2021aa} reported HR 7672 B {\vsini} $42.6 \pm 0.8$~{\kms}, fully consistent at $\sim$1~$\sigma$ with our measurement.
Therefore, the discrepancy is likely due to the different treatment of pressure-temperature profiles, but we leave resolving this discrepancy to a future investigation as our {\vsini} measurements are generally consistent with the literature.

Finally, our HD 984 B {\vsini} (14.3$^{+0.17}_{-0.15}$~{\kms}) is highly discrepant at 9.2$\sigma$ compared to the {\vsini} of 12.72 $\pm$ 0.03 from \cite{Costes:2024aa}, who used a \texttt{petitRADTRANS} retrieval framework \citep{Molliere:2019ab}.
Their PT profiles were adopted from \cite{Molliere:2020aa}, which model the pressure levels from 10$^{-4}$ to 10$^{2}$~bar into three regimes from high altitudes, photosphere, and troposphere (low altitudes).
However, the absolute {\vsini} difference is small at 1.6~{\kms} and has little effect on deriving the associated rotational velocities for this study.
Similar to HR 7672 B, the {\vsini} difference of 1.6~{\kms} could be attributed to the pressure-temperature profile deviation from the Sonora Bobcat equilibrium model \citep{Marley:2021aa} (around 0.09--0.5~bar; see their Fig. 6).
While the Sonora PT profiles also have non-smooth PT profile gradients near the radiative-convective boundary around 1600-1800~K (where the second-order derivative of the profiles is positive), their steeper retrieved PT profile gradient is deeper in the atmosphere from 2000-2400~K  (where the second-order derivative of the PT profile is negative).
We speculate that these deviated retrieved pressure levels correspond roughly to the photosphere and affect the spectral line profiles the most, but a more detailed characterization of these systems is needed in the future to validate our hypothesis.

\section{Mass and Age Dependence with Rotation} \label{sec:rotation}

We place our {\vsini} measurements at a population level to examine rotation as a function of mass and age and compare the literature spin measurements of stellar/substellar objects ($M \lesssim$0.1~{\msun}) to the broader context.

\subsection{Spin Samples and Derived Parameters}\label{subsec:spin_sample_and_method}

We construct two samples, our benchmark sample using our KPIC {\vsini} measurements and literature spin measurements for 12 planets and brown dwarf companions, along with a much larger literature spin sample. 
The benchmark sample typically has more reliable ages from the host star or young clusters, and masses determined from their orbits. 
The latter sample consists mostly of isolated stellar/substellar objects, allowing us to glimpse a bigger picture of the angular momentum evolution at a population level.

Our benchmark spin sample consists of benchmark stellar/substellar companions ($\sim$12--90~{\mjup}) spanning from the deuterium mass burning limit to the stellar/substellar boundary, along with bonafide giant exoplanets ($\sim$2--7~{\mjup}) (Section~\ref{sec:sample}).
Additionally, to make our giant planet and low-mass brown dwarf companion sample more complete, we also include beta Pic b \citep{Landman:2024aa}, AB Pic b \citep{Gandhi:2025aa}, YSES 1 b and c \citep{Zhang:2024ac} {\vsini} measurements, and planetary mass companions 2MASS J1207$-$3932 b \citep{Zhou:2016aa} and Ross 458 C \citep{Manjavacas:2019ab} as well as GU Psc B \citep{Naud:2017aa}, 2MASS J0249$-$0557 c \citep{Bryan:2020ab}, VHS J1256$-$1257 b \citep{Zhou:2020ab, Poon:2024aa}, HD 203030 B \citep{Miles-Paez:2019aa}, LP 261-75 B \citep{Manjavacas:2018aa}, and HN Peg B \citep{Zhou:2018aa, Hsu:2021aa}.
PDS 70 b is removed in the following analysis, since its {\vsini} has only the upper limit (29~{\kms}) close to its breakup velocity \citep{Hsu:2024ac}.
The four additional CRIRES+ targets are included in Table~\ref{tab:sample} \& Table~\ref{tab:measurements}, and the remaining targets are summarized in Table~\ref{tab:literature_spin}, comprising our full benchmark spin sample of 43 targets in the following analysis.

We also compiled a literature spin sample with {\vsini} and/or photometric rotational period measurements, summarized in Table~\ref{tab:literature_spin}.
The literature compilation follows the sample compiled in \cite{Hsu:2024ab}, with the majority of the measurements compiled from \cite{Crossfield:2014aa, Bryan:2020ab, Popinchalk:2021aa, Tannock:2021aa, Wang:2021aa, Hsu:2021aa, Vos:2022aa, Hsu:2024aa} for a total of 1812 objects.
We estimated their masses via \cite{Baraffe:2003aa} evolutionary models using their photometry (or spectral types for sources younger than 10~Myr) and ages (see below). The masses of the full sample span from planetary mass objects, to brown dwarfs, to low-mass stars.
A significant fraction of the full sample has masses above 0.1~{\msun}, outside the mass range of \cite{Baraffe:2003aa} models, and 336 sources have inferred masses below 0.1~{\msun}, 212 sources of which have known ages in young clusters or young moving groups.
Among these young sources, Upper Scorpius and Praesepe comprise two-thirds of the young sample from \cite{Popinchalk:2021aa}.
Other than the aforementioned literature benchmark planets/companions, the notable isolated lowest-mass objects in the full sample are young free-floating objects TWA 41 \citep{Schneider:2018aa}, TWA 42AB \citep{Schneider:2018aa}, and PSO J318.5338$-$22.8603 \citep{Biller:2018aa, Bryan:2018aa}.
While this sample may suffer from inaccurate masses due to the models' physical assumptions and their opacities (e.g., \citealp{Filippazzo:2015aa, Tannock:2022aa, Hsu:2024aa}) or unresolved binaries (e.g. \citealp{Burgasser:2010aa, Bardalez-Gagliuffi:2014aa, Hsu:2021aa, Hsu:2023aa, Xuan:2024ac}), this sample can still provide the full picture in substellar rotational evolution.

We infer the rotational velocity ($v_\mathrm{rot}$) to compare the two types of spin measurements, following \cite{Hsu:2024ab}. 
In this study, we assume rigid body rotation and ignore the oblateness due to their rotation \citep{Marley:2011aa, Tannock:2021aa, Wang:2021aa}.
For {\vsini} measurements, we derive $v_\mathrm{rot} = v\sin{i}/\sin{i}$ by assuming random orientations of inclinations ($i$) (abbreviated as random inclination) or inclinations aligned with their orbits for companions with orbit inclination measurements (abbreviated as orbit inclinations).
For photometric rotational periods ($P$), we compute $v_\mathrm{rot} = 2\pi R/P$, and by assuming theoretical radius $R$ from the \cite{Baraffe:2003aa} evolutionary models with their absolute $H$ magnitudes and ages (Table~\ref{tab:measurements} and Table~\ref{tab:literature_spin}).
For very young objects with ages $\leq$ 10~Myr, inaccurate reddening correction could lead to skewed absolute magnitude, potentially biasing the mass determination.
Deriving masses using effective temperatures and the cluster ages could mitigate this issue (e.g., Orion Nebular Cluster; \citealp{Wei:2024aa}), and effective temperatures can be inferred from spectral classifications using low-resolution optical or near-infrared spectra available in the literature.
We used the objects with spectral classifications of M5 and later ({\teff} $\lesssim$3000~K), to infer the effective temperatures using \cite{Pecaut:2013aa} (for M5) and \cite{Filippazzo:2015aa} (for M6 and later) empirical relations. 
Using this empirical spectral type, while increasing the mass uncertainties (propagated via the Monte Carlo method) compared to absolute magnitudes, ensures more conservative mass estimates for objects at very young ages.
The masses for these very young objects were then derived using the \cite{Baraffe:2003aa} models from {\teff}s and ages.
The ages are assumed as field ages (500~Myr--10~Gyr) under uniform distribution unless identified in young associations, clusters, or moving groups in the literature and verified the membership using BANYAN $\Sigma$ \citep{Gagne:2018ab}.
BANYAN $\Sigma$ assesses the membership probability of $\sim$30 nearby young moving groups within 150~pc via kinematics, including astrometry, proper motions, parallax, and (not always available) radial velocity, and bona fide members typically require full 6-dimensional kinematics with $>$99\% probability.
We also consider fractional breakup velocity (i.e. the ratio of rotational velocity and breakup velocity $v_\mathrm{breakup} = \sqrt{\frac{GM}{R}}$, where $G$ is the gravitational constant, $M$ is the mass, and $R$ is the radius), as unmagnetized planets can spin up to 60--80\% of their breakup speed \citep{Dong:2021aa}.
Finally, when comparing the spin versus mass, we define the ``initial'' fractional breakup velocity by evolving the spin, including the rotational and breakup velocities, to 10~Myr, assuming constant angular momentum evolution, to remove the age factor (Section~\ref{subsec:spin_age}).
All of these derived parameters have uncertainties propagated through Monte Carlo sampling with 1000 samples.

This assumption of 10~Myr is determined by the typical disk age $<$10~Myr \citep{Hillenbrand:2005aa, Mamajek:2009aa, Bouvier:2014aa}.
The spin-up trend under constant angular momentum evolution after disk dissipation is generally valid, as verified by spin surveys of substellar objects \citep{Zapatero-Osorio:2006aa, Bryan:2020ab, Vos:2022aa}, while some angular momentum could be lost for high-mass brown dwarfs ($\sim$10,000 times weaker than solar-type stars) shown in \cite{Bouvier:2014aa}. 
Some objects at the age of 10~Myr could still have a disk, for example, roughly 30\% of the low-mass stars and high-mass brown dwarfs ($\gtrsim$50~{\mjup}) in the $\sim$10~Myr Upper Scorpius \citep{Moore:2019aa} still possess a stellar/substellar disk, and the objects with disks generally have slower rotation due to disk braking.
Uniformly fitting the mid-infrared photometry to assess if the circumplanetary or circumsubstellar disk around our sample is beyond the scope of this work.
In our KPIC sample (Table~\ref{tab:sample}), PDS 70 b \citep{Christiaens:2019ab}, DH Tau B \citep{van-Holstein:2021aa}, ROXs 42 Bb \citep{Martinez:2022aa}, GQ Lup B \citep{Stolker:2021aa}, and GSC 6214-210 B \citep{Bowler:2011aa} are known to have circumplanetary/circumcompanion disks, while 10~Myr ROXs 12 B does not possess a disk \citep{Bowler:2017aa}, which affirms that 10~Myr is a reasonable age for disk dissipation.
While kap And B has an uncertain age from 5--100~Myr, it is not known to have a disk and is most likely a member of $\sim$42~Myr Columba.

\begin{deluxetable*}{lcccccccccc}
\tablewidth{700pt}
\tablecaption{Literature Spin Measurements \label{tab:literature_spin}} 
\tabletypesize{\scriptsize} 
\tablehead{ 
\colhead{Name} & 
\colhead{RA} & 
\colhead{Dec} & 
\colhead{{\vsini}} & 
\colhead{$P_\mathrm{rot}$} & 
\colhead{Mass} & 
\colhead{Age} & 
\colhead{Parallax} & 
\colhead{2MASS $H$} & 
\colhead{$v_\mathrm{rot}$\tablenotemark{a}} & 
\colhead{References}
 \\
\colhead{} & \colhead{(deg)} & \colhead{(deg)} &
\colhead{({\kms})} & \colhead{(hr)} & 
\colhead{({\mjup})} & \colhead{({Myr})} & 
\colhead{(mas)} & \colhead{(mag)}& \colhead{({\kms})} & \colhead{}
} 
\startdata
\hline
J0045+1634 & 11.339167 & $+$16.579083 & $31.6 \pm 1.2$ & $2.4 \pm 0.1$ & $19.5 \pm 0.8$ & $45.0$ & $65.41 \pm 0.18$ & $12.06 \pm 0.04$ & 73.3 $\pm$ 3.5 & 27, 41, 7, 47, 88, 89, 102 \\
\enddata
\tablerefs{(1) \cite{Allers:2016aa}; (2) \cite{Alves-de-Oliveira:2010aa}; (3) \cite{Alves-de-Oliveira:2012aa}; (4) \cite{Apai:2017aa}; (5) \cite{Artigau:2009aa}; (6) \cite{Bailey:2014aa}; (7) \cite{Baraffe:2003aa}; (8) \cite{Barnes:2007aa}; (9) \cite{Barrado-y-Navascues:2004aa}; (10) \cite{Bell:2015aa}; (11) \cite{Best:2018aa}; (12) \cite{Best:2020aa}; (13) \cite{Best:2021aa}; (14) \cite{Biller:2015aa}; (15) \cite{Biller:2018aa}; (16) \cite{Boro-Saikia:2015aa}; (17) \cite{Bowler:2014aa}; (18) \cite{Bowler:2016aa}; (19) \cite{Bowler:2020aa}; (20) \cite{Brandt:2015aa}; (21) \cite{Bryan:2018aa}; (22) \cite{Bryan:2020ab}; (23) \cite{Chauvin:2005aa}; (24) \cite{Chauvin:2005aa}; (25) \cite{Croll:2016aa}; (26) \cite{Crossfield:2014aa}; (27) \cite{Cutri:2003aa}; (28) \cite{Dahm:2007aa}; (29) \cite{Dahm:2015aa}; (30) \cite{Delorme:2011aa}; (31) \cite{Dittmann:2014aa}; (32) \cite{Dupuy:2012aa}; (33) \cite{Dupuy:2018aa}; (34) \cite{Eriksson:2019aa}; (35) \cite{Faherty:2012aa}; (36) \cite{Faherty:2016aa}; (37) \cite{Gagne:2015ab}; (38) \cite{Gagne:2017aa}; (39) \cite{Gagne:2018ab}; (40) \cite{Gaia-Collaboration:2018ab}; (41) \cite{Gaia-Collaboration:2021ac}; (42) \cite{Gauza:2015aa}; (43) \cite{Gennaro:2012aa}; (44) \cite{Getman:2002aa}; (45) \cite{Goldman:2010aa}; (46) \cite{Hsu:2021aa}; (47) \cite{Hsu:2024aa}; (48) \cite{Kao:2018aa}; (49) \cite{Kenyon:1995aa}; (50) \cite{Kirkpatrick:2021aa}; (51) \cite{Kurosawa:2006aa}; (52) \cite{Kuzuhara:2011aa}; (53) \cite{Lawrence:2013aa}; (54) \cite{Lew:2016aa}; (55) \cite{Lew:2020aa}; (56) \cite{Liu:2016aa}; (57) \cite{Lodieu:2008aa}; (58) \cite{Lodieu:2012aa}; (59) \cite{Looper:2010aa}; (60) \cite{Luhman:2006ab}; (61) \cite{Luhman:2007aa}; (62) \cite{Manjavacas:2018aa}; (63) \cite{Manjavacas:2019ab}; (64) \cite{Meeus:2005aa}; (65) \cite{Metchev:2006aa}; (66) \cite{Metchev:2015aa}; (67) \cite{Miles-Paez:2019aa}; (68) \cite{Mohanty:2005aa}; (69) \cite{Mohanty:2007aa}; (70) \cite{Morales-Calderon:2006aa}; (71) \cite{Nardiello:2020aa}; (72) \cite{Naud:2017aa}; (73) \cite{Newton:2016aa}; (74) \cite{Pecaut:2016aa}; (75) \cite{Popinchalk:2021aa}; (76) \cite{Poon:2024aa}; (77) \cite{Posch:2023aa}; (78) \cite{Radigan:2014aa}; (79) \cite{Rebull:2018aa}; (80) \cite{Reid:2006ab}; (81) \cite{Rice:2010aa}; (82) \cite{Schneider:2018aa}; (83) \cite{Shkolnik:2017aa}; (84) \cite{Stone:2016aa}; (85) \cite{Tinney:2014aa}; (86) \cite{Vos:2017ab}; (87) \cite{Vos:2018aa}; (88) \cite{Vos:2019aa}; (89) \cite{Vos:2020aa}; (90) \cite{Vos:2022aa}; (91) \cite{Watson:2006aa}; (92) \cite{Wilking:2008aa}; (93) \cite{Yang:2016aa}; (94) \cite{Zhang:2021aa}; (95) \cite{Zhou:2016aa}; (96) \cite{Zhou:2018aa}; (97) \cite{Zhou:2019aa}; (98) \cite{Zhou:2020aa}; (99) \cite{Zhou:2020aa}; (100) \cite{Zuckerman:2006aa}; (101) \cite{Zuckerman:2013aa}; (102) \cite{Zuckerman:2019aa}; (103) \cite{Luhman:2023aa}.}
\tablecomments{%
This table shows only one example row of J0045$+$1634 (2MASS J00452143+1634446) in the sample and is available in its entirety in machine-readable form in the online article. 
This table excludes objects presented in Table~\ref{tab:measurements}. See Section~\ref{sec:rotation} for details.}
\tablenotetext{a}{Derived in this work.}
\end{deluxetable*}

\subsection{Age Dependent Trend with Rotation}\label{subsec:spin_age}

\begin{figure}
    \centering
    \includegraphics[width=0.5\textwidth]{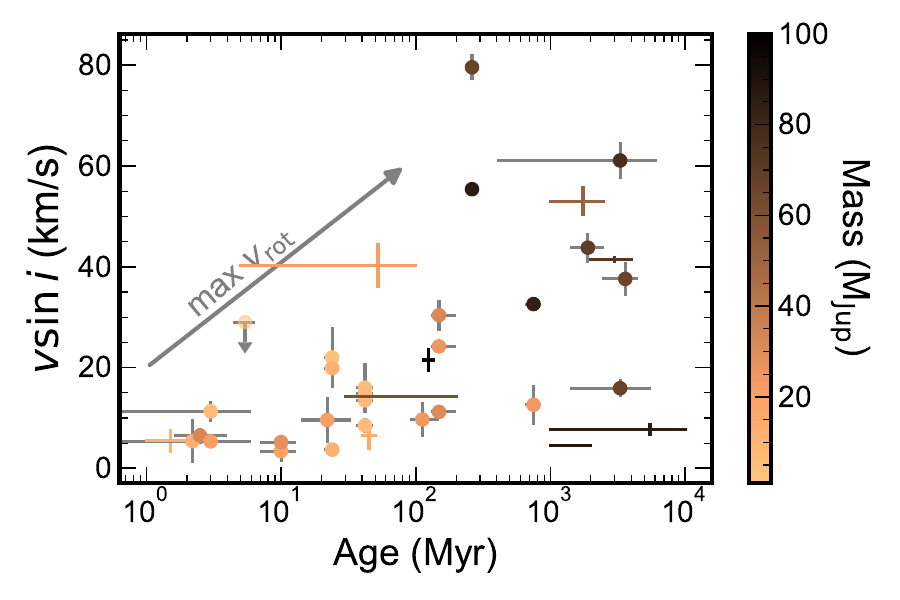}
    \caption{Rotation ({\vsini}) versus age in our KPIC sample, color-coded by mass. Generally our {\vsini} increases toward older ages, but the mass in our sample is biased due to survey sensitivity and overall detection yields for field-age planetary mass objects.
    It is noted that the spin-up trend should be interpreted using the maximum {\vsini} as {\vsini} is dependent on inclination, depicted with the grey arrow.
    See Section~\ref{subsec:spin_age} for details.
    }
    \label{fig:vsini_age}
\end{figure}

Figure~\ref{fig:vsini_age} shows our KPIC {\vsini} across ages, color-coded with their masses.
While the substellar objects in our sample are biased toward young ages (Section~\ref{sec:sample}), the spin-up trend over time can already be seen for objects (4--33~M$_\mathrm{Jup}$) younger than 200~Myr in Figure~\ref{fig:vsini_age}.
This spin-up trend is generally consistent with {\vsini} surveys conducted by \cite{Bryan:2020ab} for 27 objects from planetary mass to brown dwarfs (5--20~{\mjup}; majority of 10--20~{\mjup}) and with the {\vsini} survey for 19 intermediate- to high-mass brown dwarfs (30--70~{\mjup}) by \cite{Zapatero-Osorio:2006aa}, as well as the photometric rotational period survey by \cite{Vos:2020aa, Vos:2022aa} ($<$80~{\mjup}).
The consensus among these spin surveys across ages is that objects below $\sim$70~{\mjup} near the hydrogen mass burning limit spin up over time after age at $\sim$10~Myr \citep{Burrows:1997aa, Chabrier:2023aa}, roughly following constant angular momentum evolution.
The surveys mentioned above were targeted mostly toward young substellar objects or companions.
For field age objects, the {\vsini} survey across spectral types for 275 field low-mass stars and brown dwarfs by \cite{Hsu:2021aa} shows faster rotation from mid-M to T dwarfs, consistent with the general spin-up picture, while the interpretation of spectral type (as a proxy for temperature) is more complex.
The mass, age, and temperature are folded together as spectral type is not a time-invariant quantity:
T dwarfs (all brown dwarfs, generally old at field ages) show faster rotation as they age compared to that of young brown dwarfs (some early to mid-L dwarfs and all late-L dwarfs; \citealp{Hsu:2021aa}).
Late-M dwarfs, mostly low-mass stars, are generally old and more massive and lose their angular momentum much more effectively through magnetized winds than L and T dwarfs.

The general interpretation of this effect is that the weakened magnetized winds associated with these substellar objects make the angular momentum loss inefficient compared to that of low-mass stars \citep{Reiners:2009aa, Irwin:2011aa, Newton:2018aa, Popinchalk:2021aa, Hsu:2024aa}.
Low-mass stars exhibit a spin-down trend toward older ages, albeit the time scale of spin-down is already much longer than that of Sun-like stars \citep{Skumanich:1972aa, Irwin:2011aa, Stassun:2024aa}.
We will revisit the spin-up trend for brown dwarfs and planetary mass objects and explore the transition between the spin-up and spin-down trends toward older ages in Section~\ref{subsec:spin_transition}.

\subsection{Mass Dependent Trend with Rotation}\label{subsec:spin_mass}

As our main goal for this work, built upon previous works by \cite{Wang:2021aa} and \cite{Hsu:2024ab}, we seek to identify rotational signatures of giant planets and brown dwarfs if these two populations share similar or distinct spins.
Two metrics are used to define whether or not a companion is a giant planet: the mass ($<$10~{\mjup}) and the mass ratio ($<$0.8\%).
Only six objects in our sample obey both constraints with sufficient spin measurements, including AF Lep b, HR 8799 bcde, and YSES 1 c, which we regard as our gold standard giant planet sample.
These planets have masses of $\sim$2--7~{\mjup} and mass ratios of $<$0.8\%\footnote{While HIP 79098 AB spectroscopic binary has a total mass of 3.52$\pm$0.06~{\msun} \citep{Zorec:2012aa}, the primary component HIP 79098 A has a mass of 2.5~{\msun} \citep{Janson:2019aa}, making the mass ratio of 1.07\% for HIP 79098 AB b.}.
Other than YSES 1 c, the remaining 5 planets have mass ratios of $<$0.5\%.
Four objects satisfy only one of the two criteria.
Kap And B and beta Pic b, while having higher masses ($M$ $>$10~{\mjup}), share mass ratios $<$0.8\%. 2MASS J1207$-$3932 b and Ross 458 C have masses of $\sim$5--9~{\mjup} with mass ratios of 25\% and 1.5\%, respectively.
The remaining benchmark companions satisfy neither criterion, including ROXs 42 Bb, DH Tau B, HIP 99770 B, GU Psc B, 2MASS J0249$-$0557 c, AB Pic b, VHS J1256$-$1257b, and HD 203030 B.

We compare the spin (parameterized as fractions of breakup velocity {\vovb}; see Section~\ref{subsec:spin_sample_and_method}) versus mass for our benchmark spin sample, where `spin` refers to {\vovb} and {\vovb} at 10~Myr in this subsection.
As giant planets and brown dwarfs are found to spin up over time (Section~\ref{subsec:spin_age}), it is necessary to place our sample at the same age.
The spins at 10 Myr are derived assuming constant angular momentum evolution after their disks dissipate, as the disk ages are typically $<$10~Myr (\citealp{Hillenbrand:2005aa, Mamajek:2009aa}; but see also counter examples in \citealp{Pfalzner:2014aa}). 
The results are summarized in Table~\ref{tab:measurements} and Table~\ref{tab:literature_spin}.

Regardless of assumptions of random or orbit-aligned inclinations, our sample shows some objects with much faster rotation in terms of {\vovb} at their current ages compared to those of \cite{Bryan:2020ab}, who reported that young companions of masses 5--20~{\mjup} (with the majority of their sample being 10--20~{\mjup}) have {\vovb} $\lesssim$0.2, with the only exception being beta Pic b ({\vovb} $\sim$ 0.24).
Additionally, kap And B exhibits a very fast rotation of {\vovb} $\sim$0.47 \citep{Morris:2024aa}, along with our gold sample giant planets HR 8799 bde and AF Lep b show {\vovb} $\gtrsim$0.3.

\begin{figure*}
    \centering
    \includegraphics[width=0.9\textwidth]{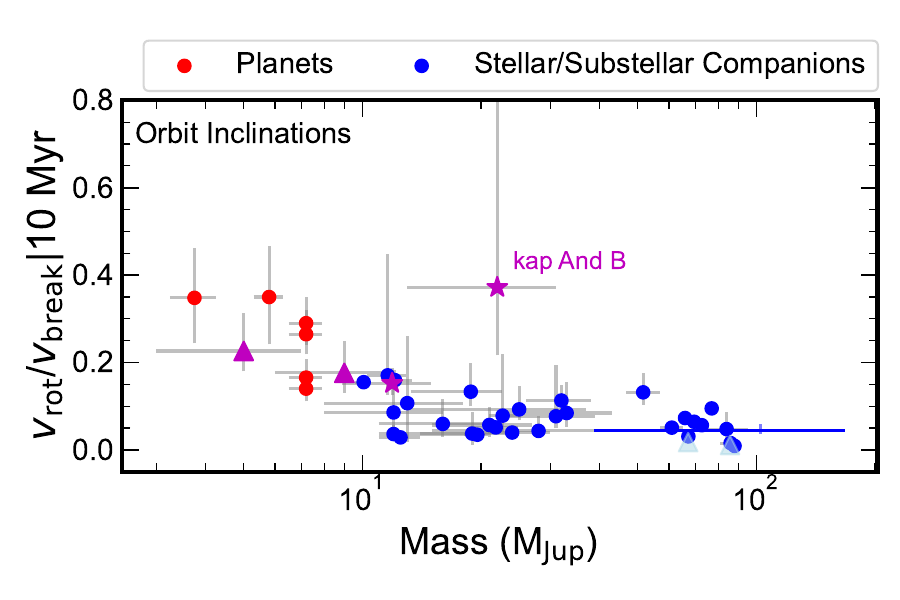}
    \includegraphics[trim=0 0.5cm 0 0.6cm, width=0.9\textwidth]{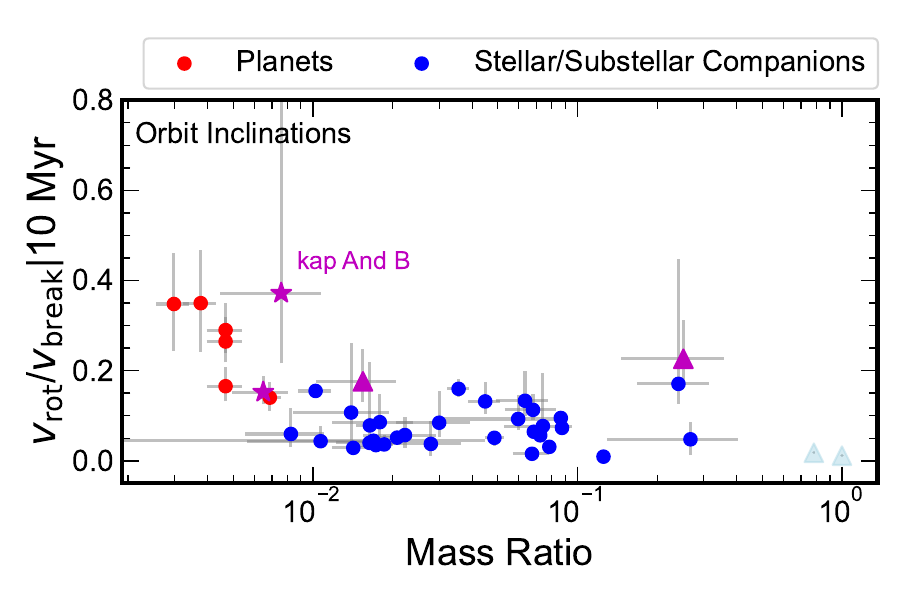}
    \caption{Rotation versus mass (top) and rotation versus mass ratio (bottom) in our benchmark companion spin sample, assuming the inclinations of companions and planets aligned with their orbits.
    The rotation is parameterized by fractions of breakup velocity (i.e., rotational velocity over breakup velocity) at 10~Myr under angular momentum conservation from their current ages.
    The bona fide giant planets are labeled with red dots.
    The companions with mass above 10~{\mjup} and mass ratio $<$0.8\% ($\kappa$ And B and $\beta$ Pic b) are depicted in magenta stars, while the companions with mass below 10~{\mjup} and mass ratio $>$0.8\% (2MASS J1207$-$3932 b and Ross 458 C) are plotted in magenta triangles.
    LP 349-25 AB are depicted in light blue.
    The remaining targets (mass above 10~{\mjup}) are labeled in blue dots. 
    See Section~\ref{subsec:spin_mass} for details.
    }
    \label{fig:vrot_over_vbreak_mass_orbit}
\end{figure*}

\begin{figure*}
    \centering
    \includegraphics[width=0.9\textwidth]{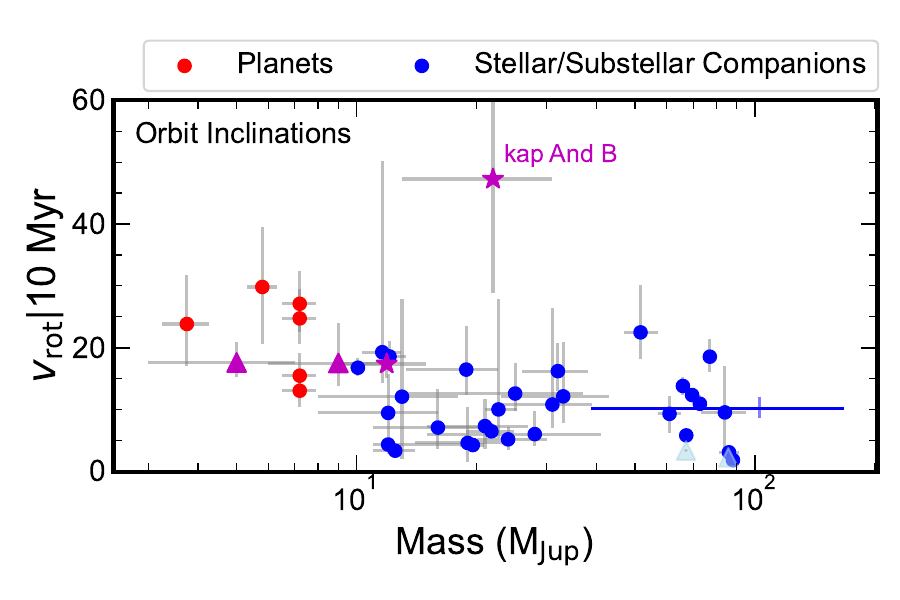}
    \includegraphics[trim=0 0.5cm 0 0.6cm, width=0.9\textwidth]{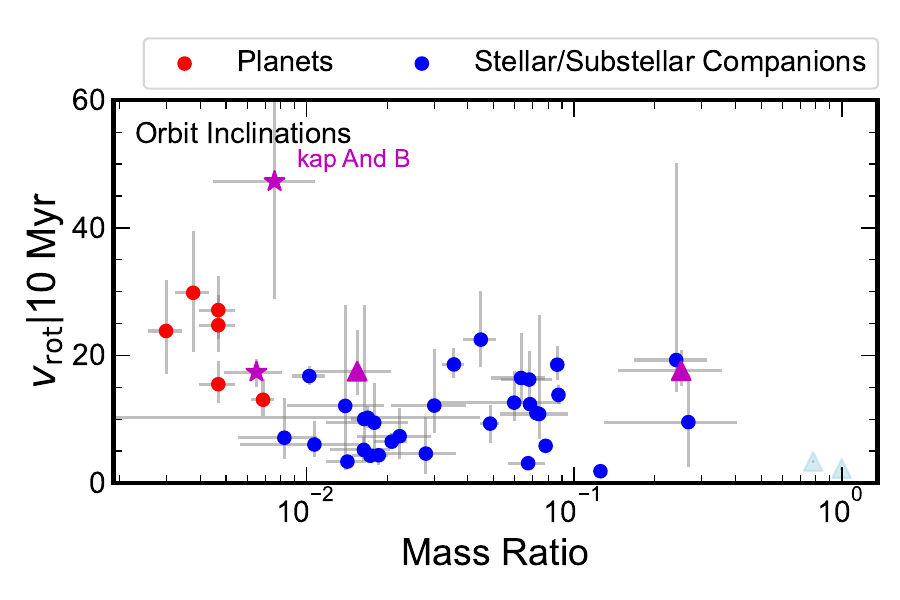}
    \caption{Same as Figure~\ref{fig:vrot_over_vbreak_mass_orbit} with the rotation parameterized by rotational velocities (instead of fractional breakup velocities) at 10~Myr (in the units of {\kms}) under angular momentum conservation since their current ages.
    See Section~\ref{subsec:spin_mass} for details.
    }
    \label{fig:vrot_mass_orbit}
\end{figure*}

We further look into {\vovb} at 10~Myr for our sample.
Figure~\ref{fig:vrot_over_vbreak_mass_orbit} shows the {\vovb} at 10~Myr versus mass and mass ratio assuming no planetary/companion obliquity (i.e., spins are aligned to their orbits).
In the {\vovb} at 10~Myr versus mass space, HR 8799 bcde and AF Lep b ($\sim$2--7~{\mjup}) show a faster rotation compared to companions of masses from 10--30~{\mjup}.
Similar distributions are found when comparing rotational velocities versus mass and mass ratio (Figure~\ref{fig:vrot_mass_orbit}).
Kap And B, mentioned above, is an exception to the slow rotation trend among 10--30~{\mjup} companions, while beta Pic b is transitional among both groups.
When comparing the {\vovb} at 10~Myr in the mass ratio space of 0.8\%, kap And B and beta Pic b with mass ratio $<$0.8\% share similar spins with our gold sample giant planets (Figure~\ref{fig:vrot_over_vbreak_mass_orbit}).
The rotation of our sample is separated more clearly in the mass ratio than in the mass, where objects with mass ratios below 0.8\% appear to show faster rotation than objects with mass ratios above 0.8\%.
For wide planetary mass companions with higher mass ratios, 2MASS J1207$-$3932 b and Ross 458 C have spins consistent with the gold planet sample, while on the higher end of the other group.
This implies that the rotation of the giant planets and companions depends not only on their own properties (mass) but also on their host stars (mass ratios) and their protoplanetary disk properties.

The fast rotation of kap And B was already reported in \cite{Morris:2024aa} and \cite{Xuan:2024ab}. 
When we compute its {\vovb}, it stands out among the companions of similar masses between 10--30~{\mjup}.
It appears that the rotation and mass ratio of kap And B make it more consistent with a planet-like formation scenario.
However, its formation (through gravitational instability or core accretion outside of the CO snowline) is not clear through chemistry or orbital dynamics, as it exhibits solar metallicity and C/O (similar to its host star) \citep{Xuan:2024ab} and high eccentricity (0.88$^{+0.03}_{-0.04}$; \citealp{Morris:2024aa, Bowler:2020aa}).
Similar to kap And B, beta Pic b can be at the higher-mass end of planet formation based on its rotation and mass ratio of 0.65\% \citep{Brandt:2021ad, Dupuy:2019ab, GRAVITY-Collaboration:2020aa}.

As our sample is rather small, careful statistical assessment is required to support the significance of our findings.
Here we consider three criteria to define (superjovian gas giant) planets: (1) a companion that has mass $<$10~{\mjup} and mass ratio $<$0.8\% (our gold sample of giant planets), (2) mass ratio $<$0.8\% (gold planet sample plus beta Pic b and kap And B), and (3) mass $<$10~{\mjup} (gold planet sample plus 2MASS J1207$-$3932 b and Ross 458 C).
Under these definitions, the remaining companions with masses below 40~{\mjup} belong to companion brown dwarf subsamples.
Since the circumplanetary disks are expected to slow down planets or companions, isolated brown dwarfs between 10--40~{\mjup} are compared to companion brown dwarfs to examine whether they share similar or different rotations.
In companions with orbital inclination measurements, these companions with random inclinations and inclinations aligned with their orbits are considered (Section~\ref{subsec:spin_sample_and_method}).
While observations show evidence that the giant planets are more likely to have orbits aligned with their orbits compared to those of brown dwarf companions from stellar obliquity and companion obliquity measurements \citep{Bryan:2020aa, Bowler:2023aa, Poon:2024aa}, no planet obliquity has been measured for planets in our gold standard giant planet sample.
To quantify the significance of the distinct rotation between planets and substellar companions, we drew 1000 samples for our {\vovb} at 10~Myr for each object of each subsample, computed the means of {\vovb} over the target sample, and fit a Gaussian to the resulting distributions of means.

\begin{figure*}
    \centering
    \includegraphics[width=0.45\textwidth]{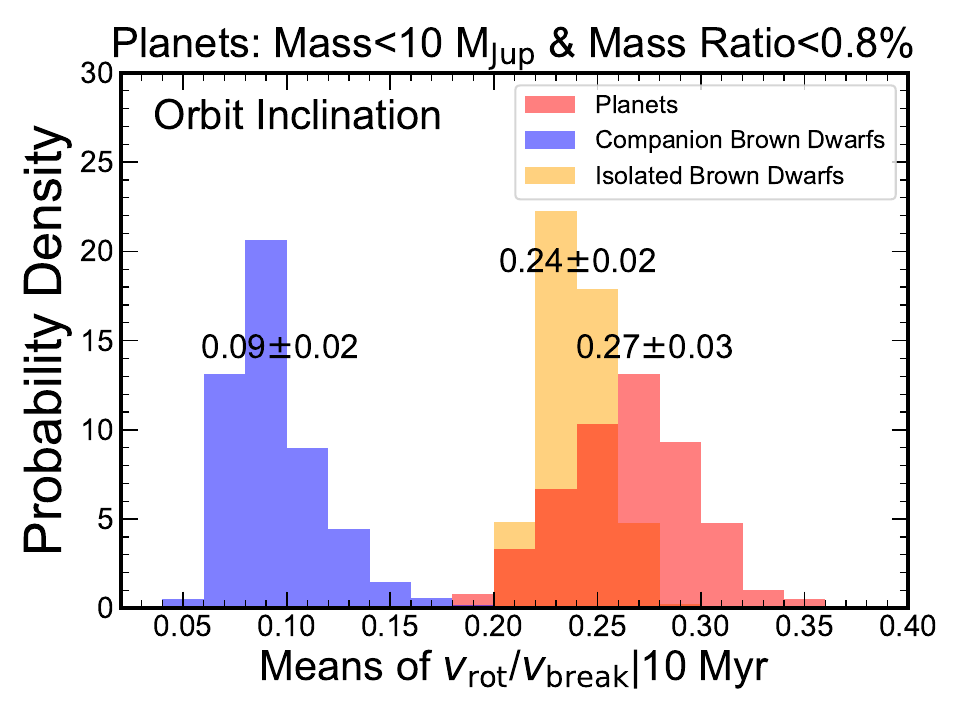}
    \includegraphics[width=0.45\textwidth]{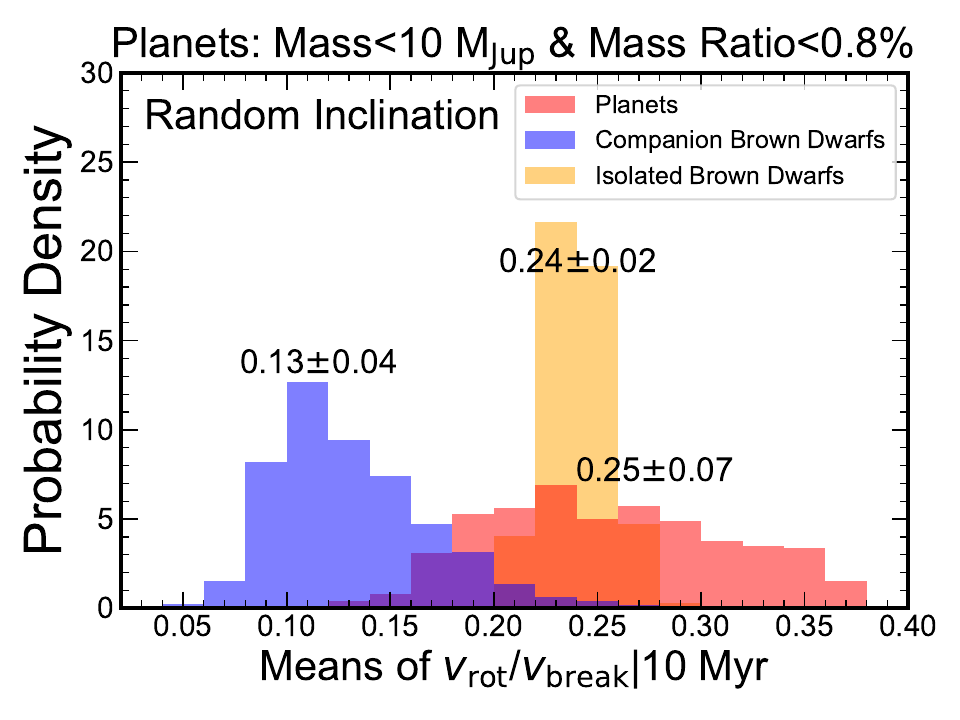}
    \includegraphics[width=0.45\textwidth]{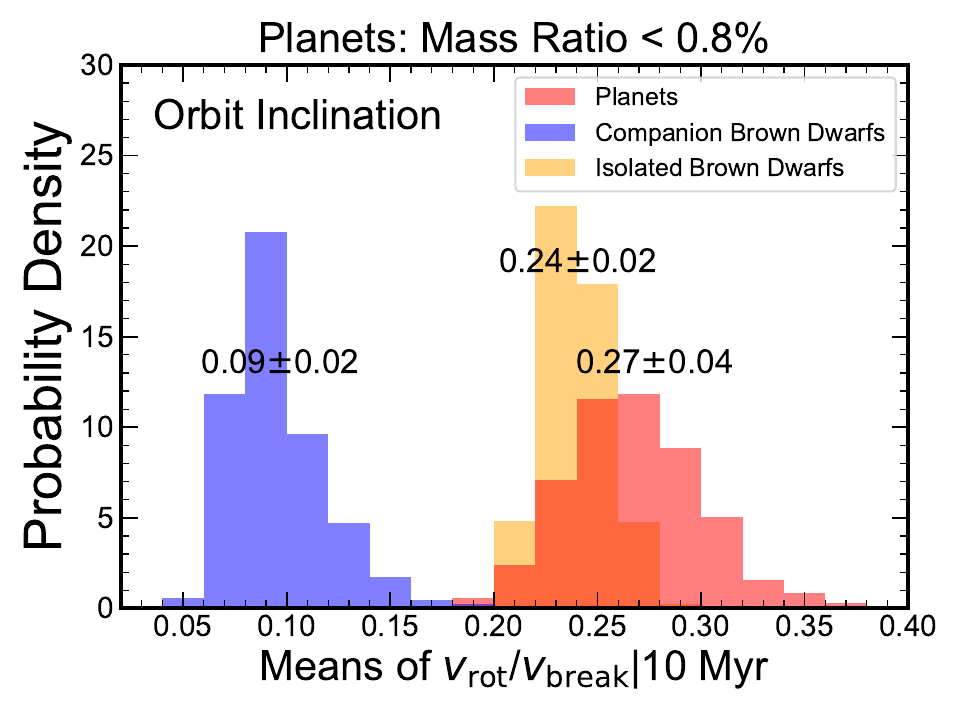}
    \includegraphics[width=0.45\textwidth]{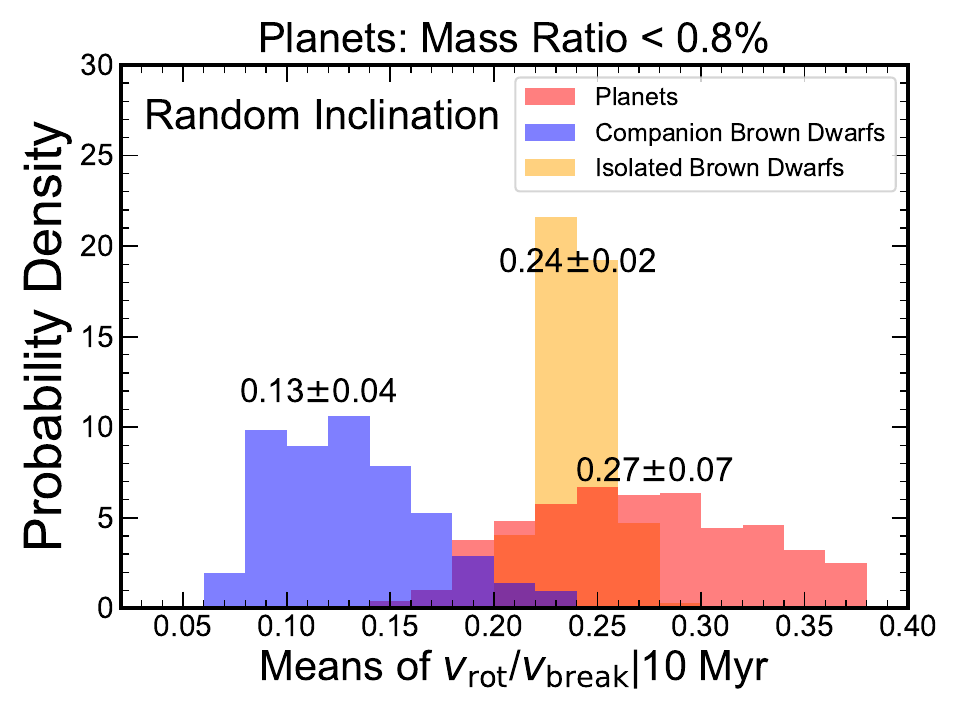}
    \includegraphics[width=0.45\textwidth]{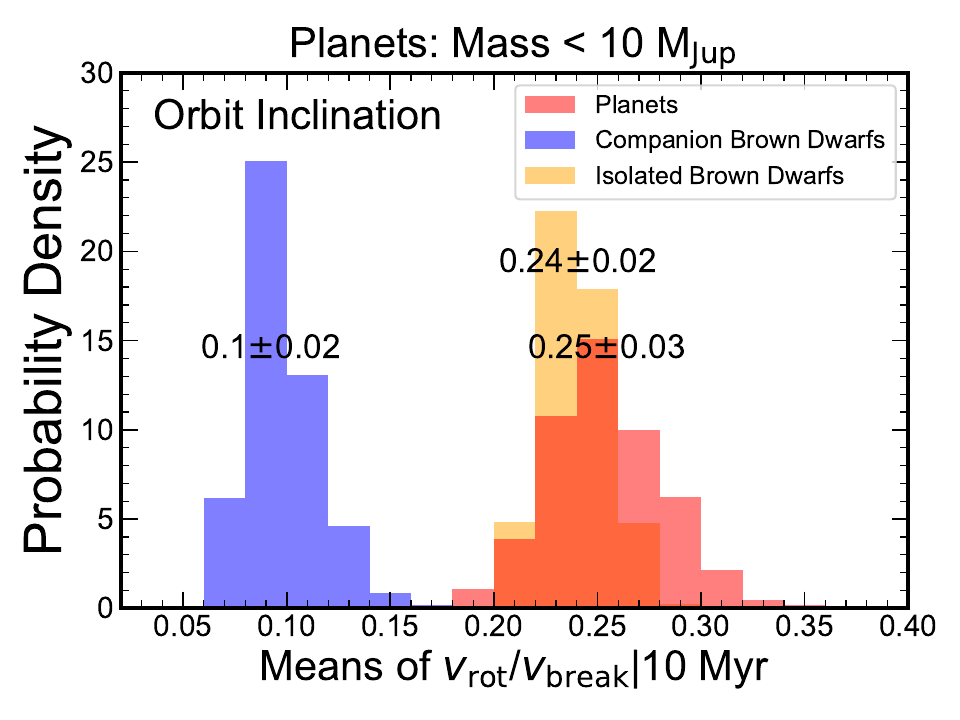}
    \includegraphics[width=0.45\textwidth]{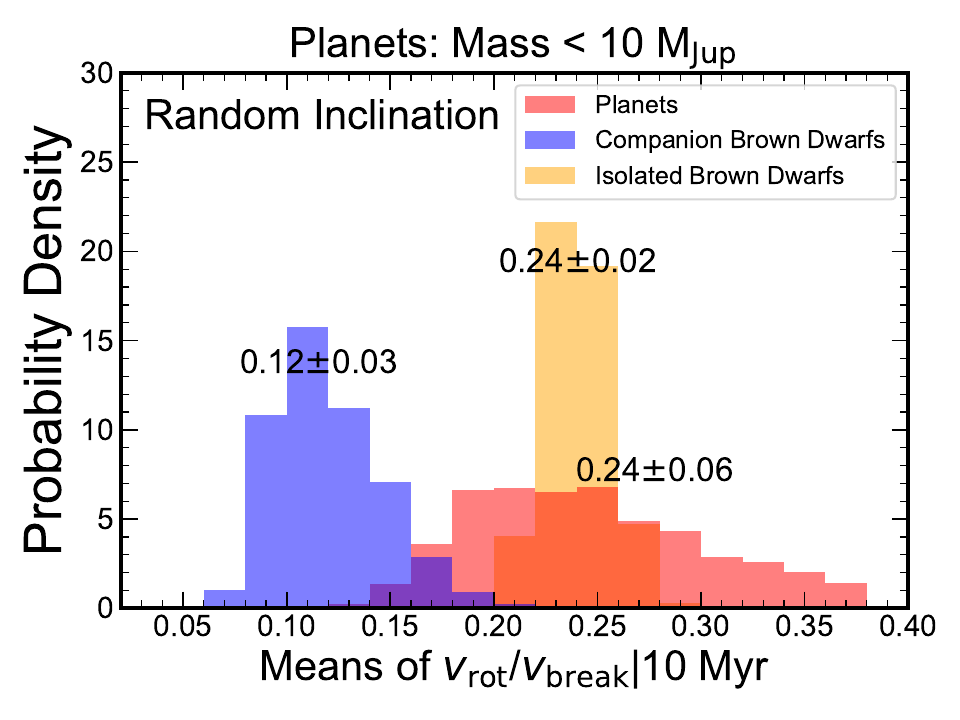}
    \caption{Probability density distributions of the means of rotation for our giant planets (red), companion brown dwarfs (blue), and isolated brown dwarfs (orange) sub-samples.
    The rotation here is parameterized by fractional breakup velocity (i.e., rotational velocity over breakup velocity) at 10~Myr under angular momentum conservation from their current ages, computed assuming inclinations aligned with their orbits (left panels) or random orientations of inclinations (right panels).
    The subsamples that define `planets' are (1) top panels: mass $<$ 10~{\mjup} and mass ratio $<$ 0.8\%, (2) middle panels: mass ratio  $<$ 0.8\%, and (3) bottom panels: mass $<$ 10~{\mjup}, also shown in the title in each panel.
    The best-fit Gaussian values of the means of rotation for the sub-samples are denoted.
    See Section~\ref{subsec:spin_mass} for details.
    }
    \label{fig:occurrence_spin}
\end{figure*}

The resulting distributions of means under these three definitions are shown in Figure~\ref{fig:occurrence_spin}.
The significance of three definitions of planets considering mass $<$10~{\mjup} and mass ratio $<$0.8\% is discussed below.
Under random inclinations, the means of companion brown dwarfs and giant planets have an average {\vovb} at 10~Myr of 0.13 $\pm$ 0.04 and 0.25 $\pm$ 0.07, respectively, whereas
under the assumption of inclinations aligned to their orbit inclinations (orbit inclinations), the means of companion brown dwarfs and giant planets are 0.09 $\pm$ 0.02 and 0.27 $\pm$ 0.03, respectively.
The significance of distinct spins among our low-mass brown dwarfs and giant planets is at 1.6~$\sigma$ for random inclinations and 4.5~$\sigma$ for inclinations aligned with orbits.
The significance of random inclinations is generally lower than that of inclinations aligned with their orbits due to the larger uncertainty of inclinations.
We found that all of our definitions of planets and companion brown dwarfs give moderate significance (1.6--2.1~$\sigma$) under random inclinations and high significance (4--4.5~$\sigma$) under inclinations aligned with their orbits\footnote{Note that our results are highly significant ($p$-value $<$ 0.05) under both assumptions of inclinations using the Kolmogorov–Smirnov test, Anderson–Darling test, and Welch's t-test.}.
We also assessed if the four objects with disks in our KPIC sample could bias our results (Section~\ref{subsec:spin_sample_and_method}). Removal of these four objects gives fully consistent results (4.3~$\sigma$ and 1.7~$\sigma$ for random inclinations and inclinations aligned with their orbits, respectively).

Comparing free-floating companions can provide insights into disk braking in different environments.
Using an additional sample of 54 isolated brown dwarfs of 10--40~{\mjup}, the means of isolated brown dwarfs (0.24$\pm$0.02) show different distributions from companion brown dwarfs, at 2.8~$\sigma$ and 6.3~$\sigma$ significance under random inclinations and no obliquity, respectively, indicating that companion brown dwarfs experience a different rotational history compared to that of isolated brown dwarfs.
This is possibly due to more angular momentum loss during the disk phase in circumplanetary versus proto(sub)stellar disks (e.g., different disk temperatures and ionization levels and thus magnetic torques).
We also compare three free-floating planetary mass objects (2MASS J11193254-1137466AB (2MASS J1119$-$1137AB) \footnote{2MASS J1119$-$1137AB is a confirmed resolved binary of separation of 0$\arcsec$.14 (3.6 $\pm$ 0.9~au; \citealp{Best:2017aa}).}, WISEA J114724.10-204021.3, and PSO J318.5338-22.8603) to our gold planet sample, and found similar levels of {\vovb} at 10 Myr.
This could imply that the disk braking of these two classes has a similar impact on the terminal spins for masses $<$10~{\mjup}.
However, the sample of young (10--100~Myr) planetary-mass objects is very small, especially given that one of the three planetary-mass objects in our sample (2MASS J1119$-$1137AB) is a binary with a fast rotation close to its breakup rotation ({\vovb} of 0.90$^{+0.10}_{-0.08}$ at 10~Myr).
Excluding 2MASS J1119$-$1137AB, the remaining two planetary mass objects still exhibit {\vovb} at 10~Myr of 0.114$^{+0.027}_{-0.018}$ (PSO J318.5338-22.8603) and 0.177$\pm$0.007 (WISEA J114724.10-204021.3), slightly faster than the bulk rotation of brown dwarf companions ({\vovb} of 0.09$\pm$0.02) at 10~Myr.
The binary planetary-mass object 2MASS J1119$-$1137AB is unlikely to lose significant angular momentum over time due to its low temperature at $\sim$1200--1400~K through magnetized winds, compared to the very-low-mass ultrashort-period binary LP 413-53 AB at 2600--2900~K \citep{Hsu:2023aa}.
However, our sample size of free-floating planetary-mass objects is too small, hindering any robust statistical analysis. Future discoveries and characterizations of more single and binary young planetary mass objects will allow us to assess their angular momentum evolution compared to gas giant planets as well as how the circumbinary disk around planetary mass objects affect their terminal rotation.
Finally, while not included in our previously defined planet sample to compute the significance, Jupiter and Saturn are fully consistent with our planet spin, with {\vovb} at $\sim$0.3 and $\sim$0.4, respectively.

Our sample, despite being the largest spin survey for giant planets and benchmark stellar/substellar companions since \cite{Bryan:2020ab}, of 43 stellar/substellar companions, requires more objects to fully validate the trend, especially for the gold giant planets sample, which currently has only 6 objects (AF Lep b, HR 8799 bcde, YSES 1 c), as well as three isolated planetary mass objects $<$10~{\mjup}.
Additionally, the substellar objects ($\lesssim$40~{\mjup}) in our sample are highly biased toward young ages (Figure~\ref{fig:sample}).
While substellar objects are shown to be qualitatively consistent with constant angular momentum evolution over time, their rotational evolution exhibits a significant spread and does not strictly follow constant angular momentum evolution tracks \citep{Vos:2022aa}.
Indeed, only the epsilon Indi Ab and 14 Her c are confirmed imaged giant planets at field ages \citep{Matthews:2024aa, Bardalez-Gagliuffi:2025aa}, and these field age giant planets generally require next-generation instruments and Extremely Large Telescopes to measure the spins.

\subsection{Interpretation: Mass-dependent Angular Momentum Loss During Disk Phase}\label{subsec:interpret}

As we observationally identify the distinct spins between giant planets and brown dwarfs, we seek to interpret what fundamentally sets different spins between these two populations.

The most obvious interpretation is the mass-dependent angular momentum loss since their formation, as substellar objects appear cooler and thus less coupled with the magnetic field of the circumplanetary disk, resulting in less angular momentum loss \citep{Batygin:2018aa}.
The mass, which sets the energy flux budget, governs how hot giant planets and brown dwarfs start and stay at a certain temperature, and ultimately determines the efficiency of the magnetic field coupling to the disk and resulting angular momentum loss.
The magnetic field strength of a giant planet or brown dwarf, driven by metallic hydrogen as a conducting fluid, scales up with mass \citep{Christensen:2009aa}.
This scaling relation holds from Jupiter, to brown dwarfs, to fully convective stars \citep{Christensen:2009aa}, but \cite{Kao:2016aa} reported that the 10--30~{\mjup} brown dwarfs are found to have $>2.5$ kG magnetic fields could be higher than the predicted magnetic field strengths from the \cite{Christensen:2009aa} relation.
While there exist no magnetic field strength measurements of extrasolar giant planets $<$10~{\mjup}, the expected magnetic field strengths are expected to be below $\sim$1 kG and decrease over time using \cite{Reiners:2010ab} and \cite{Baraffe:2003aa} models.
The circumplanetary disks at sufficiently high temperatures ($\gtrsim$1000~K) ionize alkali metals to generate magnetic induction and thus torque to slow down planets or brown dwarf companions in the first tens of thousands of years, well before the typical disk dissipation timescale of $\lesssim$10~Myr \citep{Batygin:2018aa}.
Under this picture, brown dwarf companions could experience further angular momentum loss during the disk phase around circumplanetary disks compared to isolated brown dwarfs with their proto(sub)stellar disks of different disk properties, resulting in slower rotation compared to free-floating brown dwarfs, as observed in our sample (Section~\ref{subsec:spin_mass}). 

After disks dissipate, the most efficient way of losing angular momentum is through substellar winds. 
Based on \cite{Baraffe:2003aa} models, planetary mass objects cool under $\sim$1500~K after 10~Myr, while 10--30~{\mjup} brown dwarfs have {\teff} $\sim$ 1800--2700~K at the first 10~Myr.
Additionally, more massive objects around 20--30~{\mjup} can stay hotter than 1500~K for the first few hundred million years to enable more active (substellar) magnetic winds to further remove angular momentum, as objects below 1500~K are rarely reported to have radio detections (e.g., \citealp{Kao:2016aa, Kao:2018aa, Rose:2023aa}).
While more massive and thus hotter objects can have more efficient angular momentum loss through substellar winds, brown dwarfs exhibit lower rates of observed radio emission/flares compared to those of low-mass stars.
Thus, the rotational evolution for brown dwarfs (and likely for giant planets) generally spins up over time because their radii contract and do not lose much angular momentum after disks dissipate \citep{Zapatero-Osorio:2006aa, Vos:2022aa}.

As discussed in Section~\ref{subsec:spin_mass}, two targets in our sample have mass ratios $<$0.8\%, including beta Pic b and kap And B.
The rotation of beta Pic b ($v_\mathrm{rot}$ = 17$\pm$2~{\kms} and {\vovb} = 0.15$^{+0.04}_{-0.03}$ at 10 Myr) is consistent with the bulk population in mass and mass ratio.
Specifically, beta Pic b is on the fast rotation side among 10--30~{\mjup} but on the slower rotation side among the mass ratio $<$0.8\% in the planet sample, reaffirming its transitional in mass and mass ratio.
However, the picture depicted above cannot explain the fast rotation of kap And B with a high mass of 22$\pm$9~{\mjup} \citep{Wilcomb:2020aa, Currie:2018aa}.
We speculate that the fast rotation of kap And B is due to a weaker magnetic braking of its circumplanetary disk, as the protoplanetary disk temperature and density models are a function of mass ratio expressed in the Hill radius \citep{Armitage:2010aa} used in \cite{Batygin:2018aa}.
As its host star kap And A is massive (2.78$\pm$0.1~{\msun}) compared to typical stars that host giant planets (HR 8799, AF Lep, and YSES 1) in our sample, the magnetic fields of its circumplanetary disk are similar to those of other giant planets.
This could lead to weaker magnetic field coupling and torque and thus weaker disk braking for kap And B.

Thus, we find the spins (versus mass and mass ratio) to be a useful probe of planet formation in addition to orbital eccentricities \citep{Bowler:2020aa}.
While the fundamental reasons of potentially distinct populations of spins and eccentricities are not identical, one is due to mass (and temperature) and the other is due to orbital dynamics of protoplanetary (or protostellar) disks.

\subsection{Angular Momentum Evolution of Very-low-mass Objects}\label{subsec:spin_transition}

As we showed in Section~\ref{subsec:spin_age}, brown dwarfs and planetary mass objects show spin-up relations over time \citep{Zapatero-Osorio:2006aa, Bryan:2020ab, Vos:2022aa}.
The consensus is that low-temperature objects do not lose angular momentum efficiently compared to fully convective M stars from 0.1--0.35~{\msun} \citep{Irwin:2011aa}, consistent with the expectation of constant angular momentum evolution after their disks dissipate within the first 10~Myr of their ages.

The spin-up trend has been identified via both {\vsini} and photometric rotational periods in the literature.
The first evidence of such a spin-up trend was identified in \cite{Zapatero-Osorio:2006aa} using Keck/NIRSPEC with a sample of 19 M6.5 to T8 dwarfs, whose {\vsini} values spin up between 100~Myr to field age.
Similar results were identified in the photometric rotational period survey by \cite{Vos:2020aa, Vos:2022aa} with a sample size of $\sim$140 from 1 to $>$1000~Myr, and the {\vsini} survey in \cite{Bryan:2020ab} for masses between 5--20~{\mjup} (majority of 10--20~{\mjup}) with a sample size of 27 between 1--1000~Myr, as well as in this work (Section~\ref{subsec:spin_age}).

For the low-mass stars, the spin evolution is the opposite; the low-mass stars spin down over time, albeit the time scale is much longer than the sun-like stars (\citealp{Irwin:2011aa}; 0.1~{\msun} $< M \leq$ 0.35~{\msun}).
Recently, the APOGEE {\vsini} survey conducted by \cite{Hsu:2024aa} also showed an initial spin up due to contraction to $\sim$10~Myr, followed by a spin-down trend (see their Figure~16) in their sample (with the majority of their objects being stars, from 80~{\mjup} to $>$0.1~{\msun}) but their targets are mostly members of Taurus, Upper Scorpius and Hyades clusters.

\begin{figure*}
    \centering

    \includegraphics[width=0.49\textwidth]{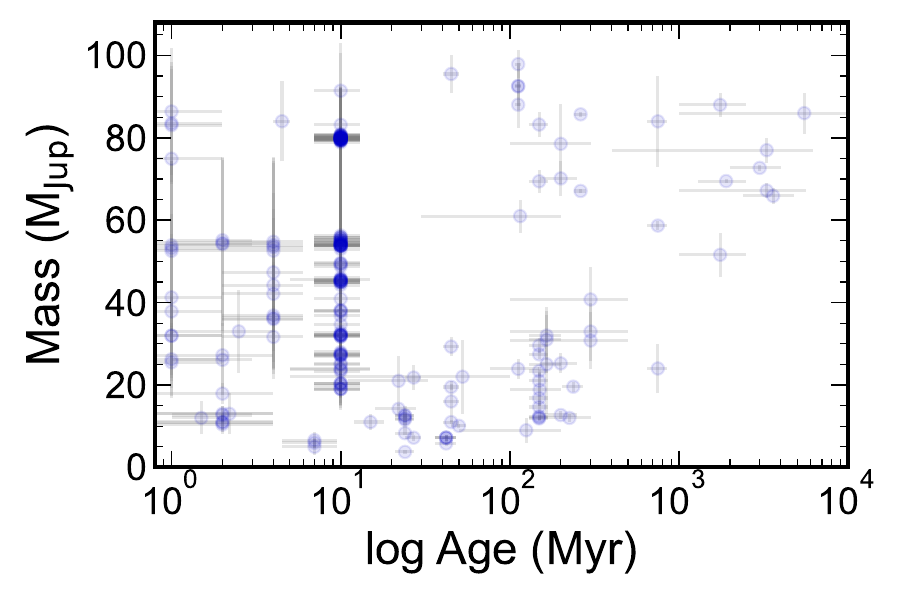}
    \includegraphics[width=0.49\textwidth]{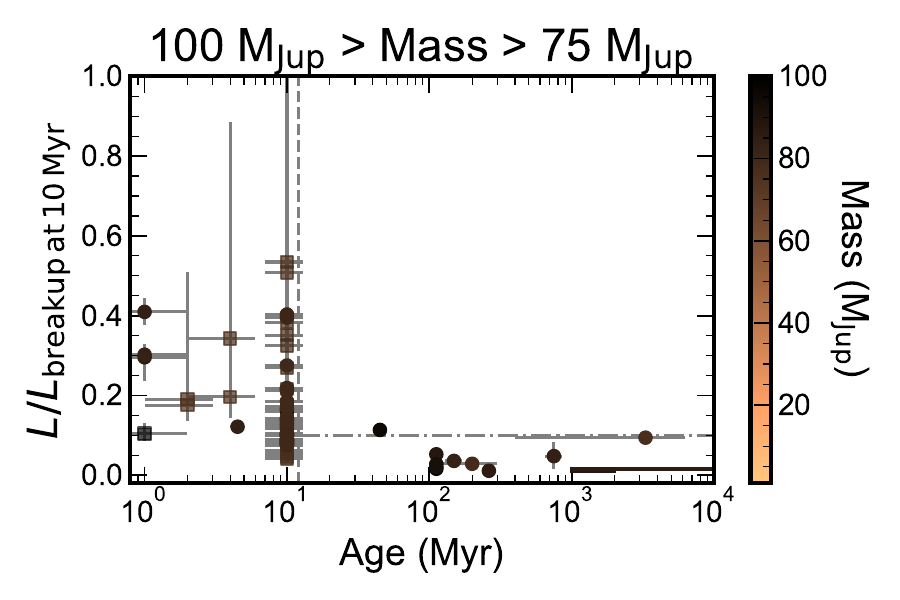}

    \includegraphics[width=0.49\textwidth]{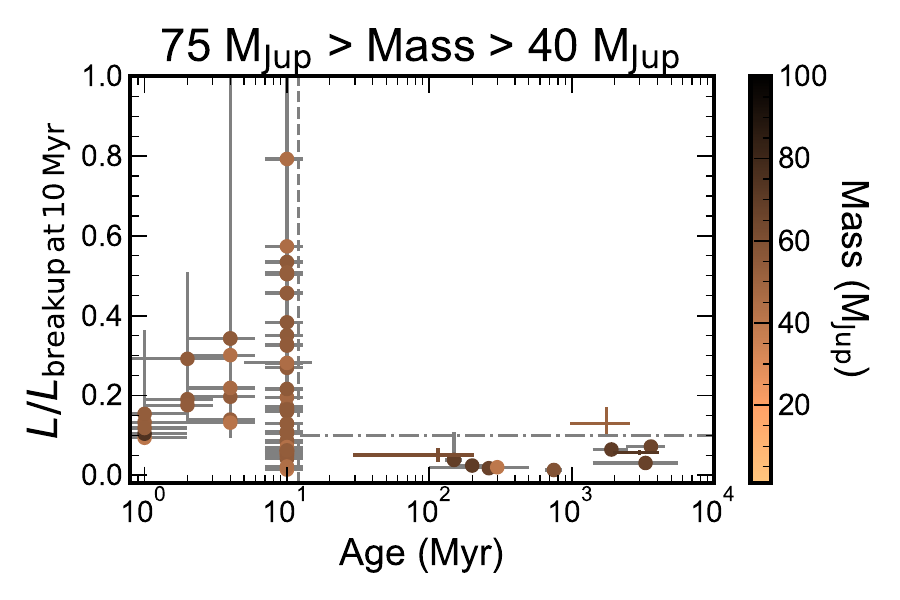}
    \includegraphics[width=0.49\textwidth]{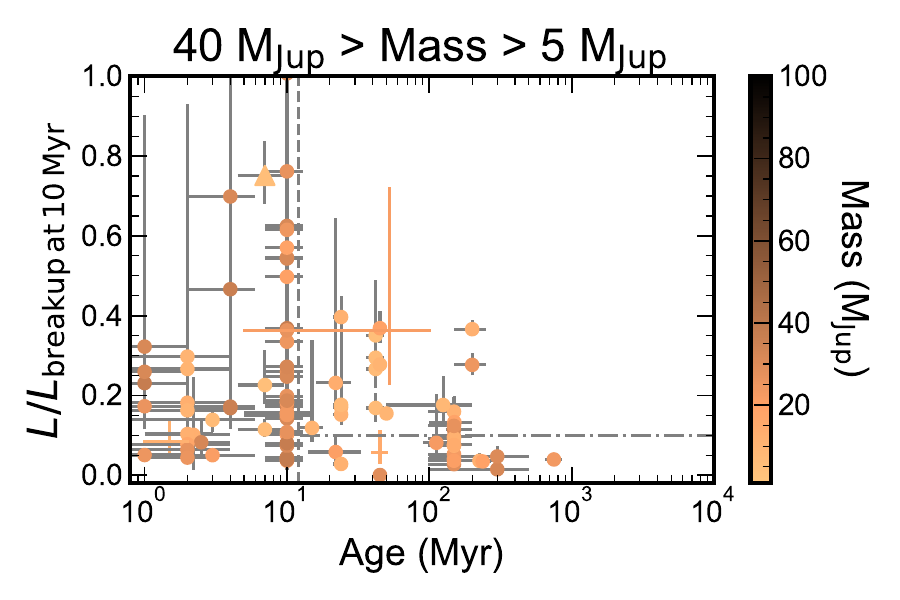}

    \caption{Mass versus age distributions and normalized specific angular momentum versus age for various mass bins in our sample.
    \textit{Upper left panel}: Mass versus log age (Myr) distributions for our full spin sample (blue), for a total of 221 objects.
    \textit{Remaining panels}: Specific angular momentum ratio ($L/L_\mathrm{breakup \, at \, 10 \, Myr}$) versus log age (Myr) across various mass bins for our spin sample, color-coded by mass, for lowest-mass stars of masses from 75--100~{\mjup} (upper right panel), high-mass brown dwarfs of masses from 40--75~{\mjup} (lower left panel), planets and low-mass brown dwarfs of masses from 5--40~{\mjup} (lower right panel).
    For the 100--75~{\mjup} plot, we included the young objects ($\leq$10~Myr) with mass and uncertainties above 75~{\mjup} in squares, while the rest of the objects are labeled in circles.
    The specific angular momentum ($L$) of an object is computed using its (present-day) rotational velocity multiplied by the model radius.
    The denominator, specific angular momentum of its breakup speed at 10 Myr ($L_\mathrm{breakup \, at \, 10 \, Myr}$) assumes the same mass (i.e., no significant mass loss over time) and the radius inferred from the model at 10 Myr.
    The vertical dashed line (at 12~Myr for better visualization) on these three panels indicates the typical disk lifetime ($<$10~Myr).
    The horizontal dashed-dotted line denotes the specific angular momentum ratio of 10\%.
    This 10\% specific angular momentum ratio separates the angular momentum evolution of two distinct populations:
    Objects below 40~{\mjup} retain much higher angular momenta compared to objects above 40~{\mjup} after 10 Myr, when their initial angular momenta were set.}
    See Section~\ref{subsec:spin_transition} for details.
    \label{fig:spin_transition}
\end{figure*}

These spin-up and spin-down trends imply differences in the stellar/substellar wind-driven angular momentum loss efficiency as a function of mass and temperature.
We seek to identify where the angular momentum loss becomes inefficient for objects with masses ranging from 5~{\mjup}--0.1~{\msun}.
Our inferred rotational velocities are derived using the method in Section~\ref{subsec:spin_sample_and_method}, assuming random distributions of inclinations.
For the literature spin sample, we consider objects younger than 1 Gyr except for companions whose ages can be independently measured from their host stars, as the field objects lack robust ages, making it more difficult to identify the trend without knowing the objects' masses and ages.
Our selection criteria result in a full sample of 221 objects, including 34 objects in our benchmark spin sample (which excludes PDS 70 b due to its not well-constrained spin and HD 1160 B due to its uncertain mass) and 187 objects in the literature, with the mass and age distributions shown in the upper-left panel of Figure~\ref{fig:spin_transition}.
In the literature spin measurements, the more massive objects at young ages are filled between 1--10~Myr in addition to our benchmark spin sample, as well as between 100~Myr--1~Gyr.

To study the angular momentum evolution (or more specifically, angular momentum loss) of our sample, we consider the normalized specific angular momentum, instead of the common parameters such as rotational velocities or periods.
The main reason is that the measured rotational velocities or periods are not time invariant, as a function of radius and age, making the inference of angular momentum loss very challenging (Appendix~\ref{sec:literature_spin_appendix}).
Assuming the rotating objects act as rigid bodies and the density profiles do not change over time (i.e., a polytropic index of $n=1.5$ for fully convective objects), the specific angular momentum ($L = v_\mathrm{rot} r$) is a function of rotational velocity $v_\mathrm{rot}$ and radius $r$ at its current age. 
As the maximum angular momentum of an object is limited by the accretion phase and how much angular momentum is lost during the disk phase, we assume the maximum possible angular momentum as the specific angular momentum of its breakup velocity at 10~Myr ($L_\mathrm{breakup \, at \, 10 \, Myr}$).
The normalized specific angular momentum is the ratio of specific angular momentum to its specific angular momentum of breakup velocity at 10~Myr.

Figure~\ref{fig:spin_transition} shows a representative set of our analysis.
Three regimes are discussed below, including the lowest-mass stars (75 to 100~{\mjup}), high-mass brown dwarfs (40 to 75~{\mjup}), and low-mass brown dwarfs and planets (5 to 40~{\mjup}).
For all three mass bins, the normalized specific angular momenta before 10~Myr are high normalized specific angular momenta ($L/L_\mathrm{breakup \, at \, 10 \, Myr} >$0.2), as these objects were still obtaining angular momenta (the initial spin-up phase).
However, after the initial 10~Myr, objects above 40~{\mjup} (high-mass brown dwarfs and lowest-mass stars) show a much lower normalized specific angular momenta ($L/L_\mathrm{breakup \, at \, 10 \, Myr} < 0.1$), whereas several objects below 40~{\mjup} (low-mass brown dwarfs and planets) exhibit a much higher normalized specific angular momenta ($L/L_\mathrm{breakup \, at \, 10 \, Myr}$) between 0.2 and 0.4.
The average (0.14) and median (0.12) $L/L_\mathrm{breakup \, at \, 10 \, Myr}$ values for the 5 to 40~{\mjup} bin are $>$3.5 times higher than the other two mass bins (with the average and median values of 0.04 and 0.03 for both 40 to 75~{\mjup} and 75 to 100~{\mjup} bins).
This implies that objects below 40~{\mjup} are more likely to retain much more angular momentum after their dissipation.

We attribute the distinction of the two populations at 40~{\mjup} to the result of two effects, the magnetic field of the object and the frequency of flares for cooler objects.
The magnetic field \citep{Christensen:2009aa, Reiners:2010ab} is a function of mass, luminosity, and radius, which determines the angular momentum budget in the disk phase. 
After disk dissipation, the most effective way to remove their angular momentum is through stellar/substellar winds and flares, which are more inefficient for lower mass objects and cooler temperatures due to the low occurrence rates of H$\alpha$, radio emission, and flares for mid-L and T dwarfs \citep{Pineda:2016aa, Kao:2016aa, Paudel:2018ab}.
However, several questions remain unanswered due to the limitations of our sample.
First, the exact boundary of the two populations might occur around 30--40~{\mjup}, but the small sample size of 4 objects with ages older than 10~Myr within 30--40~{\mjup} precludes further analysis.
Next, as shown in Figure~\ref{fig:spin_transition}, there are very few objects above 40~{\mjup} within 10--100~Myr. To assess the efficiency and timescale of angular momentum loss for objects above 40~{\mjup}, we require more objects within this mass range between 10--100~Myr.
For objects below 40~{\mjup}, it is not clear if they indeed sustain their rotation on a Gyr timescale, which requires older and evolved benchmark companions and planets.
Finally, our sample includes objects formed under various mechanisms, including binaries \citep{Bate:1997aa}, cloud fragmentation, disk instability, and core accretion, which set different angular momentum budgets and possibly different angular momentum evolution histories.

While our study is the first attempt to distinguish the angular momentum loss using specific angular momentum with the largest rotation sample from 5 to 100~{\mjup} in the literature to date, we stress that a future larger sample is required to validate this transition, especially objects of masses between 40--100~{\mjup} at 10--200~Myr, and companions in the brown dwarf mass regime with low-mass ratios (e.g., kap And B).
Other secondary issues include a closer examination of possible biases of the spins derived between rotational periods and {\vsini}, and a more sophisticated treatment of radius inflation of very-low-mass stars and brown dwarfs \citep{Carmichael:2023aa, Hsu:2024aa, Kiman:2024aa, Cao:2025aa}.
To assess these effects, a much larger sample is needed across different ages that are underrepresented in our sample, requiring more discoveries and characterizations of benchmark brown dwarfs and giant planets \citep{Rothermich:2024aa, Matthews:2024aa, Bardalez-Gagliuffi:2025aa, Lammers:2025aa}.

\section{Conclusion}\label{sec:summarize}

In this work, we present {\vsini} measurements of 32 stellar/substellar objects and directly imaged giant planets, including 25 stellar/substellar companions (12--88~{\mjup}) and 6 bona fide giant exoplanets (2--7~{\mjup}; mass ratios $<$0.5\%) using Keck/KPIC high-resolution spectroscopy.
Seven of our targets presented in this work are the first spin ({\vsini}) measurements in the literature, and three additional targets (AL Lep b and LP 349-25 AB) are the first measurements with KPIC.
Other objects in the sample are either re-measurements in a uniform way or improved measurements with better epochs (e.g., HR 8799 bcde planets).
We employed a forward-modeling method by fitting the host star contribution and companion emission using the BT-Settl or Sonora Bobcat self-consistent atmospheric model grids (see \citealp{Hsu:2024ab}), to uniformly constrain their {\vsini}.
Our median {\vsini} uncertainty is 2.4~{\kms}, largely consistent with the literature measurements.
The only two outliers, HR 7672 B and HD 984 B, can be attributed to fitting or sampling methods or systematic errors that are not accounted for.

Combining additional substellar companions and giant planets with literature spin measurements, we construct the largest spin sample for 43 benchmark substellar companions and giant planets.
We infer their rotational velocities considering both random distributions of inclinations and inclinations aligned with their orbits, and compare their spins as fractional breakup velocities at 10~Myr assuming constant angular momentum evolution.

For the first time, we find solid evidence that giant planets exhibit distinct spin distributions ({\vovb} = 0.27 $\pm$ 0.03) compared to 10--40~{\mjup} brown dwarf companions ({\vovb} = 0.09 $\pm$ 0.02) at 4.5~$\sigma$ under orbit inclinations. Under random inclinations, the spins between giant planets ({\vovb} = 0.25 $\pm$ 0.07) and brown dwarf companions ({\vovb} = 0.13 $\pm$ 0.04) are distinct at 1.6~$\sigma$.
The distinct spins between giant planets hold under various definitions of planets, which we assume combinations of mass $<$10~{\mjup} and/or mass ratio $<$0.8\%, with moderate significance assuming random distributions of inclinations (1.6--2.1~$\sigma$) and high significance assuming inclinations aligned with their orbits (4.0--4.5~$\sigma$).
Kap And B and beta Pic b, while having higher masses compared to the rest of the bona fide giant planets (masses = 2--7~{\mjup} and mass ratios $<$0.8\%), show spins consistent with the planet values and mass ratios consistent with giant planets below 0.8\%.
Our finding indicates that spins can separate giant planets and low-mass brown dwarf companions, and a mass ratio $<$0.8\% provides a cleaner cut than purely mass $<$10~{\mjup}.

The faster spins of planets (along with kap And B and beta Pic b) can be explained by angular momentum loss via circumplanetary disk braking \citep{Batygin:2018aa}.
We also find evidence (at 2.8~$\sigma$ under random distributions and 6.3~$\sigma$ under no obliquity of companion brown dwarfs) that the brown dwarf companions have lower spins ({\vovb}) at 10 Myr compared to isolated brown dwarfs, implying different levels of angular momentum loss during their disk phases.
We find similar spins of free-floating planetary mass objects (3 objects) compared to our giant planet sample (6 objects), indicating that disk braking reduces similar levels of angular momentum for objects below $\lesssim$10~{\mjup}, but more objects are needed to validate our finding.

Finally, using a large sample of 221 stellar/substellar objects and giant planets with spin measurements in the literature, for the first time, we attempt to assess angular momentum evolution using the normalized specific angular momentum for objects from 5 to 100~{\mjup}.
We identify a sharp distinction between the two populations around 40~{\mjup} for objects older than 10~Myr.
After 10~Myr, objects above 40~{\mjup} lost much more angular momentum ($\sim$0.04), while objects below 40~{\mjup} retained $\sim$3.5 times higher angular momentum on average ($\sim$0.14).
While our finding is consistent with our expectation that lower-mass objects have less disk braking and wind-driven angular momentum loss after disks dissipate, more spin measurements across various masses and ages are warranted to validate our results in future work.

Our findings in this work highlight the unique power of diffraction-limited high-resolution spectroscopy mounted on ground-based 10-m class telescopes, which offers {\vsini} measurements of giant planets, including Keck/KPIC \citep{Mawet:2017aa}, VLT/CRIRES+ \citep{Follert:2014aa}, VLT/HiRISE \citep{Vigan:2024aa}, and Subaru/REACH \citep{Kotani:2020aa}, as their high-resolution spectroscopy complements space-based telescopes such as \textit{JWST}.
Next-generation high-resolution spectrometers on 30-m class telescopes such as Keck/HISPEC and TMT-MODIS \citep{Mawet:2024aa} will enable a deeper and larger sample for studying rotational evolution of giant planets, brown dwarfs, and very-low-mass stellar objects.

\facilities{Keck:II (KPIC), Keck:II (NIRSPEC), Keck:II (NIRC2)}

\software{
\texttt{Astropy} \citep{Astropy-Collaboration:2013aa, Astropy-Collaboration:2018aa}, 
\texttt{DYNESTY} \citep{Speagle:2020aa}, 
\texttt{corner} \citep{Foreman-Mackey:2016aa},
\texttt{emcee} \citep{Foreman-Mackey:2013aa},
\texttt{kpic\_pipeline} \citep{Wang:2021aa},
\texttt{Matplotlib} \citep{Hunter:2007aa}, 
\texttt{Numpy} \citep{Harris:2020aa},
\texttt{Scipy} \citep{Virtanen:2020aa}, 
\texttt{SMART} \citep{Hsu:2021aa, Hsu:2021ab},
\texttt{SPLAT} \citep{Burgasser:2017ac}
          }

\section*{Acknowledgments}
The authors thank the anonymous referee for his/her/their insightful reviews, which significantly improved the original manuscript.
The authors thank Brendan Bowler, Jonathan Roberts, and Christopher Theissen for helpful feedback on this work. 
The authors thank the Keck support astronomers and observing assistants for their help in obtaining the Keck/KPIC spectra.
W. M. Keck Observatory access was supported by Northwestern University and the Center for Interdisciplinary Exploration and Research in Astrophysics (CIERA).
J.W.X. was supported by the NASA Future Investigators in NASA Earth and Space Science and Technology (FINESST) award \#80NSSC23K1434. J.W.X is also thankful for support from the Heising-Simons Foundation 51 Pegasi b Fellowship (grant \#2025-5887).
K.H. is supported by the National Science Foundation Graduate Research Fellowship Program under Grant No. 2139433.
This research was supported in part through the computational resources and staff contributions provided for the Quest high-performance computing facility at Northwestern University which is jointly supported by the Office of the Provost, the Office for Research, and Northwestern University Information Technology.
This work used computing resources provided by Northwestern University and the Center for Interdisciplinary Exploration and Research in Astrophysics (CIERA).
Funding for KPIC has been provided by the California Institute of Technology, the Jet Propulsion Laboratory, the Heising-Simons Foundation (grants \#2015-129, \#2017-318, \#2019-1312, and \#2023-4598), the Simons Foundation (through the Caltech Center for Comparative Planetary Evolution), and the NSF under grant AST-1611623.
This work has benefitted from The UltracoolSheet at \url{http://bit.ly/UltracoolSheet}, maintained by Will Best, Trent Dupuy, Michael Liu, Aniket Sanghi, Rob Siverd, and Zhoujian Zhang, and developed from compilations by \cite{Dupuy:2012aa}, \cite{Dupuy:2013aa}, \cite{Deacon:2014aa}, \cite{Liu:2016aa}, \cite{Best:2018aa}, \cite{Best:2021aa}, \cite{Sanghi:2023aa}, and \cite{Schneider:2023ab}.		
This research has made use of the NASA Exoplanet Archive, which is operated by the California Institute of Technology, under contract with the National Aeronautics and Space Administration under the Exoplanet Exploration Program.
Data presented herein were obtained at the W. M. Keck Observatory, which is operated as a scientific partnership among the California Institute of Technology, the University of California, and the National Aeronautics and Space Administration. The Observatory was made possible by the generous financial support of the W. M. Keck Foundation. 
The authors recognize and acknowledge the significant cultural role
and reverence that the summit of Maunakea has with the
indigenous Hawaiian community, and that the W. M. Keck
Observatory stands on Crown and Government Lands that the
State of Hawai'i is obligated to protect and preserve for future
generations of indigenous Hawaiians. 

\appendix

\restartappendixnumbering

\section{Rotation Under Randomly Distributions of Inclinations} \label{sec:spin_appendix}

\begin{figure}[ht!]
    \centering
    \includegraphics[trim=0 0 0 0.0cm, width=0.49\textwidth]{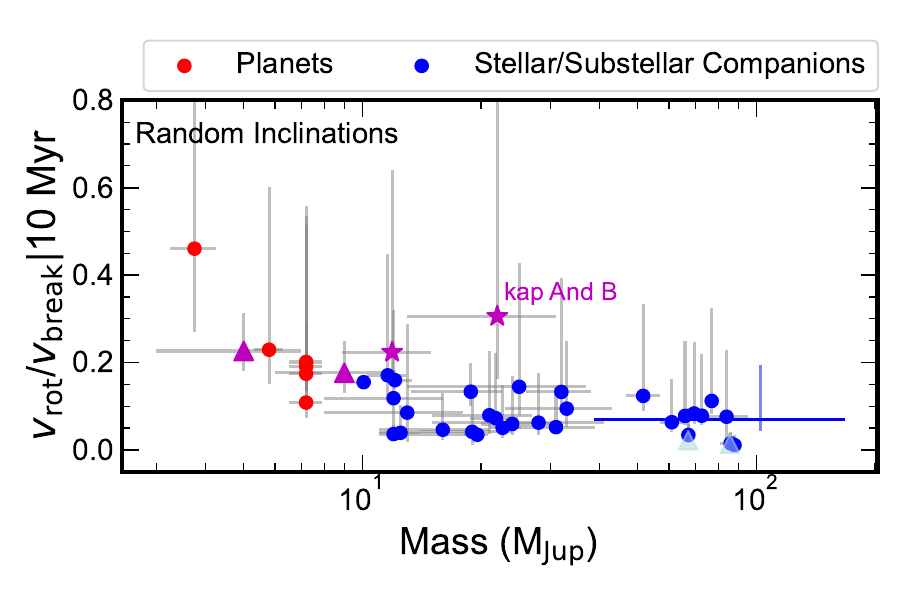}
    \includegraphics[trim=0 0 0 0.0cm, width=0.49\textwidth]{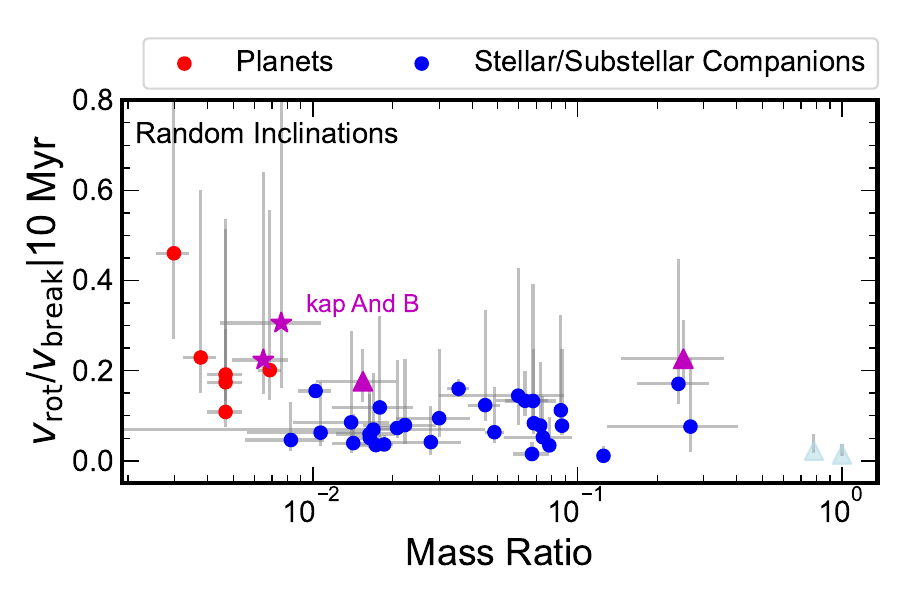}
    \caption{Same as Figure~\ref{fig:vrot_over_vbreak_mass_orbit}.
    The rotation is defined as fractional breakup velocities at 10~Myr under angular momentum conservation from their current ages, assuming random orientations of inclinations.
    }
    \label{fig:vrot_over_vbreak_initial_mass_orbit_uniform}
\end{figure}

\begin{figure*}[ht!]
    \centering
    \includegraphics[trim=0 0 0 0.0cm, width=0.49\textwidth]{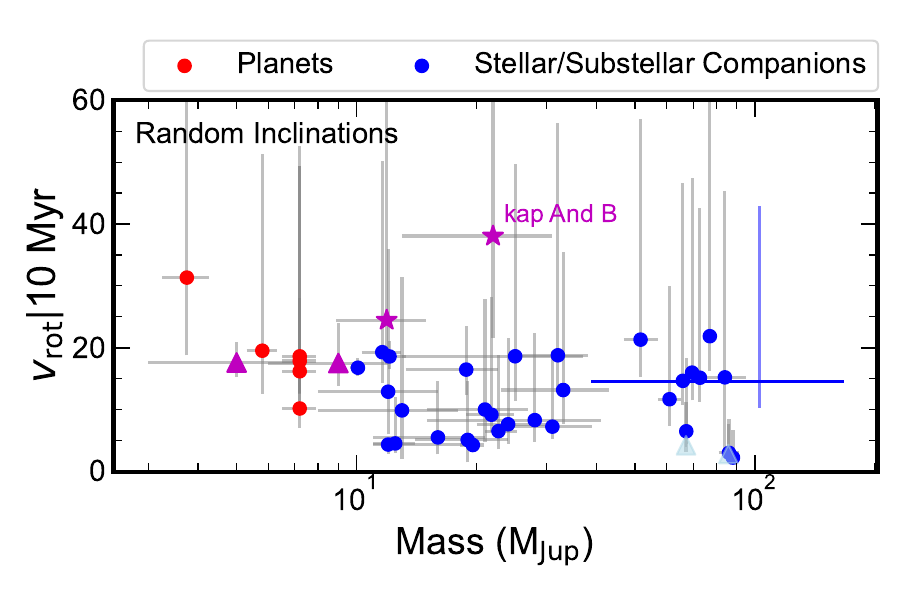}
    \includegraphics[trim=0 0 0 0.0cm, width=0.49\textwidth]{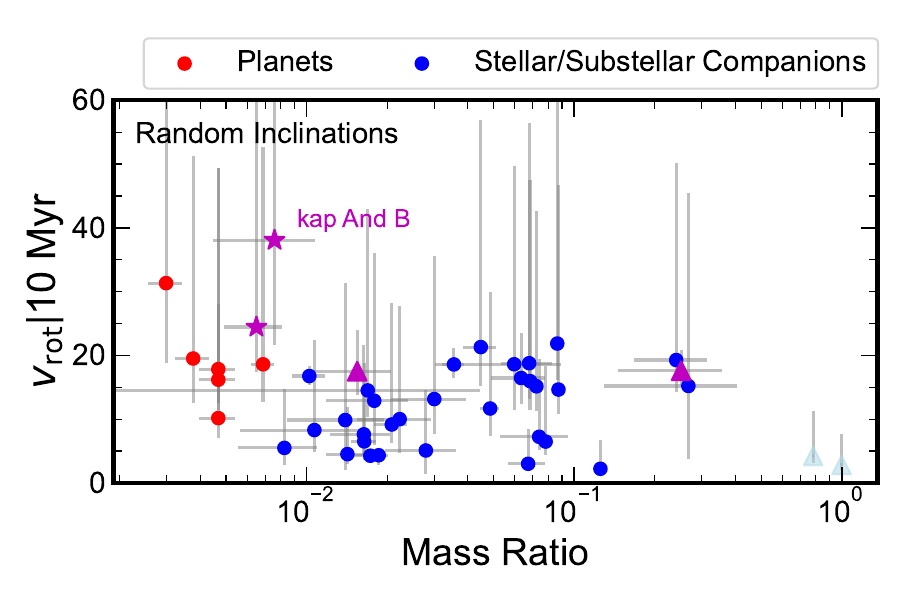}
    \caption{Same as Figure~\ref{fig:vrot_mass_orbit}.
    The rotation is defined as rotational velocities at 10~Myr (in the units of {\kms}) under angular momentum conservation from their current ages, assuming random orientations of inclinations.
    }
    \label{fig:vrot_initial_orbit_uniform}
\end{figure*}

In Section~\ref{sec:rotation} we showed the rotation, parameterized by fractions of breakup velocity at 10~Myr, versus mass and mass ratio, assuming the inclinations of planets and companions in the sample aligned with their orbits (`orbit inclinations'). 
For completeness, we present rotation versus mass and mass ratio assuming random distributions of inclinations.
The fractions of breakup velocity and rotational velocities at 10~Myr, versus mass and mass ratio, assuming their random orientations of inclinations (`random inclinations') are shown in Figure~\ref{fig:vrot_over_vbreak_initial_mass_orbit_uniform} and Figure~\ref{fig:vrot_initial_orbit_uniform}, respectively.

\clearpage
\restartappendixnumbering

\section{HD 72780~B Orbit Fit} 

\begin{figure}[ht!]
    \centering
    \includegraphics[width=0.7\textwidth]{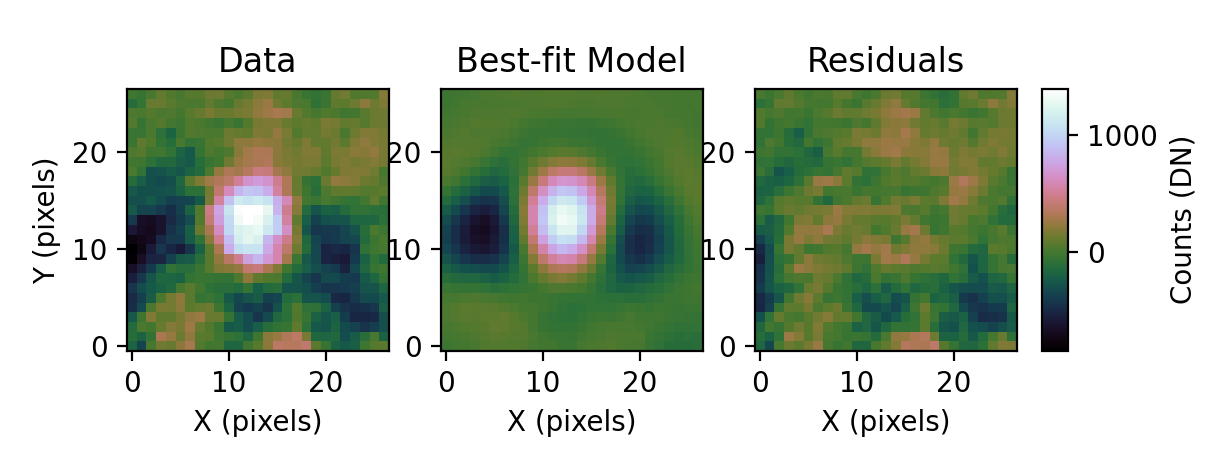}
    \caption{Detection of HD 72780 B from NIRC2/vortex imaging data from UT 2022 April 20 with the $L^\prime$ filter. The companion PSF after ADI subtraction of the stellar PSF is shown on the left. The \texttt{pyKLIP} forward model is shown in the middle, and the residuals of the fit are shown on the right.
    }
    \label{fig:HD72780B_pyklip_fit}
\end{figure}

\begin{figure}[ht!]
    \centering
    \includegraphics[trim=0 0.5cm 0 0.5cm, width=\textwidth]{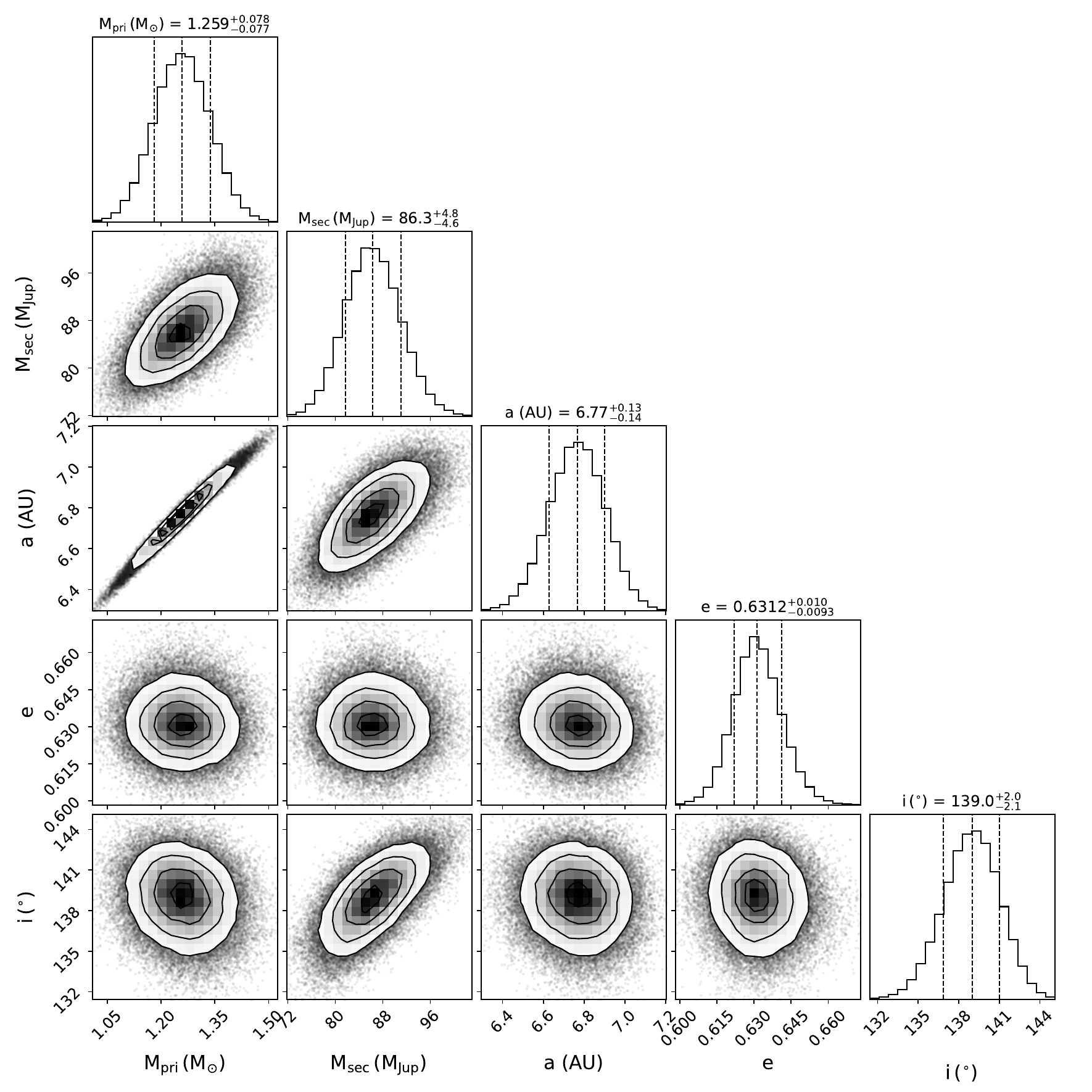}
    \caption{Posterior probability distributions of the major orbital parameters (primary mass (M$_\mathrm{pri}$), secondary mass (M$_\mathrm{sec}$), semi-major axis ($a$), eccentricity ($e$), inclination $i$) of the orbit fit for HD 72780 B.
    }
    \label{fig:HD72780B_orbit_fit}
\end{figure}

In this work, we present the direct detection of HD 72780~B using Keck/NIRC2 vortex imaging in the $L^\prime$-band \citep{Xuan:2018aa}.
HD 72780 B was first discovered from the Doppler technique using stellar radial velocities from Keck and Lick in \cite{Patel:2007aa}. 
HD 72780 A is a solar-metallicity (0.15$\pm$0.03; \citealp{Valenti:2005aa}), field (2.7$\pm$0.9~Gyr) F8 star \citep{Cannon:1993aa}, with an M8 stellar companion of mass 86$\pm$5~{\mjup} at a separation of 180.2$\pm$4.7~mas on UT 2022 April 20.
The NIRC2 data reduction procedure is done in the same way as \cite{Xuan:2024aa}, and we use \texttt{pyKLIP} \citep{Wang:2015aa} to perform angular differential imaging (ADI) subtraction of the host star point-spread function (PSF). We also extract the astrometry and photometry of the companion using the \texttt{pyKLIP} forward model \citep{Pueyo:2016aa} in ADI mode (see Fig.~\ref{fig:HD72780B_pyklip_fit}). After obtaining the relative astrometry, we perform an orbit fit using the host star RVs, NIRC2 relative astrometry, and Gaia-Hipparcos astrometric acceleration \citep{Brandt:2021aa} using \texttt{orvara} \citep{Brandt:2021ab}. The posteriors of the orbit fit are shown in Figure~\ref{fig:HD72780B_orbit_fit}.
We provide the system properties, including the host star HD 72780~A and companion HD 72780~B in Table~\ref{table:hd72780_system}.

\begin{deluxetable}{lcc}[ht!]
\tablecaption{HD~72780 System Properties \label{table:hd72780_system}}
\tablecolumns{3}
\tablehead{
\colhead{Property (unit)} &  \colhead{Value}  & \colhead{Ref.}
}
\startdata
\multicolumn{3}{c}{HD~72780~A}\\
\hline
R.A. (J2000) & 08:35:04.20 & (1) \\
Dec. (J2000) & $+$11:17:01.4 & (1) \\
$\mu_{\alpha}$ (mas yr$^{-1}$) & $-52.51 \pm 0.16$ & (1) \\
$\mu_{\delta}$ (mas yr$^{-1}$) & $-50.87 \pm 0.13$ & (1) \\
Mass (M$_{\odot}$) & 1.26$\pm$0.08 & (19) \\
Age (Gyr) & 2.7$\pm$0.9 & (1), (2)--(11), (17) \\
SpT & F8 & (12) \\
\textit{Gaia} $G$ & 7.348$\pm$0.003 & (1) \\
$J_\mathrm{\, MKO}$ (mag) & 6.462 $\pm$ 0.018 & (13) \\
$H_\mathrm{\, MKO}$ (mag) & 6.29 $\pm$ 0.04 & (13) \\
$K_{\rm S, \, MKO}$ (mag) & 6.236 $\pm$ 0.017 & (13) \\
$\pi$ (mas) & 19.06 $\pm$ 0.15 & (1) \\
distance (pc) & 52.5 $\pm$ 0.4 & (1) \\
RV ({\kms}) & $28.31 \pm 0.15$ & (14) \\
{\vsini} ({\kms}) & $6.4 \pm 0.4$ & (4), (15), (16), (17) \\
{[}Fe\slash H{]} & 0.15$\pm$0.03 & (4) \\
\hline
\multicolumn{3}{c}{HD~72780~B}\\
\hline
Mass (M$_\mathrm{Jup}$) & 86 $\pm$ 5 & (17) \\
SpT & M8$\pm$1 & (17) \\
$M_{L^{\prime}}$ (mag) & 10.06 $\pm$ 0.13 & (17) \\
$L^{\prime}$ (mag) & 13.66 $\pm$ 0.13 & (17) \\
{\teff} (K) & $2527^{+40}_{-30}$ & (17) \\
{\logg} (cm s$^{-2}$ & $5.47^{+0.02}_{-0.04}$ & (17) \\
{\vsini} ({\kms}) & $7.7^{+1.0}_{-1.2}$ & (17) \\
RV ({\kms})\tablenotemark{a} & $25.8^{+0.2}_{-0.3}$ & (17) \\
$a$ (au) & 6.77$^{+0.13}_{-0.14}$ & (17) \\
$e$ & 0.6312$^{+0.010}_{-0.009}$ & (17) \\
$i$ (deg) & 139$\pm$2 & (17) \\
$P$ (yr) & 15.2$^{+0.7}_{-0.6}$ & (17) \\
\hline
\multicolumn{3}{c}{NIRC2 Astrometry and Photometry} \\
\hline
Sep. (mas)\tablenotemark{b} & $180.2\pm4.7$ & (17) \\
PA (deg)\tablenotemark{b} & $356.03\pm1.56$ & (17) \\
$\Delta m$\tablenotemark{b} & $7.46\pm0.10$ & (17) \\
\hline
\enddata
\tablenotetext{a}{Barycentric RV measured on MJD 59897.627}
\tablenotetext{b}{measured on MJD 59689.2 (UT 2022-04-20)}
\tablerefs{(1) \cite{Gaia-Collaboration:2023aa}; (2) \cite{Marsakov:1995aa}; (3) \cite{Feltzing:2001aa}; (4) \cite{Valenti:2005aa}; (5) \cite{Takeda:2007aa}; (6) \cite{Holmberg:2009aa}; ; (7) \cite{Casagrande:2011aa}; (8) \cite{Gontcharov:2012aa}; (9) \cite{Pace:2013aa}; (10) \cite{Aguilera-Gomez:2018aa}; (11) \cite{Llorente-de-Andres:2021aa}; (12) \cite{Cannon:1993aa}; (13) \cite{Cutri:2003aa}; (14) \cite{Nidever:2002aa}; (15) \cite{Rice:2020aa}; (16) \cite{Llorente-de-Andres:2021aa}; (17) This work}
\end{deluxetable}

\clearpage
\restartappendixnumbering

\section{Literature Sample Rotation}
\label{sec:literature_spin_appendix}

In Section~\ref{subsec:spin_transition}, we identified the distinct angular momentum loss after 10~Myr for objects from 3--40~{\mjup} and from 40--100~{\mjup}.
To place the literature spin surveys to the same context, we compare the rotational velocity and angular velocity versus age for subsamples binned at different masses.

\begin{figure*}[h!]
    \centering

    \includegraphics[width=0.49\textwidth]{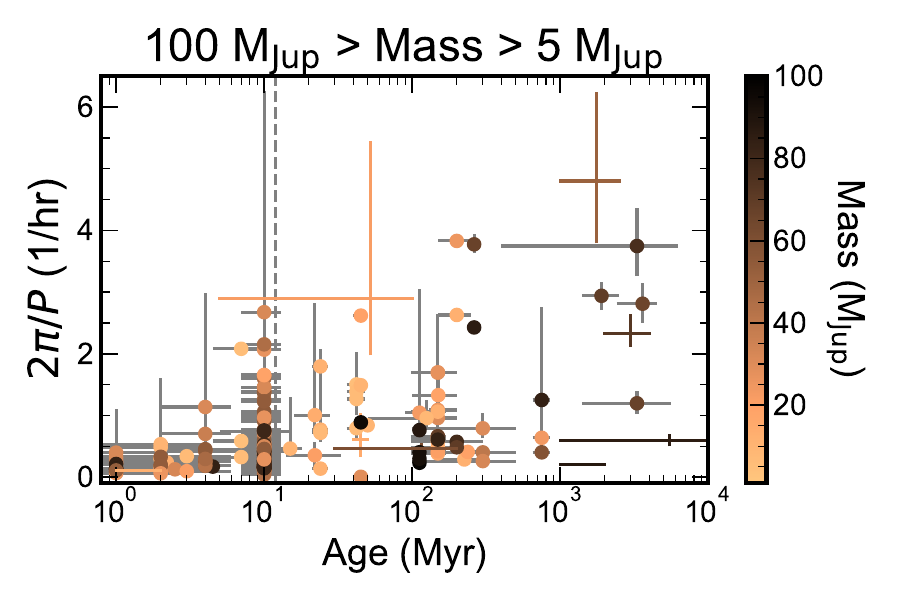}
    \includegraphics[width=0.49\textwidth]{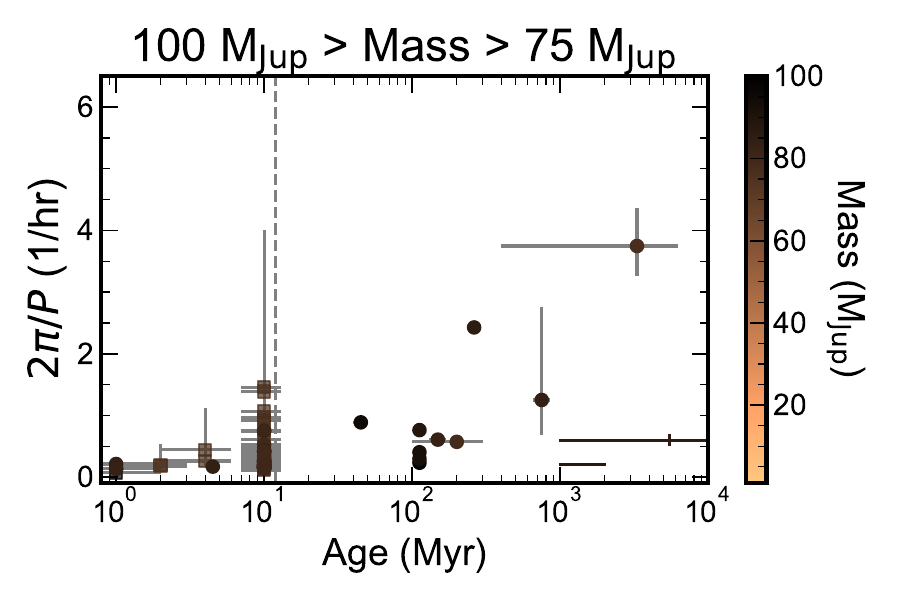}

    \includegraphics[width=0.49\textwidth]{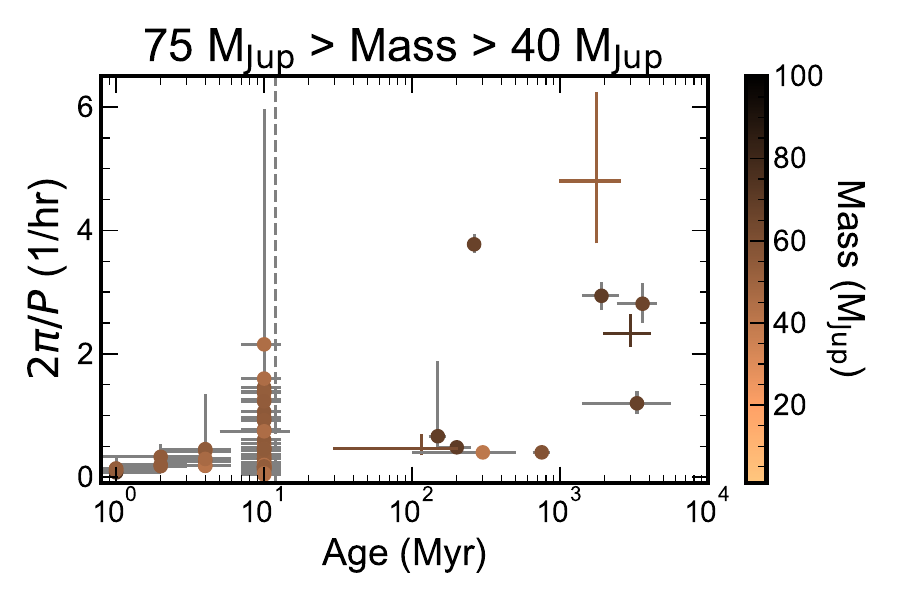}
    \includegraphics[width=0.49\textwidth]{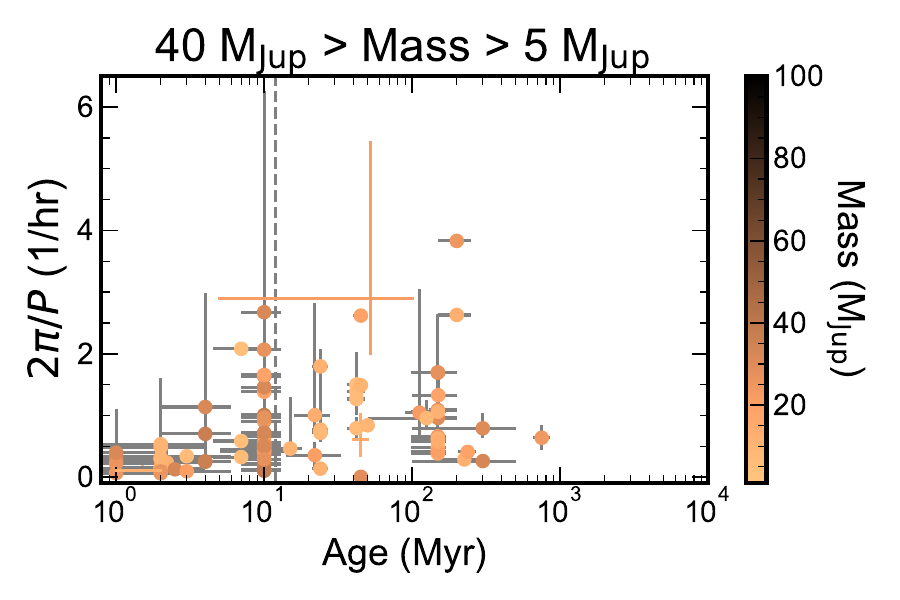}

    \caption{Angular velocity ($\omega = 2\pi/P$ in hr$^{-1}$) versus log age (Myr) across various mass bins, color-coded by mass, for our full sample (upper left panel), objects of masses from 100--75~{\mjup} (upper right panel), objects of masses from 75--40~{\mjup} (lower left panel), and objects of masses from 40--5~{\mjup} (lower right panel).
    For the 100--75~{\mjup} plot, we included the young objects ($\leq$10~Myr) with mass and uncertainties above 75~{\mjup} in squares, while the rest of the objects are labeled in circles.
    The vertical dashed line (at 12~Myr for better visualization) on these three panels indicates the typical disk lifetime ($<$10~Myr).
    The general angular velocity trends for all mass bins all appear to become faster over time, which is the known ``spin-up'' trend found in \cite{Vos:2022aa}.
    See Section~\ref{subsec:spin_transition} for details.}
    \label{fig:spin_transition_rotational_omega}
\end{figure*}

We first look at the distributions of angular velocity, which is effectively the inverse of the rotational period.
The angular velocity is computed as $\omega = v_\mathrm{rot}/R = 2\pi/P$, where $v_\mathrm{rot}$ is the rotational velocity, $R$ is the radius, and $P$ is the period. The methods to derive these parameters are described in Section~\ref{subsec:spin_sample_and_method}.
Figure~\ref{fig:spin_transition_rotational_omega} shows the angular velocities for the full sample and masses of 3--40~{\mjup}, 40--75~{\mjup}, and 75--100~{\mjup}.
These distributions generally show an increasing trend over time, which is consistent with the finding from \cite{Vos:2022aa}.

\begin{figure*}[h!]
    \centering

    \includegraphics[width=0.49\textwidth]{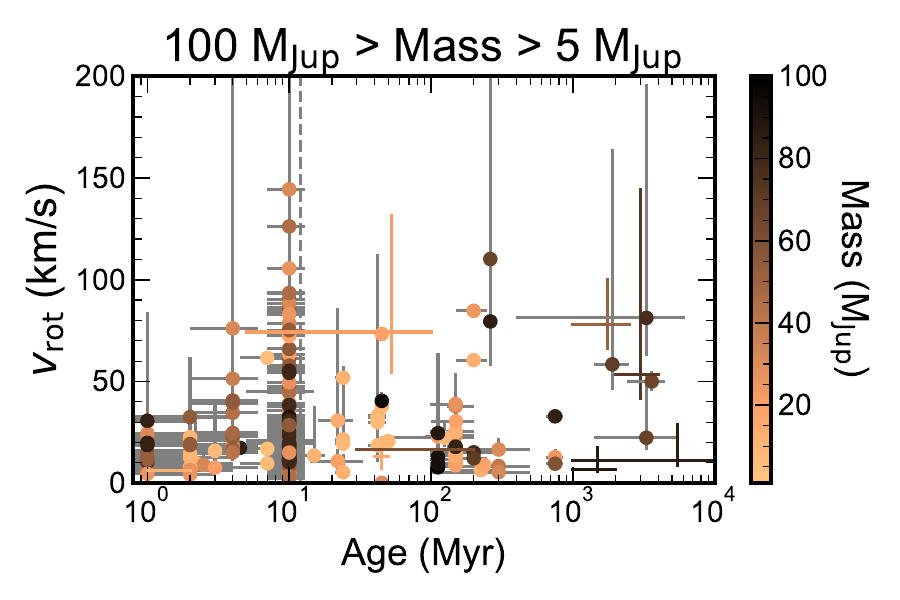}
    \includegraphics[width=0.49\textwidth]{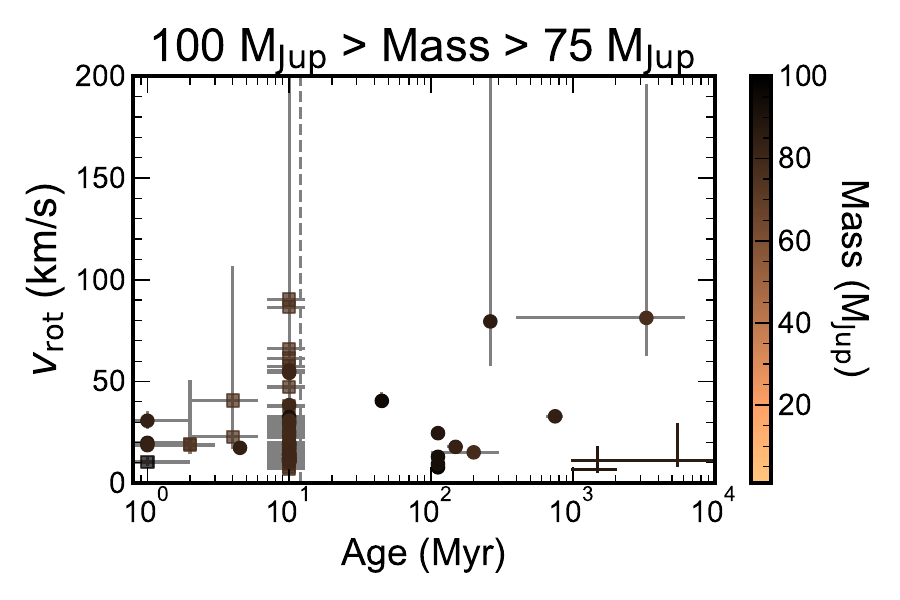}

    \includegraphics[width=0.49\textwidth]{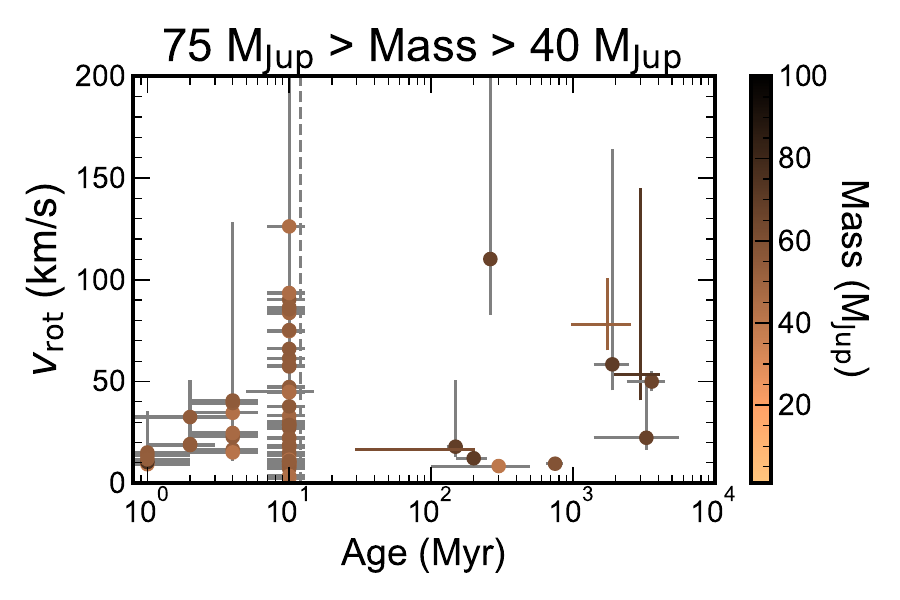}
    \includegraphics[width=0.49\textwidth]{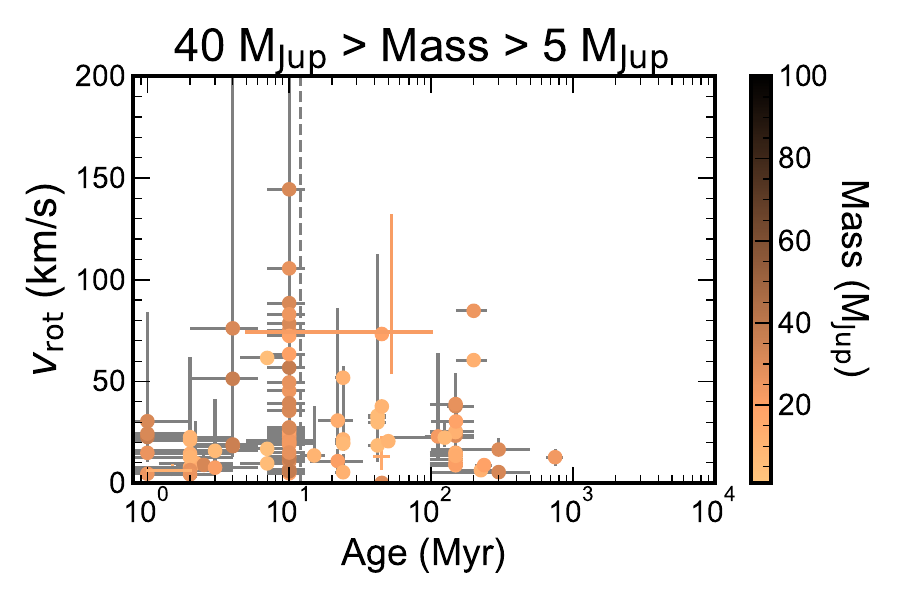}

    \caption{Rotational velocities versus log age (Myr) across various mass bins, color-coded by mass, for our full sample (upper left panel), objects of masses from 100--75~{\mjup} (upper right panel), objects of masses from 75--40~{\mjup} (lower left panel), and objects of masses from 40--5~{\mjup} (lower right panel).
    For the 100--75~{\mjup} plot, we included the young objects ($\leq$10~Myr) with mass and uncertainties above 75~{\mjup} in squares, while the rest of the objects are labeled in circles.
    The vertical dashed line (at 12~Myr for better visualization) on these three panels indicates the typical disk lifetime ($<$10~Myr).
    See Section~\ref{subsec:spin_transition} for details.} \label{fig:spin_transition_rotational_velocity}
\end{figure*}

Next, we examine the distributions of rotational velocities over time for the same mass bins, illustrated in Figure~\ref{fig:spin_transition_rotational_velocity}.
Before 10~Myr, all mass bins show an increasing trend, which is the initial spin-up phase before disks dissipate.
After 10~Myr, the trends for all mass bins are unclear, compared to the angular velocities because the constant evolution of radius along with the rotation, complicates the distributions in rotational velocities.

\clearpage

\bibliography{mylibrary}{}
\bibliographystyle{aasjournal}

\end{document}